**Anna Knezevic**

**Greg Cohen**

**Marina Domanskaya**

# Some Facts on Permanents in Finite Characteristics

**Abstract:**

*The permanent's polynomial-time computability over fields of characteristic 3 for k-semi-unitary matrices (i.e. n×n-matrices A such that* $rank(AA^T - I_n) = k$*) in the case* $k \leq 1$ *and its* $\#_3P$*-completeness for any k > 1 (Ref. 9) is a result that essentially widens our understanding of the computational complexity boundaries for the permanent modulo 3. Now we extend this result to study more closely the case k > 1 regarding the (n-k)×(n-k)-sub-permanents (or permanent-minors) of a unitary n×n-matrix and their possible relations, because an (n-k)×(n-k)-submatrix of a unitary n×n-matrix is generically a k-semi-unitary (n-k)×(n-k)-matrix.*

*The following paper offers a way to receive a variety of such equations of different sorts, in the meantime extending (in its second chapter divided into subchapters) this direction of research to reviewing all the set of polynomial-time permanent-preserving reductions and equations for a generic matrix's sub-permanents they might yield, including a number of generalizations and formulae (valid in an arbitrary prime characteristic) analogical to the classical identities relating the minors of a matrix and its inverse. Moreover, the second chapter also deals with the Hamiltonian cycle polynomial in characteristic 2 that surprisingly demonstrates quite a number of properties very similar to the corresponding ones of the permanent in characteristic 3, while in the field GF(2) it obtains even more amazing features that are extensions of many well-known results on the parity of Hamiltonian cycles.*

*Besides, the paper's third chapter is devoted to the computational complexity issues of the permanent and some related functions on a variety of Cauchy matrices and their*

*certain generalizations, including constructing a polynomial-time algorithm (based on them) for the permanent of an arbitrary square matrix in characteristic 5 and conjecturing the existence of a similar scheme in characteristic 3.*

*Throughout the paper, we investigate various matrix compressions and transformations preserving the permanent and related functions in certain finite characteristics. And, as an auxiliary algebraic tool supposed for an application when needed in all the constructions we're going to discuss in the present article, we'll introduce and utilize a special principle involving a field's extension by a formal infinitesimal and allowing, provided a number of conditions are fulfilled, to reduce the computation of a polynomial over a field to solving a system of algebraic equations in polynomial time.*

# Introduction

Historically the computation of polynomials over finite fields was considered as quiet a special area related to the general theory of computational complexity. It's known that the existence of a polynomial-time algorithm for computing the number of solutions of an NP-complete problem modulo p (i.e. the statement that the complexity class $\#_p P$ is a subset of P) implies the equality RP = NP for any prime p. This fact can be demonstrated via considering, for instance, the Hamiltonian cycle polynomial ham(Z) over a finite field F, where Z is an n×n-matrix, that is a homogeneous polynomial in Z's entries such that each variable's degree is 0 or 1 in each monomial. In the meantime, any polynomial over F in m variables such that each variable's degree is 0 or 1 in each monomial can have no more than $m|F|^{m-1}$ roots over F (it's easy to prove by the induction on m). Hence if we take an n×n-matrix $W = \{w_{i,j}\}_{n\times n}$ over F and, given a digraph G with n vertices whose adjacency matrix is $A_G$, define its weighted adjacency matrix as $A_G \star W$, where $\star$ denotes the Hadamard (entry-wise) product, then the equation for the variables $w_{i,j}$ $\operatorname{ham}(A_G \star W) = 0$ can have no more than $n^2|F|^{n^2-1}$ roots over $|F|$. It implies that if we consider W random then the probability that $\operatorname{ham}(A_G \star W) = 0$ is smaller than $\frac{n^2}{|F|}$ when G is Hamiltonian and 1 otherwise. Moreover, because for all the $2^{n^2}$ possible adjacency matrices $A_G$ we can get, altogether, no more than $2^{n^2}n^2|F|^{n^2-1}$ roots of the equations $\operatorname{ham}(A_G \star W) = 0$, in case if $\frac{2^{n^2}n^2}{|F|} < 1$ it also implies the existence of a matrix W such that for any digraph G with n vertices $\operatorname{ham}(A_G \star W) = 0$ if and only if G isn't Hamiltonian, while the probability of taking such a matrix W randomly is $1-\frac{2^n n^2}{|F|}$ . Accordingly it also demonstrates the known fact that RP is a subset of P/poly.

On the other hand, given a finite field $\mathrm{F}$ of characteristic p, any computational circuit can be polynomial-time represented as a set of relations over $\mathrm{F}$ where each variable is either expressed as the sum or product of two other variables or equaled to a given constant. Such a representation hence calculates a polynomial in the set of given constants. In the meantime, when we extend $\mathrm{F}$ to a bigger field $\widehat{\mathrm{F}}$ we therefore receive an extension of this polynomial to another one over $\widehat{\mathrm{F}}$. And, in case if the initial circuit implements a correct polynomial-time algorithm for solving an NP-complete problem, we accordingly get the question whether the extended polynomial should be $\#_p\mathrm{P}$-complete.

Besides, many polynomial-time algorithms on graphs or digraphs have algebraic representations via a determinant (of a size polynomial in the graph's or digraph's number of vertices) over a finite field. A bright example is the Tutte matrix algorithm for determining the existence of a perfect matching in a graph. Therefore we may say that a wide variety of problems from the class NP can be embedded into computational problems over finite fields and fields of finite characteristics, i. e. into algebraic complexity problems.

A number of attempts to use a prime characteristic's advantages for computing a #P-complete polynomial modulo that characteristic were already performed by many mathematicians. For instance, in their paper "The Parity of Directed Hamiltonian Cycles" (https://arxiv.org/abs/1301.7250 ) Andreas Bjorklund and Thore Husfeldt compute the parity of the number of Hamiltonian cycles of a generic digraph with n vertices in O($1{,}618^{\mathrm{n}}$) time via efficiently using certain unique properties of the field GF(2) and some relevant theorems of graph theory. Likewise, the parity of the number of Hamiltonian decompositions, as the corresponding existence problem is NP-complete too, is also a subject for an intense research yielding such results as Thomason's theorem stating that in a 4-regular graph the number of its Hamiltonian decompositions where a given pair of edges doesn't belong to one cycle is even. This result was extended in (21) by one of the authors.

In the present article we're going to apply all the algebraic machinery of arbitrary fields and rings of some finite characteristics to a number of computationally complete functions over them. Actually, the area of computations in finite fields is somewhat a special field of mathematics also due to the fact that it allows a very efficient and virtually 100% empirical verification (via computer modeling) of any algorithm for computing an analytic (particularly, rational) function. It, perhaps, does distinguish attempts to prove the statement "$\#_p$P is a subset of P" (implying RP = NP) for some prime p via computing a $\#_p$P -complete analytic function (particularly, the permanent etc.) in characteristic p from any other kind of proposed polynomial-time algorithms (either heuristic or deterministic) for an NP-complete problem.

Accordingly, each theorem/lemma of our article can be viewed as "its own proof" because it's either an analytic identity or, vice-versa, non-identity (when we speak about a matrix's rank) and therefore it's easily verifiable via taking random values of its free variables for sufficiently big corresponding vector dimensions and ground fields (though we of course supply our results with "standard" mathematical proofs). It demonstrates the pretty surprising fact that there is a very significant part of mathematics resembling such natural sciences as physics and chemistry (and, perhaps, even biology) where any statement can be firmly proven via a properly arranged experiment.

Generally, we may define an **identity space** as a set of formulas written in a certain formal language and having free variables they are functions in such that for any formula from this set either it's identically zero as a function in its free variables or the portion of vectors of its free variables it's zero in doesn't exceed a constant (i.e. the set's parameter which may be defined as its **identity threshold**) smaller than unity. Hence the question of formulating a requirement a formula should satisfy for to belong to an identity space (i.e. the identity space's defining condition) and proving its identity threshold is parallel to formulating a requirement for a physical, chemical or biological experiment allowing to consider it a statement's proof. Apparently, the class of analytic (i.e. constructed via analytic functions) formulas considered in the present article is a subset of an identity space whose identity threshold is inversely proportional to the ground field's size, although the involved vector dimensions that are special natural number variables yet make it necessary, despite all the available symmetries of "continuous" variables, to prove the corresponding threshold for each formula (due to giving "standard" proofs, we leave it out of the frames of this paper). Anyhow, the notion of an identity space arises quite a number of new questions regarding the very principle of mathematical provability and one can thus see that mathematical statements may vary from non-provable (in accordance with Goedel's incompleteness theorems) to "provable by experiments" (precisely like in natural sciences), provided we prove such a statement's belongingness (upon re-formulating it as a formula) to an identity space whose identity threshold is compatible with the series of verifying experiments. On the other hand, this notion can be easily extended to natural sciences themselves and we may even conclude that many (if not all) parts of the modern physics, chemistry and biology could be formally described as identity spaces what, in turn, possibly can influence the provability/verifiability principles in these sciences as well.

In this regard, we can formulate a much wider metaphysical theory tensely related to the computational complexity issues in mathematics (as well as some other types of complexities, including the Kolmogorov one) and proposing the best possible application of a polynomial-time algorithm for an NP-complete problem.

If we have some data set represented as an m-set $D$ of n-vectors (over a certain ground set S) and an m-set of values $\mathrm{v}_1, \dots, \mathrm{v}_\mathrm{m}$ (over a certain value set) of some unknown function $f(x) = f(\mathrm{x}_1, \dots, \mathrm{x}_\mathrm{n})$ in those vector-variables, then we look for a formula $F(x) = F(\mathrm{x}_1, \dots, \mathrm{x}_\mathrm{n})$ (understood as a computational circuit) written via functions from an alphabet $A$ by function composition such that $\forall x \in D \; F(x) = f(x)$. The number of applications of functions from $A$ in the formula $F$ we'll call ***F*'s descriptive complexity over A** and will denote desc($F$,$A$).

We'll also call:

such a formula $F$ an ***A*-interpolant** of the set $D$ and the set of values $v = (\mathrm{v}_1, \dots, \mathrm{v}_\mathrm{m})$;

the formula $F$ an ***A*-law** of $D$,$v$ if $desc(F,A)$ is minimal among all the interpolants of $D$,$v$ over $A$;

the descriptive complexity of an $A$-law of $D$,$v$ the **descriptability** of $D$,$v$ via $A$ (or its $A$-descriptability) and we'll denote it $mindesc(D,v,A)$.

It could be also useful to know the ratio $|D|$/mindesc($D$,$v$,$A$) and call it the ***A*-credibility** of $D$,$v$. The bigger the $A$-credibility, the more reliable $A$ for $D$.

We yet can extend the above-stated theory through the notion of a $(\mu,\varepsilon,A)$-interpolant of $D$,$v$ defined as a formula $F(x)$ written via functions from an alphabet $A$ by function composition such that $\forall x \in D \; \mu(F(x),f(x)) < \varepsilon$ where $\mu$ is a metric on the set S and $\varepsilon$ is a precision. (All the other above definitions should be modified accordingly in such a case).

Another type of generalization the notion of interpolant can be subjected to is related to the case when instead of the value vector $v$ we consider a set $E = \{e_1, \dots, e_m\}$ of functional equations for unknown functions $f_1(x)$,…, $f_k(x)$ on $D$. In such a case we define an $A$-interpolant of $D, E$ as a set of formulas $F_1(x)$,…, $F_k(x)$ (understood as computational circuits) written via functions from the alphabet $A$ by function composition and satisfying the functional equations $e_1, \dots, e_m$ on $D$, with the corresponding modifications of the notions of minimal interpolant, $A$-law, $A$-descriptability, $A$-credibility and their above-given generalizations (in the case of an infinite $D$ which is a subset of a measurable space its cardinality may be replaced by its measure).

The fundamental principle of this theory is that all the discovered natural (physical, chemical, biological etc.) laws could be well-received via the above-stated scheme upon collecting a necessary data, and we can also assume any law that is to be discovered in the future could be as well.

Let's also notice that in most of practical cases any data sets and unknown functions' values can be represented as binary vectors (up to a certain precision, if necessary). Hence we may reduce the generic case to the Boolean one what allows to choose $A$ equal to the whole set of two unary and 16 binary functions as the most complete (in the descriptability sense) alphabet. In such a case we may conclude that the Boolean representation of data somewhat provides the most perfect way of interpolation and finding "natural laws".

Generically, for each data type there are several reasonable candidates for the alphabet $A$ and for each chosen alphabet we can actually try to find not the shortest formula, but just a short one (when choosing its special shape as well), while also searching for it not as for an $A$-interpolant, but as for an ($\mu$,$\varepsilon$,$A$)-interpolant. For instance, we can choose $A$ to be a set of arithmetical functions over a field and try to find $F$ as a polynomial of a bounded degree whose number of monomials would be as small as possible. Hence, while the problem of finding an "absolute" Boolean law (for the binary representation of data) is naturally NP-complete, for small $n$ it's yet quite possible for many practical needs.

One can notice that the notion of descriptive complexity introduced by the above-stated metaphysical approach does differ from both the Kolmogorov and computational complexities (understood via the Turing machine model), though some of its partial cases (restrictions) may coincide with one of those classical complexities. Moreover, we may try to interpret the computational complexity as the descriptive one applied to mathematics itself, although such a topic would be far out of the frames of our present research whose primary goal is still studying the computational applications of finite fields (while, nevertheless, researching the descriptive complexity over them could be an interesting topic too).

And, when returning to our chief subject of research, it would also be worth noting that the so called "generalized Polya permanent problem" of determining, for a given field F and natural n, the minimal number r(n) such that there exists a linear mapping $M_n(F)$ → $M_{r(n)}(F)$ turning the permanent into the determinant over F generates the conjecture (supported by most of algebraists) that r(n) cannot be bounded from above by a linear (at least) function in n for any fields whose characteristic isn't 2. Although this conjecture wasn't yet rigorously studied for fields of finite characteristics, especially for finite fields, there is a plenty of works concerning so many aspects of the permanent-to-determinant convertibility; for instance, in http://www.law05.si/law14/presentations/Guterman.pdf Alexander Guterman reviews the state-of-art in this area. So, when thinking about any permanent-preserving matrix compressions and other transformations, we should take into account the "self-obligation" that they aren't to lead to a counterexample to that conjecture, even over finite fields. Therefore, when we look for a useful compression of this kind, we usually suppose it can be neither too "fast" nor too "simple" unless it becomes such a counterexample and we thus may conclude that the class of permanent-preserving polynomial-time matrix compressions/transformations is sophisticated enough not to compress/transform too "rudely" for turning the permanent into the determinant.

While the infinite fields of finite characteristics have a certain "smoothness" allowing to use a metric-like mechanics (introduced in the paper) and methods of functional analysis for efficient computations, the finite idempotent rings of characteristic 2 (including GF(2)), though deprived of "analytic smoothness", appear to possess very special useful properties unavailable in any other types of rings. For instance, the classical theorem that the number of Hamiltonian cycles through any given edge is even in any odd-degreed graph receives, under our approach, the following generalization:

over an idempotent ring $\mathcal{R}$ of characteristic 2, in a weighted graph with n+2 vertices whose weighted adjacency matrix is $\begin{pmatrix} A & b & c \\ b^T & 0 & 1 \\ c^T & 1 & 0 \end{pmatrix}$ the sum of the products of the edge weights of Hamiltonian cycles through the edge (n+1,n+2) is zero if

$$c = hb + A\vec{1}_n + \vec{1}_n + (A + \mathrm{Diag}(d))\mathrm{diag}(A + \mathrm{Diag}(d))^{-1} + d$$

for some $d \in \mathcal{R}^n$ such that $\det(A + \mathrm{Diag}(d)) = 1$ and $h \in \{0,1\}$, where $\mathrm{diag}(A + \mathrm{Diag}(d))^{-1}$ is the n-vector of diagonal entries of $(A + \mathrm{Diag}(d))^{-1}$ and $\vec{1}_n$ is the n-vector all whose entries are unity.

In the case when the ground ring is GF(2), d = $\vec{0}_n$ (where $\vec{0}_n$ is the n-vector all whose entries are zero), h = 1, $\vec{1}_n^T b = c^T\vec{1}_n = 0$ and the graph has no loops (i.e. A is a matrix with the zero diagonal) this relation yields an equality for a non-weighted odd-degreed graph that is precisely the above-mentioned classical theorem on the parity of the number of Hamiltonian cycles through a given edge.

Besides, the approach we propose is also able to yield such easily verifiable results as the statement that, for any Boolean symmetric n×n-matrix $X$ with the zero diagonal and n-vector $y$, the number of Hamiltonian cycles of the graph with the adjacency matrix $\begin{pmatrix} X & y \\ y^T & 0 \end{pmatrix}$ equals, modulo 2, the sum of these numbers of the graphs with the adjacency matrices $\begin{pmatrix} X & X\vec{1}_n + \vec{1}_n + y \\ \vec{1}_n^T X + \vec{1}_n^T + y^T & 0 \end{pmatrix}$) and $nX$, as well as the statement that the equality modulo 2 also holds between these numbers of the two graphs with the adjacency matrices $X$ and $(X + D)^{-1}$ for any diagonal Boolean n×n-matrix $D$ when $\det(X + D)$ is odd. The two latter statements (proven in the article's second chapter as generalized for any idempotent ring of characteristic 2) provide quite a powerful instrument for polynomial-time transforming a graph whose parity of the number of Hamiltonian cycles we wish to know into another graph on the same set of vertices and with the same parity of the number of Hamiltonian cycles.

## The neighboring computation principle (for any characteristic).

We'll denote by $\delta(q)$ the natural function in the natural variable q that equals unity if q is zero and equals zero otherwise, and by $\mathfrak{J}(b(t), t)|_{t=g}$ the Jacobian matrix of the vector-function b(t) on the vector-variable t computed in the point t = g. For a field $F$, we'll denote by $F(\varepsilon)$ $F$'s extension by the formal infinitesimal variable $\varepsilon$, i.e. $F(\varepsilon) := \{a = \sum_{k=\mathrm{order}_\varepsilon(a)}^{\infty} a_k \varepsilon^k, \mathrm{order}_\varepsilon(a) \in \mathbb{Z},\ a_k \in F \text{ for } k = \mathrm{order}_\varepsilon(a), \dots, \infty\}$ .

Let $f(u)$ be a polynomial in $u_1, \dots, u_{\dim(u)}$ of degree $d$ over a field F,

$\hat{u} = \hat{u}(h_1, \dots, h_{\dim(h)}),\ v = v(h_1, \dots, h_{\dim(h)})$ be two analytic vector-functions in the parameter vector-variable $h$ such that $\dim(\hat{u}) + \dim(v) \le \dim(h)$ and $h^{[0]}$ be a value of this vector-variable such that $\hat{u}(h^{[0]})$ exists, $v(h^{[0]}) = \vec{0}_{\dim(v)}$, $\mathfrak{J}(\begin{pmatrix} \hat{u}(h) \\ v(h) \end{pmatrix}, h)\Big|_{h=h^{[0]}}$ exists and is nonsingular.

Then, given a value $u^{[0]}$ of the vector-variable $u$, over $F(\varepsilon)$ (where $\varepsilon$ is a formal infinitesimal)

$$f\left(u^{[0]}\right) = \sum_{i=0}^{d} \mathrm{coef}_{\varepsilon^i}(f(\hat{u}(\sum_{k=0}^{d} \varepsilon^k h^{[k]})))$$

where the dim(h)-vectors $h^{[1]}, \dots, h^{[d]}$ satisfy the equations

$$\mathfrak{J}(\begin{pmatrix} \hat{u}(h) \\ v(h) \end{pmatrix}, h)\Big|_{h=h^{[0]}} h^{[k]} = \begin{pmatrix} \delta(k-1)\left(-\hat{u}\left(h^{[0]}\right) + u^{[0]}\right) - \mathrm{coef}_{\varepsilon^k}(\hat{u}(\sum_{i=0}^{k-1} \varepsilon^i h^{[i]})) \\ -\mathrm{coef}_{\varepsilon^k}(v(\sum_{i=0}^{k-1} \varepsilon^i h^{[i]})) \end{pmatrix}$$

$$\text{for } k = 1, \dots, d$$

(and thus $\hat{u}\left(\sum_{k=0}^{d} \varepsilon^k h^{[k]}\right) = \hat{u}\left(h^{[0]}\right) + \varepsilon\left(-\hat{u}\left(h^{[0]}\right) + u^{[0]}\right) + O\left(\varepsilon^{d+1}\right)$, $v\left(\sum_{k=0}^{d} \varepsilon^k h^{[k]}\right) = O\left(\varepsilon^{d+1}\right)$ )

This method will be called **the neighboring computation of the polynomial $\mathbf{f}(\mathbf{u})$ in the point $\mathbf{u}^{[0]}$ via the parameterization $\hat{\mathbf{u}}(\mathbf{h})$ in the region $\mathbf{v}(\mathbf{h}) = \vec{\mathbf{0}}_{\mathbf{dim}(\mathbf{v})}$ from the bearing point $\mathbf{h}^{[0]}$.**

Therefore if $f(\hat{u}(h))$ is computable in polynomial time for any $h$ such that $v(h) = \vec{0}_{\dim(v)}$ (including the ε-power series $h = \sum_{k=0}^{d} \varepsilon^k h^{[k]} + O\left(\varepsilon^{d+1}\right)$ whose members of degrees higher than d are the solutions of the above equations for k > d) and there exists a bearing point $h^{[0]}$ such that $\hat{u}(h^{[0]})$ exists, $v\left(h^{[0]}\right) = \vec{0}_{\dim(v)}$, $\mathfrak{J}(\begin{pmatrix} \hat{u}(h) \\ v(h) \end{pmatrix}, h)\Big|_{h=h^{[0]}}$ exists and is nonsingular then $f(u^{[0]})$ is computable in polynomial time for any $u^{[0]}$ too.

In the further, for the purpose of simplicity, we'll call a system of functions S algebraically *absolutely* independent in a region R (given by a system of equations with a zero right part) if and only if the joint system of functions consisting of S and the left part of the system representing R is algebraically independent at some point of R.

We'll also define a computational circuit as ***arithmetically polynomial-time*** over a field if it consists of a polynomial-time number of arithmetic operations over the field.

> *The above principle hence implies that a polynomial in n variables over a field is computable in arithmetically polynomial (in n) time over the field when its calculation can be arithmetically polynomial-time (in n, over the field) reduced to finding a solution of an algebraic equation system for a polynomial (in n) number of some other variables that consists of a polynomial (in n) number of* *equations represented by arithmetically polynomial-time (in n, over the field) computable and analytic (over the field) functions in these new variables.*

# I. Equalities for the sub-permanents of a unitary matrix over fields of characteristic 3

**Definitions:**

Let A be an nxn-matrix, I, J be two subsets of {1,…,n} of an equal cardinality. Then we define its *I→J-replacement* matrix $A^{[I\to J]}$ as the matrix received from A through replacing its rows with indexes from J by those with indexes from I, i.e. through replacing its $i_k$-th row by its $j_k$-th one for k =1,...,|I|.

Analogically, given two pairs I,J and K,L of subsets of {1,…,n} such that |I|=|J| and |K|=|L|, we define its *I→J,K→ L-double-replacement* matrix $A^{[I\to J,K\to L]}$ as the matrix received from A through replacing its rows with indexes from J by those with indexes from I and its columns with indexes from L by those with indexes from K.

We also define its *I,J-repeat* matrix $A^{[I,J]}$ as the matrix received from A through repeating twice its rows with indexes from I and its columns with indexes from J (while the pairs of doubled rows or columns receive neighboring indexes. i.e. the doubled rows and columns follow each other).

By $A^{(I,J)}$ we'll denote the matrix lying on the intersection of the rows with indexes from I and the columns with indexes from J, and by $A^{(\backslash I,\backslash J)}$ we'll denote the matrix received from A through removing its rows with indexes from I and its columns with indexes from J.

For the purpose of simplicity, for a 1-set {i} we'll omit the brackets {} and write just i instead.

**Theorem I.1:**

Let U be a unitary n×n-matrix, I, J be two disjoint subsets of {1,…,n} of an equal cardinality.

Then $\mathrm{per}\left(U^{[I\to J]}\right) = (-1)^{|I|}\mathrm{per}\left(U^{[J\to I]}\right)$

Proof:

To prove this theorem, we should effectively apply the principal equality expressing the permanent of an nxn-matrix through its "principal minor convolution", i.e.

$$(1) \quad \mathrm{per}(A) = (-1)^n \sum_{L, L \subseteq \{1,\dots,n\}} \det(A^{(L,L)}) \det(A^{(\backslash L, \backslash L)})$$

First of all, as the permanent of a square matrix doesn't change after any permutation of its rows and a unitary matrix remains unitary after any permutation of its rows, we can assume I={1,3,…,2k-1}, J={2,4,…,2k} because we always can permute the rows of U so that the latter condition is fulfilled. Therefore proving the theorem for this pair of sets I, J is equivalent to proving it for the generic case. Hence, each of the two rows of the matrix $U^{[I \to J]}$ with the indexes 2q-1, 2q are the (2q-1)-th row of U, q=1,…,k.

Since the matrix $U^{[I \to J]}$ for |I|=k has k doubled rows, the sum over T in the above equality (1) can be replaced by the sum over those T that contain exactly one element from each pair 2q-1, 2q for q=1,…,k.

And now we apply the equality expressing a minor of a square matrix A through a minor of its inverse (for L,M being subsets of {1,…,n} of an equal cardinality):

$$(2) \quad \det(A^{(L,M)}) = \det(A)\det((A^{-1})^{(\backslash M, \backslash L)})(-1)^{\sum_{l \in L} l + \sum_{m \in M} m}$$

For a unitary U this formula just takes the form

$$(3) \quad \det(U^{(L,M)}) = \det(U)\det(U^{(\backslash L, \backslash M)})(-1)^{\sum_{l \in L} l + \sum_{m \in M} m}$$

while in such a case the convolution equality (1) for the matrix $U^{[I \to J]}$ yields:

$$\mathrm{per}\left(U^{[I \to J]}\right) = \\ (-1)^n \sum_{R, R \subseteq \{1,\dots,n\} \backslash \{I \cup J\}} \sum_{h \in \{0,1\}^k} \det(U^{(R \cup I, R \cup G(h))}) \det(U^{(\backslash \{R \cup J\}, \backslash \{R \cup G(h)\})})$$

where $G(h) = \{2 - h_1 \cup \dots \cup 2k - h_k\}$.

After the application of the formula (3) to the latter equality, we receive

$$\mathrm{per}\left(U^{[I \to J]}\right) = \\ (-1)^n \sum_{R, R \subseteq \{1,\dots,n\} \backslash \{I \cup J\}} \sum_{h \in \{0,1\}^k} (-1)^k \det(U^{(\backslash \{R \cup I\}, \backslash \{R \cup G(h)\})}) \det(U^{(R \cup J, R \cup G(h))})$$

as all the indexes of the involved minors are doubled except 1,…,2k each of whom appears exactly once in the corresponding sum of indexes (according to the formula (3)) and their sum is equal to k modulo 2. Hence, we get the theorem.

**Theorem I.2:**

Let U be a unitary n×n-matrix, I,J be two subsets of {1,…,n} of an equal cardinality.

Then $\mathrm{per}\left(U^{[I,J]}\right) = (-1)^{|I|}\mathrm{per}\left(U^{(\backslash I,\backslash J)}\right)$

Proof:

The proof of this theorem virtually repeats the proof of Theorem I.1, including the preliminary permutations of repeated rows and repeated columns that make their indexes belong to the set {1,…,2k}, where |I|=|J|= k. I.e. we can assume, beforehand, that I = J = {1,…,k} – for the same reason as in the proof of Theorem I.1, while preserving the degree of commonness. In such a case in the corresponding convolution sum all the indexes would be repeated twice when passaging to the inverse's minors, while each product of principal minors will have the coefficient $2^{|I|} = (-1)^{|I|}$

**Theorem I.3:** given two pairs I,J and K,L of subsets of {1,…,n} such that

|I|=|J|=|K|=|L|, $I \cap J = K \cap L = \emptyset$,

$$\mathrm{per}\left(U^{[I\to J,K\to L]}\right) = \mathrm{per}\left(U^{[J\to I,L\to K]}\right)$$

Proof:

Once again, this theorem can be easily proven in the same way as Theorems I.1, I.2, while assuming I = K = {1,3,…,2k-1}, J = L = {2,4,…,2k}.

**Definition:**

For an n×n-matrix A and $k \leq n$, let's define its k-th *permanent-minor matrix* $P(A,k)$ as a $C_n^k \times C_n^k$-matrix whose rows and columns are indexed by k-subsets of {1,…,n} and whose I,J-entry $p_{I,J}(A,k) = \mathrm{per}\left(A^{(I,J)}\right)$ for a pair of k-subsets I,J.

Let's also define its k-th *permanent-complement matrix* $F(A,k)$ as a $C_n^k \times C_n^k$-matrix whose rows and columns are indexed by k-subsets of {1,…,n} and whose I,J-entry $f_{I,J}(A,k) = \mathrm{per}\left(A^{(\backslash I,\backslash J)}\right)$ for a pair of k-subsets I,J.

Obviously, $P(U^T,k) = P^T(U,k)$ and $F(U^T,k) = F^T(U,k)$ .

**Corollary I.4**: let U be unitary. Then

(*) $$F(U,k)P^T(U,k) = (-1)^k P(U,k)F^T(U,k) \star \{(-1)^{|I\cap J|}\}_{C_n^k \times C_n^k}$$

where ⋆ denotes the Hadamard (i.e. entry-wise) product of matrices.

**Corollary I.5**:

(**) $$(-1)^{k+1}F(U,k)+P(U,k)F^{T}(U,k)P(U,k)=$$

$$=-\{\sum_{s=0}^{k-1}\sum_{\substack{\hat{I}\subset I,\hat{J}\subset J\\|\hat{I}|=|\hat{J}|=s}}(P(U,s)F^{T}(U,s)P(U,s))_{\hat{I},\hat{J}}p_{I\setminus\hat{I},J\setminus\hat{J}}(U,k-s)\}_{C_n^k\times C_n^k}=0$$

Both the above corollaries follow from Theorems I.1 and I.2 correspondingly and the Laplace expansions of the permanent for a set of rows and for a set of rows and a set of columns.

The equalities (*) and (**) are actually linear equations expressing the entries of F(U,k) through the entries of F(U,s) for s<k.

We can also notice that for a unitary U its replacement matrix for |I|=|J|= 1 is 1-semi-unitary and, therefore, we can compute its permanent in a polynomial time, while for a unitary U and four pair-wise distinct indexes i, j, s, r (where s<r)

(***) $\mathrm{per}\left(U^{[\{i,s\}\to\{j,r\}]}\right)-\ \mathrm{per}\left(U^{[\{i,r\}\to\{j,s\}]}\right)=\mathrm{per}\left(M_{s,r}U^{[i\to j]}\right)$

where $M_{s,r}$ is the identity matrix $I_n$ where the s-th and r-th rows were left-multiplied by the unitary matrix $\begin{pmatrix}\sqrt{-1} & \sqrt{-1}\\ -\sqrt{-1} & \sqrt{-1}\end{pmatrix}$ (hence $M_{s,r}U^{[i\to j]}$ is also 1-semi-unitary as a unitary row-transformation of the 1-semi-unitary matrix $U^{[i\to j]}$).

**Lemma I.6.**

Let U be unitary, i<j. Then $\mathrm{per}\left(U^{[i\to j]}\right)=\mathrm{per}\left(M_{i,j}U\right)$

**Lemma I.7.**

Let U be unitary, i<j, s<r, |{i,j,r,s}|=4. Then

$$\mathrm{per}\left(U^{[\{i,s\}\to\{j,r\}]}\right)-\ \mathrm{per}\left(U^{[\{i,r\}\to\{j,s\}]}\right)=\mathrm{per}\left(M_{s,r}M_{i,j}U\right)$$

Since, by the Laplace expansion of the permanent for a set of rows, the I,J-entry of the matrix $P(U,2)F^{T}(U,2)$ equals $\mathrm{per}\left(U^{[I\to J]}\right)$, i.e $(P(U,2)F^{T}(U,2))_{I,J}=\mathrm{per}\left(U^{[I\to J]}\right)$, the equalities (***) (together with the fact that if $|I\cap J|>0$ then $U^{[I\to J]}$ has either precisely one replaced row or no replaced rows and, accordingly, is 1-semi-unitary or unitary correspondingly) signify that the matrix $F(U,2)P^T(U,2) = (P(U,2)F^T(U,2))^T$ can be polynomial-time expressed as the sum of a known matrix and a matrix X(U) with the following properties: $x_{I,J}(U) = x_{K,L}(U)$ if $I \cup J = K \cup L$ and $x_{I,J}(U) = 0$ if $|I \cap J| > 0$. We'll call a matrix *super-symmetric* if its rows and columns are indexed by 2-subsets of {1,…,n} and it satisfies the two latter conditions. Hence, analogically, the matrix $P^T(U,2)F(U,2) = P(U^T,2)F^T(U^T,2)$ can also be expressed as the sum of a known matrix and a super-symmetric Y(U). Accordingly, if in the generic case of a unitary U the homogeneous system of linear equations for two super-symmetric matrices X and Y $P(U,2)X - YP(U,2) = 0$ is non-singular and, therefore, its only solution is zero, we can polynomial-time compute the entries of F(U,2) because the two above-mentioned expressions yield a non-singular system of linear equations for X(U), Y(U) (via expressing F(U,2) through X(U) and Y(U) correspondingly from these expressions). As an instrument for studying the equation $P(U,2)X - YP(U,2) = 0$, we can apply, for n ≡ 0 (mod m), the unitary matrix $U = \mathrm{Diag}(\{W_q\}_{q=1,\dots,\frac{n}{m}})$ where $W_1, \dots, W_{n/m}$ are unitary m×m-matrices.

Analogically, we can polynomial-time compute the differences

$$(****)\ \mathrm{per}\left(U^{[i\to j,r\to s]}\right) - \mathrm{per}\left(U^{[i\to j,s\to r]}\right) = \mathrm{per}\left(U^{[i\to j]}\right)M_{s,r}$$

because $\mathrm{per}\left(U^{[i\to j]}\right)M_{s,r}$ is also 1-semi-unitary for the same reason as $M_{s,r}U^{[i\to j]}$.

It provides us with even more equations for the entries of the matrix F(U,2). Similar unitary linear combinations of k-1 pair-wise disjoint pairs of rows in the matrix $U^{[i\to j]}$ would lead to some linear equations relating the entries of the matrices F(U,k),…,F(U,1). And the following question arises accordingly: whether it may appear that all those equations form a non-singular system for finding those matrices for some k>1 in the generic case of a unitary U. If it's so, we may easily reduce their computation in any special case of our interest (particularly, significant for proving P=NP) to computing those matrices in the most generic case -- hence implying P=NP. As a necessary tool for such a research, we can offer the neighboring computation principle.

Let's call the equation (***) the (i,j;r,s)-replacement-shift equation for the matrix U and the equation (****) the (i,j;r,s)-double-replacement-shift equation for U.

Let's also define the matrix $B(U,\alpha) = \begin{pmatrix} \alpha I_n & \sqrt{1-\alpha^2}U^T \\ \sqrt{1-\alpha^2}U & -\alpha I_n \end{pmatrix}$ which we'll call the α-block-composition of U where α is an element of a field. It's easy to see that $B(U,\alpha)$ is unitary when U is unitary. We'll now consider, for a ground field H the entries of U belong to, its α-extension H(α) whose elements are formal power series in α, i.e. having the form $h = \sum_{t=k}^{\infty} h_t \alpha^t$ where k we'll call the smallness-order of h (or just the order of h) order(h).

**Conjecture I.8.**

*For the generic case of a unitary nxn-matrix U, the set of (i,j;s,r)-replacement-shift equations for U and* $U^T$ *and (i+n,j;r+n,s)-replacement-shift equations (considered only for the power* $\alpha^2$*) for* $B(U,\alpha)$ *and* $B(U^T,\alpha)$*, where i,j,s,r are from {1,…,n}, form an algebraically complete (i.e. having a nonsingular Jacobian matrix) system of equations for the entries of F(U,2).*

Actually, while proving the above conjecture looks yet too difficult at the present time, we can try to experimentally check it via a computer modeling on a random U.

# II. Reviewing the permanent-minors and other permanents derived from a unitary matrix from a much wider point of view.

## 1. The permanent-analog of the inverse's minor formula.

Let A be an n×n-matrix over a field of a prime characteristic p, $\alpha, \beta$ be two n-vectors having all their entries from the set {0,…,p-1}, i.e. $\alpha, \beta \in \{0,...,p-1\}^n$. Then let's denote by $A^{(\alpha,\beta)}$ the matrix received from A through repeating $\alpha_i$ times its i-th row for i = 1,…,n and $\beta_j$ times its j-th column for j=1,…,n (if some $\alpha_i$ or $\beta_j$ equals zero it would mean we remove the i-th row or j-th column correspondingly). Then, in case if $A^{(\alpha,\beta)}$ is square, i.e. $\sum_{i=1}^{n} \alpha_i = \sum_{j=1}^{n} \beta_j$, the following identity holds

**Theorem II.1.1 (in characteristic p):**

$$\mathrm{per}(A^{(\alpha,\beta)}) = \det^{p-1}(A)\,\mathrm{per}((A^{-1})^{((p-1)\vec{1}_n-\beta,(p-1)\vec{1}_n-\alpha)})\frac{\prod_{i=1}^{n} \alpha_i\,!}{\prod_{j=1}^{n}(p-1-\beta_j)\,!}$$

where $\vec{1}_n$ is the n-vector all whose coordinates are equal to 1.

The above identity can be also written as

(*) $\operatorname{per}(A^{(\alpha,\beta)}) =$

$$= \det^{p-1}(A)\operatorname{per}((A^{-1})^{((p-1)\vec{1}_n-\beta,(p-1)\vec{1}_n-\alpha)})\left(\prod_{i=1}^{n}\alpha_i\,!\right)\left(\prod_{j=1}^{n}\beta_j\,!\right)(-1)^{n+\sum_{i=1}^{n}\alpha_i}$$

Proof:

First of all, let's prove that

$$(1)\qquad \left(\prod_{i=1}^{n}\frac{(p-1)!}{\alpha_i!}\right)\operatorname{per}(A^{(\alpha,\beta)}) = \operatorname{per}\left(I_n^{((p-1)\vec{1}_n,(p-1)\vec{1}_n-\alpha)}\quad A^{((p-1)\vec{1}_n,\beta)}\right)$$

where in the right side is the permanent of a matrix composed of two blocks, the first block $I_n^{((p-1)\vec{1}_n,(p-1)\vec{1}_n-\alpha)}$ being block-diagonal itself with diagonal blocks of sizes $(p-1)\times(p-1-\alpha_i)$, i=1,…,n. The identity (1) follows from the Laplace expansion of the right permanent by the columns corresponding to the first block (this expansion is the direct product of the Laplace expansions for its diagonal blocks).

Secondly, if B is an m×((p-1)m)-matrix and G is an m×m-matrix then

$$(2)\qquad \operatorname{per}((GB)^{((p-1)\vec{1}_m,\vec{1}_{(p-1)m})}) = \det^{p-1}(G)\operatorname{per}(B^{((p-1)\vec{1}_m,\vec{1}_{(p-1)m})})$$

as in characteristic p the permanent doesn't change if each row of a matrix is repeated p-1 times and we add one of its (p-1)-tuples of equal rows to another (p-1)-tuple of equal rows, and is multiplied by $d^{p-1}$ if we multiply a (p-1)-tuple of equal rows by d.

Upon applying the formula (2) to the case $B = \left(I_n^{(\vec{1}_n,(p-1)\vec{1}_n-\alpha)}\quad A^{(\vec{1}_n,\beta)}\right)$, $G = A^{-1}$, we'll receive an identity involving a two-blocked matrix with the second block being block-diagonal itself (like in the identity (1)) and hence analogous to the identity (1) what will give us the initial identity.

This identity is, first of all, a generalization (for an arbitrary prime characteristic p) of all the repeat-removal identities we received in characteristic 3, and, secondly, the permanent-analog of the classical formula for the matrix inverse's minor.

Besides, in characteristic p there is the following pair of dual identities for an n×n-matrix A:

$$\operatorname{per}(A) = (-1)^n \sum_{J_1,\dots,J_{p-1}} \det\left(A^{(J_1,J_1)}\right)\dots\det\left(A^{(J_{p-1},J_{p-1})}\right)$$

$$\det(A) = (-1)^n \sum_{J_1,\dots,J_{p-1}} \operatorname{per}\left(A^{(J_1,J_1)}\right)\dots\operatorname{per}\left(A^{(J_{p-1},J_{p-1})}\right)$$

where in both the above formulas the summation is over all the (p-1)-tuples $J_1, \ldots, J_{p-1}$ that are partitions of the set {1,…,n} into p-1 subsets, some of them possibly empty, while p isn't obliged to be prime.

# 2. Permanent-preserving compressions over fields of characteristic 3

## 2.1 The basic compression

Let A be an n×n-matrix over a field of characteristic 3 with at least one pair of equal rows. Let i,j (i<j) be the indexes of the lexicographical minimum (index-wise) of such pairs of rows. We'll define the compression of A Comp(A) as the (n-1)×(n-1)-matrix received from A through making zero (via the Gauss algorithm) all the entries of the first column of A by its i-th row (having $a_{i,1}$ as the leading entry for the column elimination) and then removing the first column and the j-th row of the received matrix. Then

$$\mathrm{per}(A) = -a_{i,1}\mathrm{per}(\mathrm{Comp}(A))$$

We'll also define the compression-closure of A $\widehat{\mathrm{Comp}}(A)$ as the limit of the compression operator's sequential application to A or, if at some stage the received matrix is incompressible, to its transpose (i.e. the limit of actions when we compress the matrix and transpose it if no rows are equal any more but there are equal columns, until it would have no equal rows or equal columns). We can also speak about applying the compression and compression-closure operators to sets of matrices that would map them into another sets of matrices. It's obvious that, if we denote by $\mathbb{U}_k$ the set of k-semi-unitary matrices, $\widehat{\mathrm{Comp}}(\mathbb{U}_0) = \mathbb{U}_0$, i.e. unitary matrices are incompressible because they are non-singular and can't have equal rows.

But, if we take a unitary matrix with one row replaced by another one (whose permanent we can polynomial-time compute) and multiply both copies of the repeated row by $\sqrt{-1}$, then such a matrix will be both 1-semi-unitary and compressible, and, though strange, its compression won't be 1-semi-unitary but will be 2-semi-unitary instead. Hence $\mathrm{Comp}(\mathbb{U}_1) \subset \mathbb{U}_2$ and the latter fact raises somewhat a hope that $\widehat{\mathrm{Comp}}(\mathbb{U}_1)$ is a set of matrices that is $\#_3P$ complete and, in such a case, we can use the neighboring computation principle to prove $\#_3$P = P and, therefore, P = NP.

## 2.2 The generalized compression.

Let A be an n×n-matrix over a field of characteristic 3 having at least one linearly dependent triple of rows, i.e. a triple of rows with pair-wise distinct indexes i, j, k such that $a_k = ga_i + ha_j$ where g and h are some elements of the field (we also assume $a_i$ and $a_j$ are linearly independent). Then adding the row $ga_i - ha_j$ multiplied by any element of the field to any row of A except the i-th, j-th and k-th ones doesn't change the permanent because of its row-wise multi-linearity and the fact that the permanent of a matrix having four rows $a_i$, $a_j$, $ga_i+ha_j$, $ga_i-ha_j$ is zero in characteristic 3. Hence we can eliminate, while assuming that the first entry of the row $ga_i-ha_j$ is non-zero (or permuting A's columns for to fulfill this condition otherwise) and using it as the Gaussian column-elimination's leading entry, the first column of A except the entries $a_{i1}, a_{j1}, a_{k1}$. Then per(A) equals the permanent of the matrix received from A through replacing its i-th, j-th and k-th rows by the pair of rows $a_{j1}a_i - a_{i1}a_j$, $ga_i - ha_j$ and removing its first column. We'll call such a compression a triple-compression (as it involves a triple of linearly dependent rows), while the case of two linearly dependent rows $a_j = da_i$ , where d is some element of the field, we'll call a pair-compression (i.e. we can divide the j-th row by d while the permanent will also be divided by d and hence we'll receive the above-described case of equal rows). In fact, the pair-compression is a partial case of the triple-compression by putting d=g, h=0, but, nevertheless, these are two cases of permanent-preserving matrix compressions we'll distinguish and in the further let's understand by Comp(A) the lexicographically least (index-wise) pair- or triple-compression of A. In the meantime, the compression-closure operator's definition won't change in this generalization of the compression-operator.

Thus the question of determining the structure of the matrix class $\widehat{\mathrm{Comp}}(\mathbb{U}_1)$ becomes even more intriguing and challenging towards the chief mystery P versus NP. Actually we can even consider, instead of $\mathbb{U}_1$, the wider class $\mathbb{U}_{(1)} \subset \mathbb{U}_2$ of matrices received from unitary ones via replacing one row by an arbitrary vector-row (by the Laplace expansion, the permanent of such a matrix is the sum of the permanents of 1-semi-unitary matrices and hence polynomial-time computable and if the replacing vector-row is a linear combination of the matrix's two other rows then such a matrix is triple-compressible) and, accordingly, study the class $\widehat{\mathrm{Comp}}\left(\mathbb{U}_{(1)}\right)$.

It would be also useful to notice that the identity given and proven in the first section of this paper that links the generalized permanent-minors of a matrix and its inverse is, if considered only for characteristic 3, merely an application of the pair-compression operator to certain matrices. Accordingly, the following questions could be raised: what family of identities may we receive in characteristic 3 when applying the most general compression, i.e. including the triple case, and, actually, what are the possible analogs of the permanent-preserving compressions we found in characteristic 3 for other prime characteristics?

An answer to the former of these two questions might be gotten via studying an arbitrary $(3n)\times(3n)$-matrix consisting of n linearly dependent triples of rows whose compression-closure would be of size at most $(2n)\times(2n)$. It could also provide us, upon permuting the matrix's columns so that a chosen n-subset of its column set will turn into $\{1,\ldots,n\}$, with an opportunity to determine a relation between all the $(2n)\times(2n)$ matrices (thus having equal permanents) we may receive in this way. Let's call two $(2n)\times(2n)$-matrices triple-conjugate if they can be received via such a procedure from one initial $(3n)\times(3n)$-matrix consisting of n linearly dependent triples of rows. By the way, if we apply the same scheme for a $(2n)\times(2n)$-matrix consisting of n pairs of equal rows and the pair-compression operator then we'll receive $n\times n$-matrices that are partial inverses to each other. Hence, while computing the permanent in char 3, we can transfer not only to the matrix's partial inverse, but to its triple-conjugate as well (however, beforehand we should actually verify that, in fact, a triple-conjugate is (generically) not a partial inverse but its genuine generalization).

## 2.3 A wider generalization of permanent-preserving compressions.

Let A be a square matrix such that its first 2k rows form a matrix of rank k, i.e., generically, its rows with the indexes $k+1,\ldots,2k$ are linear combinations of its first k rows with a coefficient $k\times k$-matrix B. Then per(A) is equal to the product of per(B) and the permanent of the matrix received from A through removing its rows with the indexes $k+1,\ldots,2k$ and doubling its rows with the indexes $1,\ldots,k$. In such a case we'll receive a matrix with k pairs of equal rows to which we can apply (k times) the pair-compression operator in order to reduce its size by k. We'll call such a compression an *even* compression.

Let A be a square matrix such that its first 2k-1 (for k>1) rows form a matrix of rank k, i.e., generically, its rows with the indexes $k+1,\ldots,2k-1$ are linear combinations of its first k rows with a coefficient $(k-1)\times k$-matrix B. Then per(A) is equal to the permanent of the matrix received from A through removing its rows with the indexes $k+1,\ldots,2k-1$, doubling its rows with the indexes 1,…,k, and, afterwards, adding one new column whose entries corresponding to both copies of the q-th row are $-\mathrm{per}(\mathrm{B}^{(\{1,\ldots,k-1\},\{1,\ldots,k\}\setminus q)})$ for q=1,…,k and all the other entries are zeros. (Please notice that the added column ensures the matrix remains square). In such a case we'll receive a matrix with k pairs of equal rows to which we can apply (k times) the pair-compression operator in order to reduce its size by k. We'll call such a compression an *odd* compression.

In both the above cases of even and odd compressions, we just should suppose, naturally, that k is fixed so that per(B) or $\mathrm{per}(\mathrm{B}^{(\{1,\ldots,k-1\},\{1,\ldots,k\}\setminus q)})$ correspondingly could be computed in a polynomial time, or that we, at least, can polynomial-time calculate these values in some way otherwise. And, apparently, we may speak about applying to a 2k- or (2k-1)-set of rows of rank k the 2k- or (2k-1)-compression operator correspondingly (as stated above) upon an appropriate permutation of the matrix's rows turning the set into {1,…,2k} or {1,…,2k-1}. We'll call k and k-1 the compression's *velocity* for an even and odd compression correspondingly as they lessen the matrix's size by k and k-1 correspondingly. We can also prove that a compression transformation doesn't change the matrix's rank (although changing, possibly, its semi-unitarity class), while transferring to a partial inverse doesn't change the matrix's semi-unitarity but, nevertheless, may change its rank. Hence, coupled together, the compression operator and all the partial inversions form a yet more perfect instrument for permanent-preservingly compressing matrix classes, first of all $\mathbb{U}_{(1)}$. If we define transferring to a partial inverse as a compression of velocity 0, we might be able to compute the permanent on a yet more reach class $\widehat{\mathrm{Comp}}\left(\mathbb{U}_{(1)}\right)$.

Actually, it's easy to notice that the earlier mentioned pair- and triple-compressions are merely partial cases of even and odd compressions correspondingly (that are, in fact, their generalizations), and, though strange, the triple-compression itself can be expressed as a double pair-compression (in a beforehand modified matrix, though). Hence since now we can start defining the compression of a matrix as its lexicographically least (index-wise) even or odd compression, with the same notion of the matrix's compression-closure.

### 2.4 A criterion of the permanent's equality to zero

Let A be a square matrix such that its rows with the indexes k+1,…,k+m are linear combinations of its first k rows with the coefficient $m\times k$-matrix B such that all its $m\times m$-subpermanents (or permanent-minors) are zero. Then per(A) = 0. In characteristic 3, an example of such a matrix B is the matrix C(x,y) where dim(y) < 2dim(x) and the joint vector $\binom{x}{y}$ is the root vector of a polynomial that is the derivative of another polynomial. This fact is based on a Lemma that is to be given further in the article.

## 2.5 The partial inverse equivalence and classification of permanent-preserving compressions.

**Lemma II.2.5.1** (on the permanent of a partial inverse):

over a field of characteristic 3, for $A_{11}$, $A_{22}$ being square,

$$(**)\quad \mathrm{per}\begin{pmatrix} A_{11} & A_{12} \\ A_{21} & A_{22} \end{pmatrix} = \det^2(A_{11})\mathrm{per}\begin{pmatrix} A_{11}^{-1} & A_{11}^{-1}A_{12} \\ A_{21}A_{11}^{-1} & A_{22} - A_{21}A_{11}^{-1}A_{12} \end{pmatrix}$$

The proof of the above formula can be received via the technique applied in proving the analog of the inverse's minor formula for permanent-minors in Part 1 of this article.

Apparently, the latter formula is a generalization of the formula for the permanent of a matrix's inverse in characteristic 3, i.e., for a square non-singular A, $\mathrm{per}(A) = \det^2(A)\mathrm{per}(A^{-1})$. In the meantime, in a generic prime characteristic p and with the same technique's usage, we can even similarly generalize the formula (*) for permanent-minors via giving to both parts of the formula (**) their row/column multiplicity degrees:

$$(***)\quad \mathrm{per}\begin{pmatrix} A_{11} & A_{12} \\ A_{21} & A_{22} \end{pmatrix}^{(\alpha,\beta)} =$$

$$= \det^{p-1}(A_{11})\mathrm{per}\begin{pmatrix} A_{11}^{-1} & A_{11}^{-1}A_{12} \\ A_{21}A_{11}^{-1} & A_{22} - A_{21}A_{11}^{-1}A_{12} \end{pmatrix}^{\left((p-1)\vec{1}_n-\beta,(p-1)\vec{1}_n-\alpha\right)} \cdot$$

$$\cdot \left(\prod_{i=1}^{n} \alpha_{1,i}!\right)\left(\prod_{j=1}^{n} \beta_{1,j}!\right)(-1)^{n_1+\sum_{i=1}^{n}\alpha_{1,i}}$$

where:

for an n×m-matrix $M$, an n-vector x and an m-vector y, both vectors having all their entries from the set {0,...,p-1}, $M^{(x,y)}$ denotes the matrix received from $M$ via repeating $x_i$ times its i-th row for i = 1,...,n and $y_j$ times its j-th column for j = 1,...,m (if some row's or column's multiplicity equals zero it would mean that the row or column was removed, and thus this notion is a generalization of the notion of submatrix);

$\begin{pmatrix} A_{11} & A_{12} \\ A_{21} & A_{22} \end{pmatrix}^{(\alpha,\beta)} = \begin{pmatrix} A_{11}^{(\alpha_1,\beta_1)} & A_{12}^{(\alpha_1,\beta_2)} \\ A_{21}^{(\alpha_2,\beta_1)} & A_{22}^{(\alpha_2,\beta_2)} \end{pmatrix}$, $A_{11}$ is of size $n_1 \times n_1$ and in the right part of the equality (***) each block-matrix is multiplicity-degreed correspondingly (while $A_{11}$ and $A^{(\alpha,\beta)} = \begin{pmatrix} A_{11} & A_{12} \\ A_{21} & A_{22} \end{pmatrix}^{(\alpha,\beta)}$ are square).

**Corollary II.2.5.2 (in characteristic 3):** over a field of characteristic 3, for $A_{11}$ of size $n_1 \times n_1$ and invertible and $A_{22}$ of size $n_2 \times n_2$, let the last $n_2$ rows of $\begin{pmatrix} A_{11} & A_{12} \\ A_{21} & A_{22} \end{pmatrix}$ be linearly expressible through its first $n_1$ rows (what implies $A_{22} - A_{21}A_{11}^{-1}A_{12} = 0_{n_2\times n_2}$). Then

$$\operatorname{per}\begin{pmatrix} A_{11} & A_{12} \\ A_{21} & A_{22} \end{pmatrix} = \det^2(A_{11})\operatorname{per}\begin{pmatrix} A_{11}^{-1} & A_{11}^{-1}A_{12} \\ A_{21}A_{11}^{-1} & 0_{n_2\times n_2} \end{pmatrix}$$

The above corollary can be used for another interpretation of the earlier introduced even and odd compressions: if we permute A's rows so that all the linearly dependent rows we refer to in the corresponding definitions would form the second block-row in the received matrix's block-decomposition then its corresponding (to the block-decomposition) partial inverse will have the form

$$\begin{pmatrix} A_{11}^{-1} & A_{11}^{-1}A_{12} \\ (B \quad 0_{k\times(n_1-k)}) & 0_{k\times k} \end{pmatrix} \text{ or } \begin{pmatrix} A_{11}^{-1} & A_{11}^{-1}A_{12} \\ (B \quad 0_{(k-1)\times(n_1-k)}) & 0_{(k-1)\times(k-1)} \end{pmatrix} \text{ correspondingly}$$

and, due to B being of size either k×k or (k-1)×k correspondingly, we can permanent-preservingly reduce this matrix via the Laplace-expansion for the second block-row (we'll call such compressions *primitive*). If there are several pair-wise disjoint sets of such linear dependencies we refer to by the definitions, they'll yield the direct product of corresponding primitive compressions. Hence even and odd compressions are equivalent to primitive ones via partial inverse reductions.

But there are yet compressions that are not (at least, so obviously) equivalent to primitive ones. For instance,

$$\operatorname{per}\begin{pmatrix} \begin{pmatrix} \alpha & \beta & \alpha+\beta \\ c_1 & c_2 & c_3 \end{pmatrix} & A_{12} \\ b_1 \quad b_2 \quad b_3 & 0 \end{pmatrix} = \operatorname{per}\left(\begin{pmatrix} r_{11}\alpha + r_{12}\beta & r_{21}\alpha + r_{22}\beta \\ d & e \end{pmatrix} \quad A_{12}\right)$$

where $r_{11}r_{21} = b_2$ , $r_{12}r_{22} = b_1$, $r_{11}r_{22} + r_{12}r_{21} = b_1 + b_2 + b_3$ ,

$$r_{11}e + r_{21}d = \operatorname{per}\begin{pmatrix} 1 & 0 & 1 \\ c_1 & c_2 & c_3 \\ b_1 & b_2 & b_3 \end{pmatrix}$$

$$r_{12}e + r_{22}d = \mathrm{per}\begin{pmatrix} 0 & 1 & 1 \\ c_1 & c_2 & c_3 \\ b_1 & b_2 & b_3 \end{pmatrix}$$

$\alpha, \beta$ are vector-columns, $A_{12}$ is a matrix (of appropriate sizes), all the other values are elements of the ground field.

All the types of compression we discussed in Chapter 2 of the present article we'll call elementary. To summarize, we may, hence, conclude that in characteristic 3 there exists a whole variety of permanent-preserving compressions of a square matrix which, together with the set of partial inverse transformations, form the set of permanent-preserving and polynomial-time computable elementary transformations of a matrix. The problem of finding and classifying all of them is yet to be solved. And, accordingly, the compression-closure operator (understood as the closure-limit of those elementary compressions) is a pretty rich opportunity to reduce the size of a matrix whose permanent we need to know. Therefore the question of studying the compression-closure of important matrix classes we can polynomial-time compute the permanent on like $\mathbb{U}_{(1)}$ still arises as one of the chief mysteries related to the ever mysterious indefiniteness of P versus NP.

Besides, the formulae (*), (**), (***) provide us, when applied to a unitary matrix in characteristic 3, with another variety of linear equations for a unitary matrix's permanent-minors of a bounded depth (i.e. its sub-permanents received via removing k-sets of their rows and columns, with a bounded k) whose non-singularity (for a given maximum of k) is to be researched as well.

And, at last, it would be worth noting the following simple construction that, actually, is applicable in any characteristic. Let's call reducing the permanent via the Laplace expansion on a set I of k rows (or columns) containing only k or k+1 non-zero columns (or rows) **the Laplace compression on I**. If we extend an n×n-matrix whose first row has only two non-zero entries, one of them equal to 1 and another to -1, by one row and column so that the new matrix's first column would contain only two non-zero entries, 1 and -1, the Laplace compression on its first column would yield the original matrix, the original matrix's first row would be involved in this extension, and transpose the received matrix afterwards, -- then each time we can involve some arbitrary row vector as a parameter, let's call it a Laplace extension vector, while such an extended matrix we'll call the elementary Laplace extension of the original matrix by a Laplace extension vector. Hence a generic sequence of elementary Laplace extensions will provide us with a matrix whose lower-right corner n×n-submatrix is the initial matrix or its transpose and whose permanent is the same as of the initial one. Actually such a sequence of elementary Laplace extensions could be started with an arbitrary matrix via extending it by one row and column so that the new matrix's first column would have only two non-zero entries, 1 and -1, and the new matrix's permanent would equal the permanent of the original one whose first row would be involved in the extension, and transposing the result afterwards (a Laplace extension vector is supposed here too, and we'll call it an initial Laplace extension). We'll also call the overall result of an elementary Laplace extension sequence the Laplace extension of a matrix by a sequence of Laplace extension vectors.

A special interest the Laplace extensions can present in characteristic 3 is the following question: what is the class of square matrices possessing 1-semi-unitary Laplace extensions? If, say, a generic square matrix can be Laplace-extended to a 1-semi-unitray one then this fact would yet imply, via the neighboring computational principle, the permanent's polynomial-time computability in characteristic 3. And, actually, even though earlier in this paper we somehow paralleled the elementary compressions based on row linear dependencies and those we've now called the elementary Laplace ones, still we also may investigate the above-defined (i.e. row linear dependency) compression-closure of the class of Laplace extensions generated by a given matrix or just a matrix class. We hence may conjecture that, despite the mutual expressibility of the elementary Laplace and row linear dependency compressions through each other, their self-generating chains (and, in fact, their compression-closures defined as the limits of action) pretty might appear to be nonequivalent. However, On the other hand, if we add the elementary Laplace compressions to the whole set of row linear dependency compressions (completed by the transpose and partial inverse transformations) we earlier introduced then we just may receive a much wider variety of elementary transformations whose compression-closure is to be studied – but, nevertheless, for the Laplace compression the structure of its corresponding extension (Laplace extension) perceived as its inverse modification (i.e. aka decompression) of a matrix is much more clear and can be expressed in a simple manifest form, while the row linear dependency compressions and their inverses (extensions) are apparently more difficult to express algebraically.

# 3. Some formulae for the hafnian of a symmetric matrix in characteristic 3 and the Hamiltonian cycle polynomial in characteristic 2.

## 3.1 The even permanent

The approaches demonstrated and applied in the present article's first chapter for proving (in characteristic 3 only) a number of dependencies between the permanents of matrices received from a unitary one via certain row/column repeat/replacement modifications were in fact overlapped by the compression techniques and associated formulas that appeared in the second chapter. Nevertheless, the first chapter's methods of proof aren't yet deprived of some independent meaning as we can also use them (in characteristic 3 only as well) for proving various facts on the hafnian of a symmetric (2n)×(2n)-matrix that is a generalization of the permanent of a square matrix. For this purpose, first of all let's define for a (2n)×(2n)-matrix A its *even-permanent* as

$$\mathrm{per}_{\mathrm{even}}(A) = \sum_{\pi \in S_{2n}^{(\mathrm{even})}} \prod_{i=1}^{2n} a_{i,\pi(i)}$$

where $S_{2n}^{(\mathrm{even})}$ is the set of 2n-permutations having only cycles of even lengths.

**Theorem II.3.1.1:** let A be a (2n)×(2n)-matrix. Then, in characteristic 3,

$$\mathrm{per}_{\mathrm{even}}(A) = \sum_{L \subseteq \{1,\dots,2n\}} (-1)^{|L|} \det(A^{(L,L)}) \det(A^{(\backslash L,\backslash L)})$$

Hence, analogically to the permanent,

$$\mathrm{per}_{\mathrm{even}}(A) = \det{}^2(A)\,\mathrm{per}_{\mathrm{even}}(A^{-1})$$

**Lemma II.3.1.2:** let A be a symmetric (2n)×(2n)-matrix. Then

$$\mathrm{haf}^2(A) = \mathrm{per}_{\mathrm{even}}(A)$$

We may also notice that the even-permanent of A doesn't depend on its diagonal entries. Secondly, in characteristic 3, if we represent a symmetric matrix in the form $\begin{pmatrix} d & b^T \\ b & M \end{pmatrix}$ where d is an element of the ground field and b is a (2n-1)-vector then $\mathrm{haf}\begin{pmatrix} d & b^T \\ b & M \end{pmatrix} = \mathrm{haf}\begin{pmatrix} d & b^T \\ b & M + \alpha b b^T \end{pmatrix}$ for any scalar coefficient α. The latter fact implies the analogical (to the permanent) relation between the hafnians and even-permanents of a symmetric matrix $A = \begin{pmatrix} A_{11} & A_{12} \\ A_{21} & A_{22} \end{pmatrix}$ and its symmetric partial inverse:

$$\mathrm{haf}^2\begin{pmatrix} A_{11} & A_{12} \\ A_{21} & A_{22} \end{pmatrix} = \det^2(A_{11})\mathrm{haf}^2\begin{pmatrix} A_{11}^{-1} & A_{11}^{-1}A_{12} \\ A_{21}A_{11}^{-1} & A_{22} - A_{21}A_{11}^{-1}A_{12} \end{pmatrix}$$

and

$$\mathrm{per}_{\mathrm{even}}\begin{pmatrix} A_{11} & A_{12} \\ A_{21} & A_{22} \end{pmatrix} = \det^2(A_{11})\mathrm{per}_{\mathrm{even}}\begin{pmatrix} A_{11}^{-1} & A_{11}^{-1}A_{12} \\ A_{21}A_{11}^{-1} & A_{22} - A_{21}A_{11}^{-1}A_{12} \end{pmatrix}$$

We can expect that, as a generalization of the permanent, the hafnian probably does possess its own types of compression, some of them analogical to certain types we've earlier found for the permanent, while others, perhaps, not. The one we would call primitive is to be applied to a symmetric matrix having the form $\begin{pmatrix} 0_{m,m} & A_{12} \\ A_{21} & A_{22} \end{pmatrix}$ where $A_{12}$'s number of non-zero columns equals m or m+1.

## 3.2 The case of characteristic 2: the Hamiltonian cycle polynomial.

**Definition:** let A be an n×n-matrix. Then its Hamiltonian cycle polynomial (or, shortly, the Hamiltonian of A) is defined as $\mathrm{ham}(A) := \sum_{\pi \in H_n} \prod_{i=1}^{n} a_{i,\pi(i)}$ where $H_n$ is the set of Hamiltonian n-permutations, i.e. n-permutations having only one cycle.

**Theorem II.3.2.1:** let A be an n×n-matrix. Then

1) in an arbitrary characteristic:
$$\mathrm{ham}(A) := \sum_{I \in \{2,\ldots,n\}} \det(-A^{(I,I)})\mathrm{per}(A^{(\{1,\ldots,n\}\setminus I, \{1,\ldots,n\}\setminus I)})$$
2) in a finite characteristic p (not necessarily prime):
$$\mathrm{ham}(A) = (-1)^{n+1} \sum_{J_1,\ldots,J_p} \det\left(A^{(J_1,J_1)}\right) \ldots \det\left(A^{(J_p,J_p)}\right) =$$
$$= \sum_{J_1,\ldots,J_p} \mathrm{per}\left(A^{(J_1,J_1)}\right) \ldots \mathrm{per}\left(A^{(J_p,J_p)}\right)$$
where the summation is over all the p-tuples $J_1, \ldots, J_p$ that are partitions of the set {1,…,n} into p subsets (some of them possibly empty) such that $1 \in J_1$.

**Theorem II.3.2.2 (in characteristic 2):**

let A be an n×n-matrix. Then

$$\mathrm{ham}(A) := \sum_{I \in \{2,\ldots,n\}} \det(A^{(I,I)})\det(A^{(\{1,\ldots,n\}\setminus I, \{1,\ldots,n\}\setminus I)}) .$$

**Theorem II.3.2.3 (in characteristic 2):**

1) let U be a unitary n×n-matrix, i. e. such that $\mathrm{UU^T} = \mathrm{I_n}$. Then $\mathrm{ham(U)} = \det^2(\mathrm{U} + \mathrm{I_n} + \mathrm{C_{1,1}})$ where $\mathrm{C_{1,1}}$ is the n×n-matrix whose 1,1-th entry is 1 and all the others are zero.
2) let A be an involutory n×n-matrix, i. e. such that $\mathrm{A}^2 = \mathrm{I_n}$.
Then $\mathrm{ham(A)} = \det^2(\mathrm{A} + \mathrm{I_n} + \mathrm{C_{1,1}}) = 0$ for n > 1.

The above theorem implies that, when n > 2, the Hamiltonian of an n×n-matrix having either three identical rows or a pair of indexes i,j such that its i-th and j-th rows are identical and its i-th and j-th columns are identical too equals zero.

While the former property generates, in this characteristic, a Hamiltonian-preserving compression of the Gaussian type (analogical to the simplest pair-compression for the permanent in characteristic 3 that possesses the same feature), the latter one (specific only for the Hamiltonian modulo 2) implies the following identity generating a type of Hamiltonian-preserving compressions applicable to certain structured unitary matrices (what makes the unitary class Hamiltonian-compressible in characteristic 2 like the 1-semi-unitary class is permanent-compressible in characteristic 3):

**Theorem II.3.2.4 (in characteristic 2):**

1) $\mathrm{ham}\left(\begin{pmatrix} \mathrm{V} & \mathrm{V+D} & \mathrm{A} \\ \mathrm{V+D^{-1}} & \mathrm{V+D^{-1}+D} & \mathrm{A} \\ \mathrm{B} & \mathrm{B} & \mathrm{U} \end{pmatrix}\right) = \det(\mathrm{D}+\mathrm{D^{-1}})\,\mathrm{ham}\left(\begin{pmatrix} \mathrm{V} & \mathrm{A} \\ \mathrm{B} & \mathrm{U} \end{pmatrix}\right)$ where D is diagonal, V, U are square;
2) if U is unitary of size n×n, V is of size m×m, $\mathrm{VD} + \mathrm{DV^T} + \mathrm{AA^T} = \mathrm{I_m} + \mathrm{D}^2$ then the matrix $\begin{pmatrix} \mathrm{V} & \mathrm{V+D} & \mathrm{A} \\ \mathrm{V+D^{-1}} & \mathrm{V+D^{-1}+D} & \mathrm{A} \\ \mathrm{UA^TD^{-1}} & \mathrm{UA^TD^{-1}} & \mathrm{U} \end{pmatrix}$ is unitary and
$$\mathrm{ham}\left(\begin{pmatrix} \mathrm{V} & \mathrm{V+D} & \mathrm{A} \\ \mathrm{V+D^{-1}} & \mathrm{V+D^{-1}+D} & \mathrm{A} \\ \mathrm{UA^TD^{-1}} & \mathrm{UA^TD^{-1}} & \mathrm{U} \end{pmatrix}\right) = \det{}^2\left(\begin{pmatrix} \mathrm{V+I_m+C_{1,1}} & \mathrm{V+D} & \mathrm{A} \\ \mathrm{V+D^{-1}} & \mathrm{V+D^{-1}+D+I_m} & \mathrm{A} \\ \mathrm{UA^TD^{-1}} & \mathrm{UA^TD^{-1}} & \mathrm{U+I_n} \end{pmatrix}\right)$$

We'll call the passage of the theorem's part (1) the (multiple) **two-sided** pair-compression**.**

As, upon multiplying the first block-column of the matrix $\begin{pmatrix} \mathrm{V} & \mathrm{A} \\ \mathrm{UA^TD^{-1}} & \mathrm{U} \end{pmatrix}$ by D, we'll receive the matrix $\begin{pmatrix} \mathrm{W} & \mathrm{A} \\ \mathrm{UA^T} & \mathrm{U} \end{pmatrix}$ where $\mathrm{W} = \mathrm{VD}$ whose Hamiltonian is $\det(D)\, ham\left(\begin{pmatrix} \mathrm{V} & \mathrm{A} \\ \mathrm{UA^TD^{-1}} & \mathrm{U} \end{pmatrix}\right)$ and is hence also polynomial-time computable and therefore we get the following generalization of the theorem that the Hamiltonian of a unitary matrix is polynomial-time computable in characteristic 2:

**Corollary II.3.2.5 (in characteristic 2):** let U be unitary of size n×n, W, D be of size m×m, D be non-singular diagonal such that $D + D^{-1}$ is non-singular, $W + W^T + AA^T = I_m + D^2$. Then

$$\mathrm{ham}\left(\begin{pmatrix} W & A \\ UA^T & U \end{pmatrix}\right) = \frac{\det(D)}{\det\left(D+D^{-1}\right)} \det^2\left(\begin{pmatrix} WD^{-1} + I_m + C_{1,1} & WD^{-1} + D & A \\ WD^{-1} + D^{-1} & WD^{-1} + D^{-1} + D+I_m & A \\ UA^TD^{-1} & UA^TD^{-1} & U+I_n \end{pmatrix}\right)$$

**Theorem II.3.2.6:** let $A_{11}$, $\begin{pmatrix} A_{11} & A_{12} \\ A_{11} & A_{12} \\ A_{21} & A_{22} \end{pmatrix}$ be square matrices, $\det(A_{11}) \neq 0$. Then

$$\mathrm{ham}\left(\begin{pmatrix} A_{11} & A_{12} \\ A_{11} & A_{12} \\ A_{21} & A_{22} \end{pmatrix}\right) = \det^2(A_{11})\mathrm{ham}\left(\begin{pmatrix} A_{11}^{-1}A_{12} \\ A_{21}A_{11}^{-1}A_{12} + A_{22} \end{pmatrix}\right)$$

We'll call the theorem's passage the (multiple) pair-compression.

We can also add that, like the even-permanent in characteristic 3, the Hamiltonian of a square matrix naturally doesn't depend on its diagonal elements and $\mathrm{ham}(A) = \det^2(A)\mathrm{ham}(A^{-1})$

This Hamiltonian-preserving compressibility and variability analogically yields the conjecture that the compression-closure (defined by the analogy with the permanent in characteristic 3) of the unitary class (which, as we've just showed above, is compressible unlike the case of the permanent in characteristic 3) is the whole set of square matrices that likewise implies the polynomial-time computability of the Hamiltonian in characteristic 2.

Besides, in characteristic 2 the Hamiltonian possesses replacement identities for a unitary matrix U similar to the earlier introduced relations for the permanent in characteristic 3.

**Definition:** for a square matrix X, a pair I,J of equally sized sets of its row-indexes with a bijection $f_1: I \overset{f_1}{\rightarrow} J$ and a pair K,L of equally sized sets of its column-indexes with a bijection $f_2: K \overset{f_2}{\rightarrow} L$, let's define ***the*** $\boldsymbol{I \overset{f_1}{\rightarrow} J, K \overset{f_2}{\rightarrow} L}$***-double-replacement matrix*** $X^{[I \overset{f_1}{\rightarrow} J, K \overset{f_2}{\rightarrow} L]}$ as the matrix received from X through replacing, for each $i \in I$, its $f_1(i)$-th row by its i-th row and, for each $k \in K$, its $f_2(k)$-th column by its k-th column.

**Theorem II.3.2.7 (in characteristic 2):**

Let A be a square matrix, I,J be sets of its row-indexes and K,L be sets of its column-indexes, |I|=|J|, |K|=|L|, $f_1, f_2$ be bijections $I \xrightarrow{f_1} J, K \xrightarrow{f_2} L$ correspondingly. Then

$$\mathrm{ham}(A^{[I \xrightarrow{f_1} J, K \xrightarrow{f_2} L]}) = \det^2(A)\mathrm{ham}(((A^{-1})^T)^{[J \xrightarrow{f_1^{-1}} I, L \xrightarrow{f_2^{-1}} K]})$$

where $f_1^{-1}, f_2^{-1}$ are the inverse bijections $J \xrightarrow{f_1^{-1}} I, L \xrightarrow{f_2^{-1}} K$ correspondingly.

Like for the permanent in characteristic 3, the proof of this identity can be received by means of using the fact that in characteristic 2 (where we have no signs +/-) any minor of a unitary matrix equals its algebraic complement, with the only essential difference that a square matrix's rows and columns can be Hamiltonian-preservingly permuted only by an arbitrary pair of identical permutations (unlike the permanent that allows independent arbitrary permutations of rows and columns).

**Corollary II.3.2.8 (in characteristic 2):**

Let U be a unitary matrix, I,J be sets of its row-indexes and K,L be sets of its column-indexes, |I|=|J|, |K|=|L|, $f_1, f_2$ be bijections $I \xrightarrow{f_1} J, K \xrightarrow{f_2} L$ correspondingly. Then

$$\mathrm{ham}(U^{[I \xrightarrow{f_1} J, K \xrightarrow{f_2} L]}) = \mathrm{ham}(U^{[J \xrightarrow{f_1^{-1}} I, L \xrightarrow{f_2^{-1}} K]})$$

where $f_1^{-1}, f_2^{-1}$ are the inverse bijections $J \xrightarrow{f_1^{-1}} I, L \xrightarrow{f_2^{-1}} K$ correspondingly.

**Definition:**

Let $A$ be an n×n-matrix, $\varepsilon$ be a formal infinitesimal. Then we'll call the matrix formal power series $U = \sum_{k=0}^{\infty} \varepsilon^k U_k$ , where each $U_k$ is an n×n-matrix over a ground field F, $U_0 = I_n$ and $U_1 = A$, an ***ε-unitarization*** of $A$ over $F(\varepsilon)$ if $U(A)$ is unitary as a matrix formal power series in $\varepsilon$.

It's easy to see that an $\varepsilon$-unitarization U of A exists in characteristic 2 if and only if $A = A^T$, while for a pair $i, j \in \{1, \dots, n\}$ $\mathrm{coef}_{\varepsilon^{n-1}}\mathrm{ham}\big(U^{(\setminus i, \setminus j)}\big) = \mathrm{ham}(A^{(\setminus i, \setminus j)})$ and thus it's $\#_2$P-complete as a function in the edge weights of the weighted digraph corresponding to A which is identically equal to zero if and only if this graph has no Hamiltonian path between the vertices i and j. Hence, taking into account the fact that for a unitary U the matrix $U^{(\setminus i, \setminus j)}$ is 1-semi-initary, we conclude that computing the Hamiltonian of a 1-semi-unitary matrix in characteristic 2 is a $\#_2$P-complete problem. It also implies, likewise, the $\#_2$P-completeness of computing the Hamiltonian of a unitary matrix over a ring of characteristic 4.

If for a an n×n-matrix A we define the matrix $H(A) = \{ham(A^{(\backslash i,\backslash j)})\}_{n\times n}$ then we'll receive, based on the above relation $ham(U^{[I\xrightarrow{f_1}J,K\xrightarrow{f_2}L]}) = ham(U^{[J\xrightarrow{f_1^{-1}}I,L\xrightarrow{f_2^{-1}}K]})$, the identity $UH^T(U) = H(U)U^T$.

And we may add that the partial inverse relation also concerns the Hamiltonian in characteristic 2:

**Lemma II.3.2.9 (in characteristic 2):**

For an $n_1 \times n_1$-matrix $A_{11}$ and an $n_2 \times n_2$-matrix $A_{22}$ ,

$$ham(\begin{pmatrix} A_{11} & A_{12} \\ A_{21} & A_{22} \end{pmatrix}) = \det^2(A_{11})ham(\begin{pmatrix} A_{11}^{-1} & A_{11}^{-1}A_{12} \\ A_{21}A_{11}^{-1} & A_{22} + A_{21}A_{11}^{-1}A_{12} \end{pmatrix})$$

Proof:

This fact can be easily proven via the identities $ham(\begin{pmatrix} I_n & A \\ I_n & A \end{pmatrix}) = ham(A)$ and $ham(\begin{pmatrix} B & BA \\ B & BA \end{pmatrix}) = \det^2(B)ham(\begin{pmatrix} I_n & A \\ I_n & A \end{pmatrix})$ for any two n×n-matrices $A, B$ (the latter relation is due to the fact that the Hamiltonian of a matrix with two equal rows isn't changed by adding one of them to a third row) when putting $B = \begin{pmatrix} A_{11}^{-1} & 0_{n_1\times n_2} \\ A_{21}A_{11}^{-1} & I_{n_2} \end{pmatrix}$ and permuting the rows and columns of $\begin{pmatrix} B & BA \\ B & BA \end{pmatrix})$ by the 2n-permutation (where $n = n_1 + n_2$) mapping $i$ and $n + i$ to each other for $i = 1, \dots, n_1$ and all the other elements from the set (1,…,2n} to themselves.

It has the following

**Corollary II.3.2.10 (in characteristic 2):**

Let X, Y, Z be n×n-matrices. Then

$$ham\begin{pmatrix} X & XZ \\ YX & YXZ \end{pmatrix} = \det^2(X)ham(\begin{pmatrix} 0_{n\times n} & Z \\ Y & 0_{n\times n} \end{pmatrix})$$

Besides, when speaking about the Hamiltonian in characteristic 2 that is a direct algebraic representation of a fundamental NP-complete problem, it would be worth noting the existence of a non-trivial class of digraphs whose arcs could be given non-zero weights over a field of characteristic 2 making their weighted adjacency matrices unitary. Let's call them ***weight-unitarazable*** over a ground field F, while such a system of arc weights we'll call a digraph's ***weight-unitarization*** over F. Let's consider several examples of digraphs weight-unitarizable over fields of characteristic 2.

One partial case is a system of pair-wise vertex-disjoint simple directed cycles whose arc set is partitioned into pairs of vertex-disjoint arcs (a,b) and (c,d) connected by two additional arcs (c,b) and (d,a) so that the four arcs form the anti-cycle $a \rightarrow b \leftarrow c \rightarrow d \leftarrow a$ and their weights satisfy the following system of equations:

weight(a,b)weight(c,b) = weight(a,d)weight(c,d),

weight(a,b) + weight(a,d) = weight(c,b) + weight(c,d) = 1.

Those systems are variable-disjoint for different anti-cycles and are solvable in linear time, while leaving, for each anti-cycle, one independent weight-variable as a parameter. In the case of its planarity, particularly, such a digraph depicts a city with a system of two way streets between one way cyclic roads around squares where the digraph's vertices correspond to the crossroads.

Another interesting example is the arc-digraph of a digraph (received via taking the initial digraph's arc set as the new vertex set, while two new vertices are connected if and only if they form, as initial arcs, a path of length 2) where some connections between new vertices (i.e. initial arcs) are removed so that for each initial vertex the remained connections form a weight-unitarizable digraph. This example generates a direct algebraic representation of a constrained Eulerian cycle problem where some passages between adjacent arcs are forbidden.

Tournaments can be conjectured to be weight-unitarizable in characteristic 2 as well.

Moreover, in characteristic 2 the Hamiltonian has a generalization that is also computable in polynomial time for unitary and involutory matrices.

**Definition:** let A be an n×n-matrix, w be an n-vector. Then its ***cycle polynomial*** is $\mathrm{cycle}(A, w) := \sum_{\pi \in S_n} \prod_{C \in \mathbb{C}(\pi)} (1 + \prod_{i \in C} w_i) \prod_{i=1}^{n} a_{i,\pi(i)}$ where $\mathbb{C}(\pi)$ is the set of $\pi$'s cycles.

**Theorem II.3.2.11:** let A be a unitary or involutory matrix, w be an n×n-vector. Then

$$\mathrm{cycle}(A, w) = \det(A^{\star 2} + \mathrm{Diag}(w))$$

As this function is polynomial-time computable for involutory matrices as well, we may analogically (with weight-unitarizable ones) define ***weight-involutorizable*** digraphs.

Hence the cycle polynomial can be considered for digraphs weight-unitarizable or weight-involutorizable in characteristic 2 and presumably serve as a direct algebraic representation of a number of problems on digraphs.

### II.3.3 The Hamiltonian cycle polynomial over idempotent rings of characteristic 2 and undirected graphs.

Another issue related to the Hamiltonian in characteristic 2 is its usage for undirected graphs. One can notice that for a symmetric n×n-matrix and two n-vectors b,c $\mathrm{ham}(\begin{pmatrix} A & b \\ c^T & 0 \end{pmatrix})$ is the sum of the products of the edge weights of Hamiltonian cycles through the edge (n+1,n+2) in the weighted undirected graph with n+2 vertices whose weighted adjacency matrix is $\begin{pmatrix} A & b & c \\ b^T & 0 & 1 \\ c^T & 1 & 0 \end{pmatrix}$. (For simplicity, further we'll call the product of the arc/edge weights of a path in a weighted digraph/graph the **path's weight**).

In this regard, over idempotent rings of characteristic 2 (whose partial case is GF(2)) the Hamiltonian obtains some additional properties unavailable over any fields of this characteristic bigger than GF(2), like the following:

**Theorem II.3.3.1 (over idempotent rings of characteristic 2):**

Let X be a symmetric n×n-matrix with the zero diagonal, y be an n-vector, n > 1. Then $\mathrm{ham}(\begin{pmatrix} X & y \\ \vec{1}_n^T X + \vec{1}_n^T + h y^T & 0 \end{pmatrix}) = 0$ for h = 0,1.

Proof:

The proof of this theorem is based on the fact that, due to the Hamiltonian's linearity on each row,

$$\mathrm{ham}(\begin{pmatrix} X & y \\ \vec{1}_n^T X + \vec{1}_n^T + h y^T & 0 \end{pmatrix}) = \mathrm{ham}(\begin{pmatrix} X & y \\ \vec{1}_n^T X & 0 \end{pmatrix}) + \mathrm{ham}(\begin{pmatrix} X & y \\ \vec{1}_n^T & 0 \end{pmatrix}) + \mathrm{ham}(\begin{pmatrix} X & y \\ y^T & 0 \end{pmatrix}) h.$$

The first summand $\mathrm{ham}(\begin{pmatrix} X & y \\ \vec{1}_n^T X & 0 \end{pmatrix})$ in the right side is, generally over any field for a weighted digraph with n+1 vertices, the sum of the weights of the graph's Hamiltonian cycles where the arc from the vertex n+1 was replaced by an arc with the same end and a beginning different from n+1. Due to characteristic 2 and the matrix X's symmetry (i.e. the symmetry of the digraph induced by the vertices 1,…,n), the first summand hence equals the sum of the weights of such transformed Hamiltonian cycles where the appearing "internal" cycle is of length 2. As any element of the ground ring is idempotent, it's exactly the sum of the weights of the digraph's Hamiltonian paths ending in the vertex n+1, i.e. the second summand $\mathrm{ham}(\begin{pmatrix} X & y \\ \vec{1}_n^T & 0 \end{pmatrix})$. And the third summand $\mathrm{ham}(\begin{pmatrix} X & y \\ y^T & 0 \end{pmatrix})h$ is zero for n > 1 because it's the Hamiltonian of a symmetric matrix with more than 2 rows.

This theorem's equality for h = 1 over GF(2), when completed by the two requirements

$\vec{1}_n^T y = (\vec{1}_n^T X + \vec{1}_n^T + y^T)\vec{1}_n = 0$ that make the corresponding undirected graph (whose adjacency matrix is $\begin{pmatrix} X & y & X\vec{1}_n + \vec{1}_n + y \\ y^T & 0 & 1 \\ \vec{1}_n^T X + \vec{1}_n^T + y^T & 1 & 0 \end{pmatrix}$ ) odd-degreed, is a generalization of the well-known theorem that any odd-degreed graph has an even number of Hamiltonian cycles through a given edge.

For an arbitrary symmetric n×n-matrix X (with an arbitrary diagonal) the relation of Theorem II.3.3.1 can be formulated, over idempotent rings of characteristic 2, as

$$\mathrm{ham}(\begin{pmatrix} X & y \\ \vec{1}_n^T X + (\mathrm{diag}(X))^T + \vec{1}_n^T + hy^T & 0 \end{pmatrix}) = 0$$

where $\mathrm{diag}(X) := \{x_{i,i}\}_n$, h = 0,1. Let's call this relation the **simple parity condition** for Hamiltonian cycles through the edge (n+1,n+2). (Actually, it's meaningful to use the word "parity" here because any idempotent ring of characteristic 2 is an extension of GF(2) that can be represented as the ring of k-variate Zhegalkin polynomials for some k. In this ring, unity is the only invertible element and accordingly the non-singularity of a matrix is equivalent to its determinant's equality to unity).

In the meantime, if we take an arbitrary matrix $\begin{pmatrix} A & b \\ c^T & 0 \end{pmatrix}$ where A is a symmetric n×n-matrix then, due to the above-given identity relating the Hamiltonians of a matrix and its partial inverse and the Hamiltonian's independence from diagonal entries, we get the following identity for an arbitrary diagonal n×n-matrix $D$:

$$\mathrm{ham}(\begin{pmatrix} A & b \\ c^T & 0 \end{pmatrix}) = \det^2(A+D)\mathrm{ham}(\begin{pmatrix} (A+D)^{-1} & (A+D)^{-1}b \\ c^T(A+D)^{-1} & 0 \end{pmatrix}).$$

Hence, when applying the simple parity condition for Hamiltonian cycles through the edge (n+1,n+2), we get the following condition implying, for any diagonal $D$ such that $A+D$ is nonsingular and h = 0,1, the equality $\mathrm{ham}(\begin{pmatrix} A & b \\ c^T & 0 \end{pmatrix}) = 0$:

$$c^T(A+D)^{-1} = \vec{1}_n^T(A+D)^{-1} + (\mathrm{diag}(A+D)^{-1})^T + \vec{1}_n^T + h((A+D)^{-1}b)^T$$

Upon right-multiplying by $(A+D)^{-1}$, it turns, due to the symmetry of $(A+D)^{-1}$, into

$$c^T = \vec{1}_n^T + (\mathrm{diag}(A+D)^{-1})^T(A+D) + \vec{1}_n^T(A+D) + hb^T$$

Upon transposing it and denoting $D = \mathrm{Diag}(d)$ where d is an n-vector, we hence obtain for h = 0,1:

$$c + hb + A\vec{1}_n + \vec{1}_n = (A + \mathrm{Diag}(d))\mathrm{diag}(A + \mathrm{Diag}(d))^{-1} + d$$

Let's call it the **diagonal parity condition** for Hamiltonian cycles through the edge (n+1,n+2).

Hence if the diagonal parity condition is solvable as an equation for $d$ then $\mathrm{ham}(\begin{pmatrix} A & b \\ c^T & 0 \end{pmatrix}) = 0.$

In the case when d = $\vec{0}_n$, h = 1 and A is a nonsingular matrix with the zero diagonal, it generates the simple parity condition, and if we also restrict this case by the two additional requirements $\vec{1}_n^T b = c^T\vec{1}_n = 0$ then over GF(2) we again receive the classical theorem about the parity of Hamiltonian cycles through a given edge in an odd-degreed graph.

And now, once more, let's use the fact that $\mathrm{ham}(\begin{pmatrix} A & b \\ c^T & 0 \end{pmatrix})$ is a linear function in the vector c. Let's express it from the diagonal parity condition, when it's fulfilled (for simplicity, further we'll always assume A with the zero diagonal):

$$c = hb + A\vec{1}_n + \vec{1}_n + (A + \mathrm{Diag}(d))\mathrm{diag}(A + \mathrm{Diag}(d))^{-1} + d$$

Let's denote by $\mathrm{DPCS}(A, b)$ (the **diagonal parity condition space** of A, b) the linear space over the ground idempotent ring $\mathcal{R}$ generated by all the vectors from the set $\left\{hb + A\vec{1}_n + \vec{1}_n + \left(A + \mathrm{Diag}(d)\right)\mathrm{diag}\left(A + \mathrm{Diag}(d)\right)^{-1} + d,\ h = 0,1,\ d \in GF^n(2), \det\left(A + \mathrm{Diag}(d)\right) = 1\right\}$

that is a subspace of $\mathcal{R}^n$ . We hence can now formulate the following statement:

**Theorem II.3.3.2 (over idempotent rings of characteristic 2):**

If $c \in \mathrm{DPCS}(A, b)$ then $\mathrm{ham}(\begin{pmatrix} A & b \\ c^T & 0 \end{pmatrix}) = 0.$

This theorem (that is also a generalization of the above-mentioned theorem about the parity of Hamiltonian cycles) provides an instrument for changing the vector $\gamma$ in the expression $\mathrm{ham}(\begin{pmatrix} A & b \\ \gamma^T & 0 \end{pmatrix})$ via adding any vector from $\mathrm{DPCS}(A, b)$, while unchanging the Hamiltonian. Hence if upon completing $\mathrm{DPCS}(A, b)$, for some $i \in \{1, \dots, n\}$, by the vector $e_i = \begin{pmatrix} \vec{0}_{i-1} \\ 1 \\ \vec{0}_{n-i} \end{pmatrix}$ it generates all the space $\mathcal{R}^n$ (i.e. when it's possible to turn $\gamma$ into $e_i$ via adding a vector from $\mathrm{DPCS}(A, b)$) then $\mathrm{ham}(\begin{pmatrix} A & b \\ \gamma^T & 0 \end{pmatrix}) = \mathrm{ham}(\begin{pmatrix} A & b \\ e_i^T & 0 \end{pmatrix})$ and in such a case this computational problem can be reduced to the same problem of a smaller size via removing the (n+1)-th row and the i-th column from the matrix $\begin{pmatrix} A & b \\ \gamma^T & 0 \end{pmatrix}$.

**Definition:**

Let A be a symmetric n×n-matrix, n > 2. Then we define $\mathrm{unham}(A) := \frac{1}{2}\mathrm{ham}(A)$ as its **undirected Hamiltonian cycle polynomial** (or, shortly, as its undirected Hamiltonian).

Like the Hamiltonian, the undirected Hamiltonian satisfies the partial inverse relation:

**Theorem II.3.3.3 (in characteristic 2):**

For a non-singular $n_1 \times n_1$-matrix $A_{11}$ and an $n_2 \times n_2$-matrix $A_{22}$,

$$\mathrm{unham}(\begin{pmatrix} A_{11} & A_{12} \\ A_{21} & A_{22} \end{pmatrix}) = \det{}^2(A_{11})\mathrm{unham}(\begin{pmatrix} A_{11}^{-1} & A_{11}^{-1}A_{12} \\ A_{21}A_{11}^{-1} & A_{22} + A_{21}A_{11}^{-1}A_{12} \end{pmatrix})$$

Proof:

For a (2n)×(2n)-matrix X, let's denote $\mathrm{per}_{-2,\mathrm{even}}(X) := \sum_{\pi\in S_{2n}^{(\mathrm{even})}} (-2)^{c(\pi)} \prod_{i=1}^{2n} x_{i,\pi(i)}$

where $S_{2n}^{(\mathrm{even})}$ is the set of 2n-permutations having only cycles of even lengths and $c(\pi)$ is the number of $\pi$'s cycles. (In characteristic 3, $\mathrm{per}_{-2,\mathrm{even}}(X) = \mathrm{per}_{\mathrm{even}}(X)$ .) Then we get, over an arbitrary ring of any characteristic, the identity

$$\mathrm{per}_{-2,\mathrm{even}}(X) := \sum_{L\subseteq\{1,\dots,2n\}} (-1)^{|L|} \det(X^{(L,L)}) \det(X^{(\backslash L,\backslash L)})$$

The proof of the theorem can be based on the following relation for a symmetric n×n-matrix A over a ring of characteristic 2 for n > 2 :

$$\mathrm{unham}(A) = \frac{\mathrm{per}_{-2,\mathrm{even}}(\begin{pmatrix} I_n & A \\ I_n & A \end{pmatrix})}{4}$$

where the right side should be understood as a quotient taken modulo 2 because $\mathrm{per}_{-2,\mathrm{even}}(\begin{pmatrix} I_n & A \\ I_n & A \end{pmatrix})$ is a multiple of 4 when n > 2. But, because of the above determinantal expansion of $\mathrm{per}_{-2,\mathrm{even}}(X)$, if a (2n)×(2n)-matrix X has two equal rows then one of them can be added to a third one, while unchanging $\mathrm{per}_{-2,\mathrm{even}}(X)$. Therefore $\mathrm{per}_{-2,\mathrm{even}}(\begin{pmatrix} B & BA \\ B & BA \end{pmatrix}) = \det^2(B)\mathrm{per}_{-2,\mathrm{even}}(\begin{pmatrix} I_n & A \\ I_n & A \end{pmatrix})$ for any non-singular n×n-matrix B. Hence putting $B = \begin{pmatrix} A_{11}^{-1} & 0_{n_1\times n_2} \\ A_{21}A_{11}^{-1} & I_{n_2} \end{pmatrix}$ and permuting the rows and columns of $\begin{pmatrix} B & BA \\ B & BA \end{pmatrix})$ by the 2n-permutation (where $n = n_1 + n_2$) mapping $i$ and $n + i$ to each other for $i = 1, \dots, n_1$ and all the other elements of the set (1,…,2n} to themselves completes the proof.

**Theorem II.3.3.4 (over idempotent rings of characteristic 2):**

Let X be a symmetric n×n-matrix with the zero diagonal, y be an n-vector, n > 1. Then

$$\mathrm{unham}(\begin{pmatrix} X & y \\ y^T & 0 \end{pmatrix}) = \mathrm{unham}(\begin{pmatrix} X & X\vec{1}_n + \vec{1}_n + y \\ \vec{1}_n^T X + \vec{1}_n^T + y^T & 0 \end{pmatrix}) + n \cdot \mathrm{unham}(X)$$

Proof:

We're going to use the following simple identity, implied by the undirected Hamiltonian's definition, for an n×n-matrix X and two n-vectors $a, b$:

$$\mathrm{unham}(\begin{pmatrix} X & a+b \\ a^T+b^T & 0 \end{pmatrix}) =$$

$$= \mathrm{unham}(\begin{pmatrix} X & b \\ b^T & 0 \end{pmatrix}) + \mathrm{ham}(\begin{pmatrix} X & a \\ b^T & 0 \end{pmatrix}) + \mathrm{unham}(\begin{pmatrix} X & b \\ b^T & 0 \end{pmatrix})$$

In our case, we obtain

$$\mathrm{unham}(\begin{pmatrix} X & X\vec{1}_n + \vec{1}_n + y \\ \vec{1}_n^T X + \vec{1}_n^T + y^T & 0 \end{pmatrix}) =$$

$$= \mathrm{unham}(\begin{pmatrix} X & X\vec{1}_n + \vec{1}_n \\ \vec{1}_n^T X + \vec{1}_n^T & 0 \end{pmatrix}) + \mathrm{ham}(\begin{pmatrix} X & y \\ \vec{1}_n^T X + \vec{1}_n^T + y^T & 0 \end{pmatrix}) + \mathrm{unham}(\begin{pmatrix} X & y \\ y^T & 0 \end{pmatrix})$$

The summand $\mathrm{ham}(\begin{pmatrix} X & y \\ \vec{1}_n^T X + \vec{1}_n^T + y^T & 0 \end{pmatrix})$ equals zero due to satisfying the diagonal parity condition.

The summand $\mathrm{unham}(\begin{pmatrix} X & X\vec{1}_n + \vec{1}_n \\ \vec{1}_n^T X + \vec{1}_n^T & 0 \end{pmatrix})$, in turn, can be further expanded as the sum $\mathrm{unham}(\begin{pmatrix} X & X\vec{1}_n \\ \vec{1}_n^T X & 0 \end{pmatrix}) + \mathrm{ham}(\begin{pmatrix} X & X\vec{1}_n \\ \vec{1}_n^T & 0 \end{pmatrix}) + \mathrm{unham}(\begin{pmatrix} X & \vec{1}_n \\ \vec{1}_n^T & 0 \end{pmatrix})$.

This sum's first summand $\mathrm{unham}(\begin{pmatrix} X & X\vec{1}_n \\ \vec{1}_n^T X & 0 \end{pmatrix})$ is the sum of the weights of the graph's Hamiltonian cycles where each of the two edges adjacent (in the cycle) to the vertex n+1 was replaced by an edge adjacent to this vertex's corresponding neighbor (in the cycle) and not adjacent to n+1. Due to characteristic 2, it's the sum of the weights of such transformed Hamiltonian cycles where both the appearing "internal" cycles are of length 2 and the sum of the weights of Hamiltonian cycles with exactly one edge taken twice (what corresponds to the case when both edges adjacent (in the cycle) to the vertex n+1 were replaced by the edge connecting them). As any element of the ground ring is idempotent, it's also the sum of the weights of Hamiltonian paths of the weighted graph induced by the vertices 1,…,n, i.e. the above sum's third summand $\mathrm{unham}(\begin{pmatrix} X & \vec{1}_n \\ \vec{1}_n^T & 0 \end{pmatrix})$ and $n \cdot \mathrm{unham}(X)$. Besides, this sum's second summand $\mathrm{ham}(\begin{pmatrix} X & X\vec{1}_n \\ \vec{1}_n^T & 0 \end{pmatrix})$ is, upon transposing, $\mathrm{ham}(\begin{pmatrix} X & \vec{1}_n \\ \vec{1}_n^T X & 0 \end{pmatrix}) = \mathrm{ham}(\begin{pmatrix} X & \vec{1}_n \\ \vec{1}_n^T X + \vec{1}_n^T + \vec{1}_n^T & 0 \end{pmatrix})$ and, accordingly, equals zero due to satisfying the diagonal parity condition. It completes the proof.

Hence Theorems II.3.3.3 and II.3.3.4 provide a couple of algebraic instruments we can change a symmetric matrix over idempotent rings of characteristic 2 by (together with adding an arbitrary diagonal matrix) while preserving its undirected Hamiltonian.

However, the introduced variety of affine Hamiltonian-preserving transformations we can subject a symmetric matrix to, as well as the above-given DPCS-algorithm for transforming a matrix when computing the sum of the weights of its Hamiltonian cycles through a given edge, is deprived, over idempotent rings of characteristic 2**,** of the core algebraic tool of infinite fields – the neighboring computation principle. It arises the question of their efficient usability for computing the undirected and directed Hamiltonians.

# III. The Schur complement compression on informationally sparse classes

Given an n×n-matrix class defined by a system of matrix-functions in a set of parameters, let's define its algebraic rank as the system's algebraic rank. And, in case if the class is defined by an algebraic equation system, we'll define its algebraic rank as the difference of $n^2$ and the algebraic equation system's algebraic rank. In both cases we'll call it the algebraic n-rank of such a matrix class and we'll also call such a matrix class **algebraically definable –** in fact, it's an exact analogy of the notion of a smooth manifold in characteristic 0. We may also assume that both the above-mentioned forms of a matrix class's definition are reducible to each other and yield the same algebraic rank.

In the present article's two previous chapters we discussed some matrix compressions that polynomial-time reduce the permanent of a matrix to the permanent of a derived matrix of a smaller size, and we dealt with either arbitrary matrices or k-semi-unitary ones etc., i.e. classes of n×n-matrices whose algebraic rank is either $n^2$ or $n^2/2$ + O(n).

Let's call an algebraically definable matrix class ***informationally dense*** if for any n the ratio of its algebraic n-rank and $n^2$ (that we'll call the ***informational n-density*** of the class) is bigger than a nonzero constant, and ***informationally sparse*** otherwise.

In this chapter we're going to study some informationally sparse matrix classes (particularly, those of informational n-density 1/O(n)) built via Cauchy and Cauchy-like matrices, as well as certain matrix compression operators (particularly, the Schur complement compression) that polynomial-time reduce one function to another on those classes, while still acting as genuine self-reducing compressions for certain introduced functions.

**Definition:**

For an n×m-matrix A we define $\mathrm{per}(A)$ as $\sum_{\substack{J, J\subseteq\{1,\dots,m\} \\ |J|=n}} \mathrm{per}(A^{(\{1,\dots,n\},J)})$ if $n \le m$ and zero otherwise.

And, once again throughout the chapter, the neighboring computation principle is supposed to serve as a chief algebraic instrument the principal below-introduced polynomial-time reductions would be impossible without. In this regard, we'll also need the following related definition:

**Definition:**

For a field F, a formal infinitesimal ε, and F's ε-extension $F(\varepsilon) := \{u = \sum_{k=\mathrm{order}_\varepsilon(u)}^{\infty} u_k \varepsilon^k, \mathrm{order}_\varepsilon(u) \in \mathbb{Z},\ u_k \in F \text{ for } k = \mathrm{order}_\varepsilon(u), \dots, \infty\}$, let's define for $u \in F(\varepsilon)$ $\lim_{\varepsilon \to 0} u := \begin{cases} u_0 & \text{if } \mathrm{order}_\varepsilon(u) \ge 0 \\ \text{a nonexistent (infinitely big) element otherwise} \end{cases}$

And we'll call $\mathrm{order}_\varepsilon(u)$ the order of u on $\varepsilon$ (or, shortly, the $\varepsilon$-order of u).

**Denotation:** for a matrix A, a subset I of its row set and a subset J of its column set, by $\mathrm{Schur}_{I,J}(A)$ we'll denote (for the purpose of simplicity) the Schur complement $A/A^{(I,J)}$.

**Definition:**

Given two matrices $A = \{a_{i,j}\}_{n\times m}$ and $B = \{b_{i,j}\}_{n\times m}$, by $A \star B := \{a_{i,j}b_{i,j}\}_{n\times m}$ we'll denote their Hadamard product.

Given a rational number k, by $A^{\star k} = \{a_{i,j}^{k}\}_{n\times m}$ we'll denote the k-th Hadamard power of A and, given a sequence of rational numbers $(k_1, \dots, k_s)$, we'll call the matrix $A^{\star(k_1,\dots k_s)} := \begin{pmatrix} A^{\star k_1} \\ \dots \\ A^{\star k_s} \end{pmatrix}$ its $(k_1, \dots, k_s)$-th Hadamard-power.

**Definition:**

Let x be a vector, k be a natural number and dim(x) $= n \equiv 0 \pmod{k}$.

Then $W^{[k]}(x) := (x^T)^{\star(0 1_k^T, \dots, (\frac{n}{k}-1) 1_k^T)}$

where $1_k^T$ denotes the k-sequence all whose entries are 1.

**Definition:** for a vector y, we denote

$$\mathrm{Van}^{[k]}(y) := \begin{pmatrix} (y^T)^{\star 0} \\ (y^T)^{\star 1} \\ \dots \\ (y^T)^{\star(k-1)} \end{pmatrix}$$

and we also denote the transposed Vandermonde matrix of y as

$\mathrm{Van}(y) := \mathrm{Van}^{[\dim(y)-1]}(y) = W^{[1]}(y)$.

**Definition:** let x, y be two vectors. Then we denote $\mathbf{pol}(\mathbf{x},\mathbf{y}) := \prod_{\mathbf{i=1}}^{\mathbf{dim(x)}} \prod_{\mathbf{j=1}}^{\mathbf{dim(y)}} (\mathbf{x_i} - \mathbf{y_j})$ .

**Definition:**

1) For two vectors x, y, let's define their Cauchy matrix $C(x,y) := \{\frac{1}{x_i - y_j}\}_{\dim(x)\times\dim(y)}$ where we'll call $x_i$ its i-th row (or left) **denominator-value** and $y_j$ its j-th column (or right) **denominator-value**.
2) For an n-vector x, let's define $\tilde{C}(x)$ as an n×n-matrix whose i,j-th entry is $\frac{1}{x_i - x_j}$ if $i \neq j$, $i, j \in \{1, \dots, n\}$, and 0 otherwise. We'll call it a ***Cauchy-wave matrix*** and $x_i$ its i-th row and column denominator-value (or just the i-th denominator-value).
3) For three vectors x, y, z, let's also define $\tilde{C}(x,y,z) := \begin{pmatrix} \tilde{C}(x) & C(x,z) \\ C(y,x) & C(y,z) \end{pmatrix}$

   We'll call it a ***Cauchy-waved matrix***.

**Definition:**

Let a be an n-vector and b be an m-vector. Then its Kronecker sum is defined as

$$a\ddot{+}b := a \otimes \vec{1}_m + \vec{1}_n \otimes b = \begin{pmatrix} a_1\vec{1}_{\dim(m)} + b \\ \dots \\ a_n\vec{1}_{\dim(m)} + b \end{pmatrix}$$

**Definition:** let $u, w, v, g$ be vectors, $\dim(v) = \dim(g)$. Then

$$\varphi_{p,h}(u, w, v, g) := \sum_{J,|J|=\dim(w)} \Big(\sum\nolimits_{j\in J} \prod\nolimits_{i=1}^{\dim(u)} \frac{1}{(u_i - v_j)^p}\Big) \det^h(C(w, v_J)) \prod_{j\in J} g_j$$

and we'll call it ***the Cauchy determinant base-sum***, while calling the vector $u$ ***the Cauchy base-vector*** and $p$ ***the Cauchy base-degree***.

**Definition:** let A, B be two skew-symmetric 2n×2n-matrices. Then

$$\xi_m(A, B) := \sum_{I\subseteq\{1,\dots,2n\},|I|=m} \mathrm{Pf}(A^{(I,I)})\,\mathrm{Pf}(B^{(I,I)})$$

**Theorem III.1:** let A, B be two skew-symmetric 2n×2n-matrices. Then for an even m

$$\xi_m(A, B) = \mathrm{coef}_{\omega^{m/2}} \mathrm{Pf}\Big(\begin{pmatrix} \omega A & I_{2n} \\ -I_{2n} & B \end{pmatrix}\Big)$$

**Definition:** for a rational number r, an even number m and vectors $x, y, z, t, \alpha, \beta, \gamma, \lambda$ such that $\dim(x) = \dim(\alpha)$, $\dim(y) = \dim(\beta)$, $\dim(z) = \dim(\gamma)$, $\dim(t) = \dim(\lambda)$, we define

$$\eta_{r,m}(t, \lambda) := \sum_{L,|L|=m} \det^r\big(\tilde{C}(t_L)\big) \prod_{l\in L} \lambda_l$$

$$\check{\eta}_{r,m}(x, y, z, t, \alpha, \beta, \gamma, \lambda) :=$$

$$:= \sum_{\substack{I,J,K,L \\ |L|=m,|I|=|J|=m/2-|K|}} \det^r\big(\tilde{C}(t_L)\big)\,\mathrm{per}\Big(C\Big(\begin{pmatrix} x_I \\ z_K \end{pmatrix} \otimes \vec{1}_2, t_L\Big)\Big)\mathrm{per}\Big(C\Big(\begin{pmatrix} y_J \\ z_K \end{pmatrix} \otimes \vec{1}_2, t_L\Big)\Big)\Big(\prod_{i\in I} \alpha_l\Big)\Big(\prod_{j\in J} \beta_l\Big)\Big(\prod_{k\in K} \gamma_l\Big) \prod_{l\in L} \lambda_l$$

We'll call the first expression **the Cauchy-wave sum** and the second one **the extended Cauchy-wave sum**.

**Theorem III.2 (in characteristic p):**

$$\eta_{\frac{p^q+1}{2},m}(t, \lambda) = \xi_m\Big(\mathrm{Diag}(\lambda^{\star(1/2)})\tilde{C}(t)\mathrm{Diag}(\lambda^{\star(1/2)}), \tilde{C}\big(t^{\star p^q}\big)\Big)$$

**Theorem III.2.1 (in characteristic 3):**

let $\dim(y) = m/2$. Then

$$\check{\eta}_{\frac{3^q-1}{2},m}\left(x, y, \emptyset, t, \alpha, \vec{1}_{m/2}, \emptyset, \lambda\right) := \det{}^2(\mathrm{Van}(y))\,\frac{\xi_m\left(B^T A B, \tilde{C}\left(t^{\star 3^q}\right)\right)}{\mathrm{Pf}(A)}$$

where $B = \begin{pmatrix} \mathrm{Van}^{[m/2]}(x)\mathrm{Diag}(\alpha)C(x,t) \\ \mathrm{Van}^{[m/2]}(t) \end{pmatrix} \mathrm{Diag}\left(\left\{\frac{\lambda_h}{\mathrm{pol}(t_h,y)}\right\}_{\dim(t)}\right)$ and A is any non-singular skew-symmetric m×m-matrix.

**Theorem III.2.2 (in characteristic 3):**

Let q > 1. Then

$$\check{\eta}_{\frac{3^q-1}{2},m}(x, y, z, t, \alpha, \beta, \gamma, \lambda) =$$

$$= \mathrm{coef}_{\omega^m} \lim_{\varepsilon \to 0} \frac{\check{\eta}_{\frac{3^q-1}{2},d}\left(\begin{pmatrix} x \\ y \\ z \end{pmatrix}, \begin{pmatrix} y \\ z \end{pmatrix}, \emptyset, \begin{pmatrix} t \\ y \overset{\cdots}{+} \begin{pmatrix} \varepsilon \\ -\varepsilon \end{pmatrix} \\ z \overset{\cdots}{+} \begin{pmatrix} \varepsilon \\ -\varepsilon \end{pmatrix} \end{pmatrix}, \begin{pmatrix} \alpha \\ \varepsilon \beta^{\star(-1)} \\ \varepsilon^{-1}\gamma^{\star(-1)} \end{pmatrix}, \vec{1}_{d/2}, \emptyset, \begin{pmatrix} \omega\lambda \\ \varepsilon^{\frac{3^q}{2}+1}\vec{1}_{2\dim(y)} \\ \varepsilon^{\frac{3^q+3}{2}}\vec{1}_{2\dim(z)} \end{pmatrix}\right)}{\left(\prod_{j=1}^{\dim(\beta)} \frac{1}{\beta_j}\right) \prod_{k=1}^{\dim(\gamma)} \frac{1}{\varepsilon\gamma_k}}$$

where $d = \dim(y) + \dim(z)$.

**Definition:** for three n-vectors x, a, d, let's define the skew-symmetric n×n-block-matrix with 2×2-blocks (i.e. 2n×2n-matrix) $K(x,d,a) := \{K_{ij}\}_{n\times n}$

where

$K_{ii} := \begin{pmatrix} 0 & a_i \\ -a_i & 0 \end{pmatrix}$ for i = 1,…,n

$$K_{ij} := \begin{pmatrix} \frac{1}{x_i - x_j} & \left(\frac{\partial}{\partial x_j} + d_j \frac{\partial^2}{\partial {x_j}^2}\right)\frac{1}{x_i - x_j} \\ \left(\frac{\partial}{\partial x_i} + d_i \frac{\partial^2}{\partial {x_i}^2}\right)\frac{1}{x_i - x_j} & \left(\frac{\partial}{\partial x_i} + d_i \frac{\partial^2}{\partial {x_i}^2}\right)\left(\frac{\partial}{\partial x_j} + d_j \frac{\partial^2}{\partial {x_j}^2}\right)\frac{1}{x_i - x_j} \end{pmatrix}$$

for $i \neq j$, $i, j \in \{1, \dots, n\}$

In the above-defined matrix, we'll call $d_i$ ***the differentiation-weight*** and $a_i$ ***the absence-weight*** corresponding to the denominator-value $x_i$.

### Theorem III.3 (The Borchardt formula, in any characteristic):

Let dim(y) = dim(z). Then $\mathrm{per}\big(C(y,z)\big) = \frac{\det(C^{*2}(y,z))}{\det(C(y,z))}$ .

**Lemma III.4 (about square Cauchy-waved matrices, in characteristic zero if not specified otherwise):**

1) for dim(x) > 2: $\mathrm{ham}\left(\tilde{C}(x)\right) = 0$

2) for dim(y) = dim(z) > 0:

$$\mathrm{ham}(\tilde{C}(x,y,z)) = \mathrm{ham}(C(y,z)) \prod_{i=1}^{\dim(x)} \Big( \sum_{j=1}^{\dim(y)} \frac{1}{y_j - x_i} - \sum_{k=1}^{\dim(z)} \frac{1}{z_k - x_i} \Big)$$

3) for dim(y)=dim(z):

$$\mathrm{per}(\tilde{C}(x,y,z)) = \mathrm{per}(C(y,z))\mathrm{per}(\tilde{C}(x) + \mathrm{Diag}(\Big\{\sum_{j=1}^{\dim(y)} \frac{1}{y_j - x_i} - \sum_{k=1}^{\dim(z)} \frac{1}{z_k - x_i}\Big\}_{\dim(x)}))$$

4) for dim(y) = dim(z):

$$\det(\tilde{C}(x,y,z)) = \det\big(C(y,z)\big)\det(\tilde{C}(x) - \mathrm{Diag}(\Big\{\sum_{j=1}^{\dim(y)} \frac{1}{y_j - x_i} - \sum_{k=1}^{\dim(z)} \frac{1}{z_k - x_i}\Big\}_{\dim(x)})) ,$$

$$\det\big(C(y,z)\big) = \frac{\det(\mathrm{Van}(y))\det(\mathrm{Van}(z))}{\mathrm{pol}(y,z)}$$

5) for dim(x) = 2n:

$$\mathrm{Pf}(\tilde{C}(x)) = \frac{\sum_{I\subset\{1,\dots,2n\},|I|=n} \det^2(\mathrm{Van}(x_I))\det^2(\mathrm{Van}(x_{\setminus I}))}{2^n \det(\mathrm{Van}(x))}$$

$$\det(\tilde{C}(x)) = (-1)^n \mathrm{per}(\tilde{C}(x)) = \frac{\sum_{I\subset\{1,\dots,2n\},|I|=n} \det^4(\mathrm{Van}(x_I))\det^4(\mathrm{Van}(x_{\setminus I}))}{2^n \det^2(\mathrm{Van}(x))} = \mathrm{haf}(\tilde{C}^{*2}(x))$$

6)

$$\mathrm{Pf}(\tilde{C}(x)) = \frac{\mathrm{per}^2(W^{[2]}(x))}{\det(\mathrm{Van}(x))} \qquad \text{in characteristic 3}$$

$$\mathrm{Pf}(\tilde{C}(x)) = \frac{\mathrm{per}(W^{[4]}(\binom{x}{x}))}{\det(\mathrm{Van}(x))} \qquad \text{in characteristic 5}$$

7) in a prime characteristic p, for dim(y) = (p-1)dim(x):

$$\mathrm{per}\left(C(x \otimes \vec{1}_{p-1}, y)\right) = \frac{\det^{p-1}(\mathrm{Van}(x))\mathrm{per}\big(W^{[p-1]}(y)\big)}{\mathrm{pol}(x,y)}$$

Proof:

The proofs of all the lemma's statements can be based on the first of them (which is well-known) and the second one for dim(x) = dim(y) = 1, as well as on the Borchardt identity.

The statement (2), provable by the induction on dim(x), is a key result that implies the statements (3) and (4): we use the fact that the determinant of a square matrix is the sum, over all its transversals, of the product of the transversal's entries multiplied by $(-1)^{|C_1|+\cdots+|C_k|+k}$ where $|C_1|, \ldots, |C_k|$ are the lengths of its cycles.

The first identity of the statement (5) follows from the relation

$\mathrm{Pf}(\tilde{C}(x)) = \frac{\sum_{I,|I|=n} \sigma(\pi(I))\det(C(x_I, x_{\setminus I}))}{2^n}$ where for $I = \{i_1, \ldots, i_n\}, 1 \le i_1 < \cdots < i_n \le 2n$, $\sigma(\pi(I))$ is the sign of the 2n-permutation $\pi(I) = \begin{pmatrix} 1 \ldots n & n+1 \ldots 2n \\ i_1 \ldots i_n & \hat{i}_1 \ \ldots \ \hat{i}_n \end{pmatrix}$ where $\hat{I} = \{1, \ldots, 2n\} \setminus I = \{\hat{i}_1, \ldots, \hat{i}_n\}$, $1 \le \hat{i}_1 < \cdots < \hat{i}_n \le 2n$ (this formula is a partial case of the identity for a skew-symmetric 2n×2n-matrix A $\mathrm{Pf}(A) = \frac{\sum_{I,|I|=n} \sigma(\pi(I))\det(A^{(I,\setminus I)})}{2^n}$) because $\sigma(\pi(I))$ is also the ratio $\frac{\det(\mathrm{Van}(x_I))\mathrm{pol}(x_I, x_{\setminus I})\det(\mathrm{Van}(x_{\setminus I}))}{\det(\mathrm{Van}(x))}$ and its second identity follows from the relation $\det(\tilde{C}(x)) = \frac{\sum_{I,|I|=n} \det^2(C(x_I, x_{\setminus I}))}{2^n}$ (that is a partial case of the identity for a skew-symmetric 2n×2n-matrix A $\det(A) = \frac{\sum_{I,|I|=n} \det^2(A^{(I,\setminus I)})}{2^n}$).

The statement (6) for characteristic 3 is due to the identity (for this characteristic)

$$\mathrm{per}^2\left(W^{[2]}(x)\right) = \mathrm{per}\left(\begin{pmatrix} 0_{n\times n} & (W^{[2]}(x))^T \\ W^{[2]}(x) & 0_{n\times n} \end{pmatrix}\right) =$$

$$= 2^n \sum_{I \subset \{1,\ldots,2n\}, |I|=n} \det^2(\mathrm{Van}(x_I))\det^2(\mathrm{Van}(x_{\setminus I}))$$

that is implied by the formula for an m×m-matrix A proven earlier in the article: $\mathrm{per}(A) = (-1)^m \sum_{I \subseteq \{1,\ldots,m\}} \det(A^{(I,I)})\det(A^{(\{1,\ldots,m\}\setminus I, \{1,\ldots,m\}\setminus I)})$ in characteristic 3.

And for characteristic 5 it follows from the fact that in this characteristic there holds the identity $\lim_{\varepsilon \to 0}\left(\varepsilon \frac{\mathrm{per}\left(W^{[4]}\left(\begin{pmatrix} y \\ u \\ u+\varepsilon \\ y \\ u \\ u+\varepsilon \end{pmatrix}\right)\right)}{\det\left(\mathrm{Van}\left(\begin{pmatrix} y \\ u \\ u+\varepsilon \end{pmatrix}\right)\right)}\right) = \frac{\mathrm{per}\left(W^{[4]}\left(\begin{pmatrix} y \\ y \end{pmatrix}\right)\right)}{\det(\mathrm{Van}(y))}$ and $\mathrm{Pf}(\tilde{C}(x))$ satisfies the same functional equation, while $\mathrm{Pf}(\tilde{C}(x))$ is a fraction whose denominator is $\det(\mathrm{Van}(x))$ and whose numerator is a homogenous polynomial in x of degree $\frac{\dim^2(x)-2\dim(x)}{2}$ that is the degree of the homogenous polynomial $\mathrm{per}(W^{[4]}(\binom{x}{x}))$.

The statement (7) can be received via multiplying the j-th column of $C(x\otimes\vec{1}_{p-1},y)$ by $\prod_{i=1}^{\dim(x)}(x_i-y_j)$ and turning this matrix, via linear operations with (p-1)-tuples of rows $C(x_i\vec{1}_{p-1},y)$, into $W^{[p-1]}(y)$.

**Lemma III.5 (about rectangular Cauchy matrices)**

1) In characteristic zero:

$$\mathrm{per}(C(y,z)) = \det(\tilde{C}(y) + \mathrm{Diag}(\left\{\sum_{k,k\neq j}\frac{1}{y_j-y_k}-\sum_{k=1}^{\dim(z)}\frac{1}{y_j-z_k}\right\}_{\dim(y)}))$$

2) In characteristic 3: $\mathrm{per}(C(y,z)) = (-1)^{\dim(y)}\mathrm{per}(C(y,x))$ for any vector $\begin{pmatrix}x\\y\\z\end{pmatrix}$ such that $\frac{\partial^2}{\partial\nu^2}\mathrm{pol}(\nu,\begin{pmatrix}x\\y\\z\end{pmatrix})\equiv 0$ (identically as a polynomial in the formal scalar variable ν)

3) in a prime characteristic p, for dim(y) = (p-1)dim(x):

$$\mathrm{per}\left(C\left(x\otimes\vec{1}_{p-1},y\right)\right) = \frac{\left(\prod_{i=1}^{\dim(x)}\frac{\partial^{p-1}}{\partial x_i^{p-1}}\right)\left(\det\left(\mathrm{Van}(x)\right)\mathrm{pol}(x,y)\right)}{\det\left(\mathrm{Van}(x)\right)\mathrm{pol}(x,y)}$$

$$\left(\prod_{i=1}^{\dim(x)}\left(1+d_i\frac{\partial}{\partial x_i}\right)\right)\mathrm{per}\left(C\left(x\otimes\vec{1}_{p-1},y\right)\right) =$$

$$= \mathrm{per}\left(C\left(x\otimes\vec{1}_{p-1},y\right)\right)\mathrm{per}(\mathrm{Diag}(d)\tilde{C}(x) + \mathrm{Diag}(\left\{1-d_i\left(\sum_{k,k\neq i}\frac{1}{x_i-x_k}-\sum_{j=1}^{\dim(y)}\frac{1}{x_i-y_j}\right)\right\}_{\dim(x)}))$$

Proof:

The statement (1) follows directly from the Borchardt formula in the case dim(y) = dim(z) because

$\mathrm{per}\left(C(y,z)\right) = \frac{\det(C^{*2}(y,z))}{\det(C(y,z))} = \det\left(C^{*2}(y,z)C^{-1}(y,z)\right) =$

$$= \det(\tilde{C}(y) + \mathrm{Diag}(\left\{\sum_{k,k\neq j}\frac{1}{y_j - y_k} - \sum_{k=1}^{\dim(z)}\frac{1}{y_j - z_k}\right\}_{\dim(y)}))$$

And when dim(y) < dim(z), in the generic case there is a dim(y)-vector $\hat{z}$ such that $\sum_{k=1}^{\dim(z)}\frac{1}{y_j - z_k} = \sum_{k=1}^{\dim(z)}\frac{1}{y_j - \hat{z}_k}$ for j = 1,…,dim(y).

The statement (2) follows from the statement (1) because in characteristic 3 the condition $\frac{\partial^2}{\partial \nu^2}\mathrm{pol}(\nu, \begin{pmatrix} x \\ y \\ z \end{pmatrix}) \equiv 0$ implies for j = 1,…,dim(y)

$$\sum_{k,k\neq j}\frac{1}{y_j - y_k} - \sum_{k=1}^{\dim(z)}\frac{1}{y_j - z_k} = -(\sum_{k,k\neq j}\frac{1}{y_j - y_k} - \sum_{k=1}^{\dim(x)}\frac{1}{y_j - x_k})$$

and

$$\det(\tilde{C}(y) - \mathrm{Diag}(\left\{\sum_{k,k\neq j}\frac{1}{y_j - y_k} - \sum_{k=1}^{\dim(z)}\frac{1}{y_j - z_k}\right\}_{\dim(y)})) =$$

$$= (-1)^{\dim(y)}\det(\tilde{C}(y) + \mathrm{Diag}(\left\{\sum_{k,k\neq j}\frac{1}{y_j - y_k} - \sum_{k=1}^{\dim(z)}\frac{1}{y_j - z_k}\right\}_{\dim(y)}))$$

due to the skew-symmetry of $\tilde{C}(y)$.

**Definition:**

Let A be a square matrix, then $\mathrm{per}_\lambda(A) := \sum_{\pi \in S_n}\lambda^{c(\pi)}\prod_{i=1}^{n} a_{ij}$ where $c(\pi)$ is the number of cycles in the permutation $\pi$ .

**Theorem III.6 (in any prime characteristic p > 2):**

$$(\prod_{i=1}^{n}(1 + d_i\frac{\partial}{\partial x_i}))\mathrm{per}_{\frac{1}{4}}(\tilde{C}^{\star 2}(x) + \mathrm{Diag}(a)) = \frac{1}{2^n}\mathrm{Pf}\big(K(x, d, a)\big)$$

Proof:

first let's prove the validness of this identity for the case of $a = d = \vec{0}_n$. In an arbitrary prime characteristic p, let's consider the expression

$$(*) \quad \lim_{\varepsilon \to 0}\frac{\sum_{I\subseteq\{1,\dots,2n\}}(-1)^{n+|I|/2}\det^{\frac{p+1}{2}}(\varepsilon\tilde{C}(\begin{pmatrix} x_1 \\ x_1+\varepsilon \\ \dots \\ x_n \\ x_n+\varepsilon \end{pmatrix})^{(I,I)})}{\varepsilon^{2n}}$$

that, due to the anti-symmetry of the matrix $\tilde{C}(\begin{pmatrix} x_1 \\ x_1+\varepsilon \\ \dots \\ x_n \\ x_n+\varepsilon \end{pmatrix})$ and the well-known fact that the determinant of a skew-symmetric matrix is the square of its Pfaffian, is identical to the expression

$$(**)\quad \lim_{\varepsilon\to 0}\frac{\sum_{I\subseteq\{1,\dots,2n\}}(-1)^{n+|I|/2}\mathrm{Pf}^{p+1}(\varepsilon\tilde{C}(\begin{pmatrix} x_1 \\ x_1+\varepsilon \\ \dots \\ x_n \\ x_n+\varepsilon \end{pmatrix})^{(I,I)})}{\varepsilon^{2n}} = \mathrm{Pf}\left(K\left(x,\vec{0}_n,\vec{0}_n\right)\right).$$

Let's show that the former expression (*) is $2^n \mathrm{per}_{\frac{1}{4}}(\tilde{C}^{\star 2}(x))$.

First of all, we know that for any vector $z$ $\det\left(\tilde{C}(z)\right) = \mathrm{haf}(\tilde{C}^{\star 2}(z))$ and, hence, we can re-write the expression (*) as

$$(***)\quad \lim_{\varepsilon\to 0}\frac{\sum_{I\subseteq\{1,\dots,2n\}}(-1)^{n+|I|/2}\mathrm{haf}^{\frac{p+1}{2}}(\varepsilon^2\tilde{C}^{\star 2}(\begin{pmatrix} x_1 \\ x_1+\varepsilon \\ \dots \\ x_n \\ x_n+\varepsilon \end{pmatrix})^{(I,I)})}{\varepsilon^{2n}}$$

Secondly, due to the summation and the limit $\lim_{\varepsilon\to 0}$ (producing a "weight" $O(\varepsilon^2)$ for each infinitely-close pair $x_i, x_i+\varepsilon$), for i = 1,…,n, among the $\frac{p+1}{2}$ multipliers each of whom is $\mathrm{haf}(\varepsilon^2\tilde{C}^{\star 2}(\begin{pmatrix} x_1 \\ x_1+\varepsilon \\ \dots \\ x_n \\ x_n+\varepsilon \end{pmatrix})^{(I,I)}$ there should be exactly one where the term $\frac{\varepsilon^2}{(x_i-x_i-\varepsilon)^2} = 1$ isn't to be taken what hence implies the appearance (exactly in one of the multipliers) of a cycle $\mathfrak{C}$ connecting (by the taken terms $\frac{\varepsilon^2}{(x_i-x_j+O(\varepsilon))^2}$) those "untaken singularities" associated with pairs $x_i, x_i+\varepsilon$. The limit $\lim_{\varepsilon\to 0}$ and the denominator of the fraction $\frac{\sum_{I\subseteq\{1,\dots,2n\}}(-1)^{n+|I|/2}\mathrm{haf}^{\frac{p+1}{2}}(\varepsilon^2\tilde{C}^{\star 2}(\begin{pmatrix} x_1 \\ x_1+\varepsilon \\ \dots \\ x_n \\ x_n+\varepsilon \end{pmatrix})^{(I,I)}}{\varepsilon^{2n}}$ turn the product of $\mathfrak{C}$'s terms into the product of the corresponding terms $\frac{1}{(x_i-x_j)^2}$. The whole expression (***) hence turns into $\sum_{I_1,\dots,I_{(p+1)/2}} \prod_{q=1}^{(p+1)/2} \mathrm{haf}(A^{(I_q,I_q)})$ where A is the matrix $\tilde{C}^{\star 2}(\begin{pmatrix} x_1 \\ x_1+\varepsilon \\ \dots \\ x_n \\ x_n+\varepsilon \end{pmatrix})$ with all the "infinitely big" entries $\frac{1}{(x_i - x_i - \varepsilon)^2}$ replaced by zeros and each $I_q$ is a subset of the set of pairs $\{(2i-1, 2i), i = 1, \dots, n\}$, while the $\frac{p+1}{2}$-tuple $I_1, \dots, I_{(p+1)/2}$ runs over all its partitions (possibly including empty sets). Besides, the cycle $\mathfrak{C}$ can be considered as directed and is, in fact, the corresponding directed cycle in $\mathrm{per}_{1/4}(\tilde{C}^{\star 2}(x))$ with its coefficient ¼ that is multiplied, in (*), by $2^l$ (where $l$ is its length considered as the number of singularities it connects) because:

(with each denominator-value $x_i$ or $x_i + \varepsilon$ we'll further associate a vertex in the corresponding weighted graph with the weighted adjacency matrix A; such a vertex is to "appear" exactly in one of $\mathrm{haf}(A)'$s (p+1)/2 copies) for $l$ > 2 $\mathfrak{C}$'s direction is determined by the connection of the vertex $x_{\min(\mathfrak{C})}$ of $\mathfrak{C}$'s lexicographically minimal ε-close pair $x_{\min(\mathfrak{C})}, x_{\min(\mathfrak{C})} + \varepsilon$ (i.e. by the ε-close pair whose vertex is connected with $x_{\min(\mathfrak{C})}$ in $\mathfrak{C}$), while for each direction (including the case of $l$ = 2 when there is only one direction) there are $2^{l-1}$ variants of $\mathfrak{C}$-forming systems of connections (as in the pair $x_{\min(\mathfrak{C})}, x_{\min(\mathfrak{C})} + \varepsilon$ we already cannot choose a vertex, while in all the other ε-close pairs of $\mathfrak{C}$ we choose one vertex from a pair when coming to it from the preceding pair while traversing $\mathfrak{C}$). Besides, there are $\frac{p+1}{2} \equiv \frac{1}{2} (\mathrm{mod}\ p)$ variants of locating $\mathfrak{C}$ (independently of other connecting cycles) in one of the multipliers what, altogether, gives the overall combinatorial coefficient $2^{l-1}\frac{1}{2} = \frac{1}{4}2^l$ for each connecting cycle. We just should add that we hence built a natural bijection between the singularity-connecting directed cycles and the directed cycles of $\mathrm{per}_{1/4}(\tilde{C}^{\star 2}(x))$.

And, as well, it's easy to realize that the expression (**) is $\mathrm{Pf}(K(x, \vec{0}_n, \vec{0}_n))$ because in the multiplier $\mathrm{Pf}^p(\varepsilon\tilde{C}(\begin{pmatrix} x_1 \\ x_1+\varepsilon \\ \dots \\ x_n \\ x_n+\varepsilon \end{pmatrix}))$ we should take all the singularity terms $\frac{\varepsilon^p}{(x_i - x_i - \varepsilon)^p}$ due to obtaining, in the numerator of (**), "at least" the "weight" $O(\varepsilon^{2p-1})$ otherwise for each singularity where it's not taken. Let's explain it even in the "best" case when in $\mathrm{Pf}(\varepsilon\tilde{C}(\begin{pmatrix} x_1 \\ x_1+\varepsilon \\ \dots \\ x_n \\ x_n+\varepsilon \end{pmatrix}))$ we take the term $\frac{\varepsilon}{x_i - x_i - \varepsilon}$ for this singularity: those untaken singularities of $\mathrm{Pf}^p(\varepsilon\tilde{C}(\begin{pmatrix} x_1 \\ x_1+\varepsilon \\ \cdots \\ x_n \\ x_n+\varepsilon \end{pmatrix}))$ would also form cycles with connecting terms of the type $O(\varepsilon^p)$ and, besides, in each untaken singularity in $\mathrm{Pf}^p(\varepsilon\tilde{C}(\begin{pmatrix} x_1 \\ x_1+\varepsilon \\ \cdots \\ x_n \\ x_n+\varepsilon \end{pmatrix}))$ the edge corresponding to $x_i+\varepsilon$ should be replaced by its differential on $\varepsilon$ to prevent the "resulting" Pfaffian from having a pair of identical rows and a pair of identical columns, while in the numerator of (**) for each singularity we're supposed to get, due to the limit and the denominator $\varepsilon^{2n}$, the "weight" not "smaller" than $O(\varepsilon^2)$ (because each singularity necessarily produces it due to the summation).

And now let's prove the theorem's identity for arbitrary $a, d$.

For i = 1,…,n, differentiating $\mathrm{Pf}(K(x, \vec{0}_n, \vec{0}_n))$ on the variable $x_i$ is equivalent, due to the Pfaffian's general nature, to differentiating (with the differentiation weight coefficient $d_i$) the corresponding (i.e. containing the term $x_i$) blocks of $K(x, \vec{0}_n, \vec{0}_n)$ on this variable as it's shown in the theorem's formula (due to receiving the Pfaffian of a matrix having a pair of identical rows and a pair of identical columns otherwise). And putting $a_i$ in the corresponding diagonal block of $K(x, \vec{0}_n, \vec{0}_n)$ generates the case of "removing" (with the absence-weight coefficient $a_i$) all the terms containing $x_i$ from the Pfaffian's sum expansion.

# Sparse compressions in characteristic 5

**Corollary III.7:**

in characteristic 5,

(@) $$(\prod_{i=1}^{n}(1+d_i\frac{\partial}{\partial x_i}))\det(\tilde{C}^{\star 2}(x)+\mathrm{Diag}(a)) = \frac{1}{2^n}\mathrm{Pf}(K(x,d,a))$$

**Theorem III.8:**

Let $G$ be an n×m-matrix of a rank unexceeding k, dim(x) = n, dim(y) = m. Then in the matrix $G \star C(x,y)$ the Schur complement of the block lying on a set of rows I and a set of columns J such that |I|=|J| is a matrix of the form $\breve{G} \star C(x_{\setminus I}, y_{\setminus J})$ where $\breve{G}$ is an (n-|I|)×(m-|J|)-matrix of a rank unexceeding k.

Proof:

This theorem can be easily proven by the induction on |I| because the Schur complement of any block can be represented as the result of a chain of consequent elementary Schur complement compressions for blocks of size 1×1.

Indeed, let's consider, for i = 2,…,dim(x) and j = 2,…,dim(y), the determinant $\det\begin{pmatrix} \frac{\alpha_1^T\beta_1}{x_1-y_1} & \frac{\alpha_1^T\beta_j}{x_1-y_j} \\ \frac{\alpha_i^T\beta_1}{x_i-y_1} & \frac{\alpha_i^T\beta_j}{x_i-y_j} \end{pmatrix}$ where $\alpha_1, \beta_1, \alpha_i, \beta_j$ are k-vectors. We can represent it as

$$\alpha_i^T\beta_1\alpha_1^T\beta_j \det\begin{pmatrix} \frac{1}{x_1-y_1} & \frac{1}{x_1-y_j} \\ \frac{1}{x_i-y_1} & \frac{1}{x_i-y_j} \end{pmatrix} + \frac{\alpha_1^T\beta_1\alpha_i^T\beta_j - \alpha_i^T\beta_1\alpha_1^T\beta_j}{x_1-y_1}\frac{1}{x_i-y_j} =$$

$$= \alpha_i^T\beta_1\alpha_1^T\beta_j \frac{(x_i-x_1)(y_j-y_1)}{(x_i-y_1)(x_1-y_1)(x_1-y_j)}\frac{1}{x_i-y_j} + \frac{\alpha_1^T\beta_1\alpha_i^T\beta_j - \alpha_i^T\beta_1\alpha_1^T\beta_j}{x_1-y_1}\frac{1}{x_i-y_j} =$$

$$= \frac{1}{(x_1-y_1)} \frac{\alpha_i^T\beta_1\frac{x_i-x_1}{x_i-y_1}\cdot\frac{y_j-y_1}{x_1-y_j}\alpha_1^T\beta_j + \alpha_i^T(\alpha_1^T\beta_1 I_k - \beta_1\alpha_1^T)\beta_j}{x_i-y_j} \; .$$

Thus we receive $$\mathrm{Schur}_{\{1\},\{1\}}\left(\begin{pmatrix} \frac{\alpha_1^T\beta_1}{x_1-y_1} & \frac{\alpha_1^T\beta_j}{x_1-y_j} \\ \frac{\alpha_i^T\beta_1}{x_i-y_1} & \frac{\alpha_i^T\beta_j}{x_i-y_j} \end{pmatrix}\right) = \frac{\det\begin{pmatrix} \frac{\alpha_1^T\beta_1}{x_1-y_1} & \frac{\alpha_1^T\beta_j}{x_1-y_j} \\ \frac{\alpha_i^T\beta_1}{x_i-y_1} & \frac{\alpha_i^T\beta_j}{x_i-y_j} \end{pmatrix}}{\frac{\alpha_1^T\beta_1}{x_1-y_1}} =$$

$$= \frac{1}{\alpha_1^T\beta_1} \frac{\alpha_i^T\beta_1\frac{x_i-x_1}{x_i-y_1}\cdot\frac{y_j-y_1}{x_1-y_j}\alpha_1^T\beta_j + \alpha_i^T(\alpha_1^T\beta_1 I_k - \beta_1\alpha_1^T)\beta_j}{x_i-y_j} =$$

$$= \frac{u_i v_j + \alpha_i^T\left(\frac{1}{\alpha_1^T\beta_1}(\alpha_1^T\beta_1 I_k - \beta_1\alpha_1^T)\right)\beta_j}{x_i-y_j}$$

where $u_i = \frac{\alpha_i^T\beta_1}{\sqrt{\alpha_1^T\beta_1}}\frac{x_i-x_1}{x_i-y_1}$, $v_j = \frac{y_j-y_1}{x_1-y_j}\frac{\alpha_1^T\beta_j}{\sqrt{\alpha_1^T\beta_1}}$

Since the rank of the matrix $\frac{1}{\alpha_1^T\beta_1}(\alpha_1^T\beta_1 I_k - \beta_1\alpha_1^T)$ doesn't exceed k-1 and we hence can represent it as $A_1B_1$ where $A_1$ is a k×(k-1)-matrix and $B_1$ is a (k-1)×k-matrix, we get $\alpha_i^T(\frac{1}{\alpha_1^T\beta_1}(\alpha_1^T\beta_1 I_k - \beta_1\alpha_1^T))\beta_j = \alpha_i^T A_1 B_1 \beta_j$ and, therefore, the compressed matrix's entries are $\frac{u_i v_j + \alpha_i^T A_1 B_1 \beta_j}{x_i - y_j}$ (where $\alpha_i^T A_1$ is a (k-1)-row and $B_1\beta_j$ is a (k-1) column) what completes the proof.

**Definition:**

We'll call two elements of a field's extension by the infinitesimal $\varepsilon$ ***infinitely close*** on $\varepsilon$ (or, shortly, $\varepsilon$-close) if their difference's order on $\varepsilon$ is bigger than zero.

**Definition:** let's call an n×m-matrix A that can be represented in the form $G \star C(x, y)$ where G is an n×m-matrix of rank k, dim(x) = n, dim(y) = m, a matrix of **Cauchy-rank k**; and we also define, for this representation, G as A's **numerator-matrix** and x,y as A's **row** and **column** (or left and right) **denominator-value** vectors **(**or sets**)** correspondingly. If G = LR, where L is an n×k-matrix and R is a k×m-matrix, we'll call the i-th row of L its i-th **numerator-row** and the j-th column of R its j-th **numerator-column**, while $x_i, y_j$ will be called its i-th **row** and j-th **column denominator-values** correspondingly (for i = 1,…n, j = 1,…,m). In case if n = m and for each i = 1,…,n its i-th row-numerator equals its transposed column-numerator right-multiplied, optionally, by an k×k-matrix M that we'll call the **multiplication matrix**, the i-th numerator-column will be further called its i-th **numerator-vector**.

The above theorem hence tells us that the class of matrices of a Cauchy-rank unexceeding k is closed under the Schur complement compression operator.

We'll also consider the following generalization of a matrix of Cauchy-rank k:

**Definition:** let $\varepsilon$ be an infinitesimal the ground field is extended by, L be a dim(x)×k-matrix and R be a k×dim(y)-matrix and in $LR \star C(x, y)$ some row and column denominator-values form $\varepsilon$-close families, hence forming row- and column-disjoint blocks whose entries have denominators of $\varepsilon$-order 1, i.e. of the type $O(\varepsilon)$ (let's call them singular entries), while all the denominator-values, numerator-rows and numerator-columns are of $\varepsilon$-order 0 or bigger. In case if the matrix $LR \star C(x, y)$ has no entries of $\varepsilon$-order smaller than zero (infinitely big entries) then we'll call its entry-wise limit on $\varepsilon$ $\lim_{\varepsilon \to 0}(LR \star C(x, y))$ a ***singularized*** matrix of Cauchy-rank k.

**Theorem III.8.1** Each Schur complement compression turns a singularized matrix of Cauchy-rank k into a singularized matrix of a Cauchy-rank unexceeding k, and in such a matrix any numerator-row and any numerator-column corresponding to equal row and column denominator-values correspondingly are orthogonal.

***A polynomial-time algorithm for computing the permanent in characteristic 5***

**Definition:**

Let A be a 2n×2n-matrix. Then its **alternate determinant** is

$$\mathrm{altdet}(A) \coloneqq \sum_{I\subseteq\{1,\ldots n\}} \det(A^{(I\cup\{n+1,\ldots,2n\}\setminus\hat{I},\{1,\ldots n\}\setminus I\cup\hat{I})})$$

where $\hat{I}$ is the subset of {n+1,…,2n} received via adding n to each element of $I$.

**Definition**: let $P, \widehat{P}$ be 4×n-matrices, z, $g^{(11)}, g^{(12)}, g^{(21)}, g^{(22)}$ be n-vectors. Then

in the matrix $\begin{pmatrix} P^TMP \star \tilde{C}(z) + \mathrm{Diag}(g^{(11)}) & P^TM\widehat{P} \star \tilde{C}(z) + \mathrm{Diag}(g^{(12)}) \\ \widehat{P}^TMP \star \tilde{C}(z) + \mathrm{Diag}(g^{(21)}) & \widehat{P}^TM\widehat{P} \star \tilde{C}(z) + \mathrm{Diag}(g^{(22)}) \end{pmatrix}$

we'll call, for j = 1,…,n, the j-th columns of $P$ and $\widehat{P}$ ($p_j$ and $\hat{p}_j$ correspondingly) the **numerator-vector** and **alternate numerator-vector** correspondingly and $g_j^{(12)} + g_j^{(21)}$ the **absence-weight** of the denominator-value $z_j$.

**Theorem III.9 (in characteristic 5):**

Let $P, \widehat{P}$ be 4×n-matrices, t, g be n-vectors, $M = \begin{pmatrix} 0_{2\times 2} & \begin{matrix} 0 & 1 \\ -1 & 0 \end{matrix} \\ \begin{matrix} 0 & 1 \\ -1 & 0 \end{matrix} & 0_{2\times 2} \end{pmatrix}$. Then

a) $$\mathrm{altdet}(\begin{pmatrix} P^TMP \star \tilde{C}(t^{\star 5}) & P^TM\widehat{P} \star \tilde{C}(t^{\star 5}) \\ \widehat{P}^TMP \star \tilde{C}(t^{\star 5}) + \mathrm{Diag}(g) & \widehat{P}^TM\widehat{P} \star \tilde{C}(t^{\star 5}) \end{pmatrix}) =$$

$$= \frac{1}{2^n} \mathrm{coef}_{\lambda^n} \mathrm{Pf}(K(\begin{pmatrix} x \\ t \end{pmatrix}, \begin{pmatrix} \vec{0}_m \\ \lambda\vec{1}_n \end{pmatrix}, \begin{pmatrix} \alpha \\ \beta \end{pmatrix}))$$

where

$\text{for } i = 1,2,3,4,\ \ j = 1, \ldots, n$

$$\begin{cases} \frac{1}{\Delta}\det\left(\tilde{C}^{\star 2}(x, v_i, t_j) + \mathrm{Diag}(\binom{\alpha}{0})\right) = r_i^T M p_j/(v_i - t_j)^5 \\ \frac{1}{\Delta}\frac{\partial}{\partial t_j}\det\left(\tilde{C}^{\star 2}(x, v_i, t_j) + \mathrm{Diag}(\binom{\alpha}{0})\right) = r_i^T M \hat{p}_j/(v_i - t_j)^5 \\ \frac{1}{\Delta}\frac{\partial}{\partial t_j}\det\left(\tilde{C}^{\star 2}(\binom{x}{t_j}) + \mathrm{Diag}(\binom{\alpha}{t_j\beta_j})\right) = g_j \end{cases}$$

for $i_1, i_2 = 1,2,3,4$

$$\frac{1}{\Delta}\det\left(\tilde{C}^{\star 2}(x, v_{i_1}, v_{i_2}) + \mathrm{Diag}(\binom{\alpha}{0})\right) = r_{i_1}^T M r_{i_2}/(v_{i_1} - v_{i_2})^5 \quad \text{if} \quad i_1 \neq i_2$$

where $p_j, \hat{p}_j$ are the j-th rows of $P, \widehat{P}$ correspondingly, $v_1, v_2, v_3, v_4$ are generic scalars, $r_1, r_2, r_3, r_4$ are some 4-vectors, $\Delta = \det\left(\tilde{C}^{\star 2}(x) + \mathrm{Diag}(\alpha)\right)$

b) the set of functions in $x, \alpha, \beta$ that are the left parts of the above system of equations is a system of functions whose algebraic rank is 11n + 6 and it implies that the entries of $p_1, \hat{p}_1 \dots, p_n, \hat{p}_n$ is a system of functions maximally algebraically independent under the condition $p_j^T M \hat{p}_j = 0$ for j = 1,…,n (fulfilled for any $x, \alpha, \beta$).

Proof:

Part (1). According to Corollary III.7, the right side of the theorem's first part's equality is $(\prod_{i=1}^{n}\frac{\partial}{\partial t_j})\det(\tilde{C}^{\star 2}(\binom{x}{t}) + \mathrm{Diag}(\binom{\alpha}{\beta})) =$

$$= \sum_{I \in \mathcal{J}} \det\begin{pmatrix} \tilde{C}^{\star 2}(x) + \mathrm{Diag}(\alpha) & C^{\star 2}(x,t) & C^{\star 2}(x,t)D \\ C^{\star 2}(t,x) & \tilde{C}^{\star 2}(t) & \tilde{C}^{\star 2}(t)D \\ DC^{\star 2}(t,x) & D\tilde{C}^{\star 2}(t) + \mathrm{Diag}(\beta) & D\tilde{C}^{\star 2}(t)D \end{pmatrix}^{(\setminus I, \setminus \bar{I})}$$

where $\mathcal{J} = \{m+1, m+n+1\} \times \dots \times \{m+n, m+2n\}$, $\bar{I}$ is the set received from I via taking the other element from each set $\{m+k, m+n+k\}$ for k = 1,…,n, $D = \mathrm{Diag}(\{\frac{\partial}{\partial t_j}\}_n)$.

and it's equal to

$$\Delta \mathrm{altdet}(\mathrm{Schur}_{\{1,\dots,m\},\{1,\dots,m\}}\begin{pmatrix} \tilde{C}^{\star 2}(x) + \mathrm{Diag}(\alpha) & C^{\star 2}(x,t) & C^{\star 2}(x,t)D \\ C^{\star 2}(t,x) & \tilde{C}^{\star 2}(t) & \tilde{C}^{\star 2}(t)D \\ DC^{\star 2}(t,x) & D\tilde{C}^{\star 2}(t) + \mathrm{Diag}(\beta) & D\tilde{C}^{\star 2}(t)D \end{pmatrix}) =$$

$$= \Delta \mathrm{altdet}(\mathrm{Schur}_{\{1,\dots,m\},\{1,\dots,m\}}\begin{pmatrix} \tilde{C}^{\star 2}(x) + \mathrm{Diag}(\alpha) & C^{\star 2}(x,t) & C^{\star 2}(x,t)D \\ C^{\star 2}(t,x) & \tilde{C}^{\star 2}(t) & \tilde{C}^{\star 2}(t)D \\ DC^{\star 2}(t,x) & D\tilde{C}^{\star 2}(t) & D\tilde{C}^{\star 2}(t)D \end{pmatrix} +$$

$$+\begin{pmatrix} 0_{n\times n} & 0_{n\times n} \\ \mathrm{Diag}(\beta) & 0_{n\times n} \end{pmatrix})$$

where $\Delta = \det(\tilde{C}^{\star 2}(x) + \mathrm{Diag}(\alpha))$.

The matrix $\begin{pmatrix} \tilde{C}^{\star 2}(x) + \mathrm{Diag}(\alpha) & C^{\star 2}(x,t) & C^{\star 2}(x,t)D \\ C^{\star 2}(t,x) & \tilde{C}^{\star 2}(t) & \tilde{C}^{\star 2}(t)D \\ DC^{\star 2}(t,x) & D\tilde{C}^{\star 2}(t) & D\tilde{C}^{\star 2}(t)D \end{pmatrix}$ is symmetric and of Cauchy-rank 4 for the denominator-value set $\begin{pmatrix} \{x_k^5\}_m \\ \{t_j^5\}_n \\ \{t_j^5\}_n \end{pmatrix}$ because for two independent indeterminates u, v there holds (for M defined in the theorem)

$$\frac{1}{(u-v)^2} = \frac{(u-v)^3}{(u-v)^5} = \frac{(1\ \ u\ \ 3u^2\ \ u^3)M\begin{pmatrix} 1 \\ v \\ 3v^2 \\ v^3 \end{pmatrix}}{(u-v)^5}$$

$$\frac{\partial}{\partial u}\frac{1}{(u-v)^2} = \frac{\frac{\partial}{\partial u}(1\ \ u\ \ 3u^2\ \ u^3)M\begin{pmatrix} 1 \\ v \\ 3v^2 \\ v^3 \end{pmatrix}}{(u-v)^5}$$

$$\frac{\partial}{\partial v}\frac{1}{(u-v)^2} = \frac{(1\ \ u\ \ 3u^2\ \ u^3)M\frac{\partial}{\partial v}\begin{pmatrix} 1 \\ v \\ 3v^2 \\ v^3 \end{pmatrix}}{(u-v)^5}$$

$$\frac{\partial}{\partial u}\frac{\partial}{\partial v}\frac{1}{(u-v)^2} = \frac{\frac{\partial}{\partial u}(1\ \ u\ \ 3u^2\ \ u^3)M\frac{\partial}{\partial v}\begin{pmatrix} 1 \\ v \\ 3v^2 \\ v^3 \end{pmatrix}}{(u-v)^5}\ .$$

Hence its Schur complement on $\{1,\dots,m\},\{1,\dots,m\}$ is also symmetric and of a Cauchy-rank unexceeding 4 for the same denominator-value set and, accordingly, has the form $\begin{pmatrix} P^T MP \star \tilde{C}(t^{\star 5}) + D_{11} & P^T M\hat{P} \star \tilde{C}(t^{\star 5}) + D_{12} \\ \hat{P}^T MP \star \tilde{C}(t^{\star 5}) + D_{21} & \hat{P}^T M\hat{P} \star \tilde{C}(t^{\star 5}) + D_{22} \end{pmatrix}$ where $D_{11}, D_{12}, D_{21}, D_{22}$ are diagonal, $D_{12} = D_{21}$, and the diagonal entries of $P^T MP$, $P^T M\hat{P}$, $\hat{P}^T MP$, $\hat{P}^T M\hat{P}$ are zeros (as it's supposed in matrices of any Cauchy-rank for entries with equal row and column denominator-values).

Part (2). As, according to Corollary III.7, the right side of the theorem's first part's equality is $(\prod_{i=1}^{n}\frac{\partial}{\partial t_j})\det(\tilde{C}^{\star 2}(\begin{pmatrix} x \\ t \end{pmatrix}) + \mathrm{Diag}(\begin{pmatrix} \alpha \\ \beta \end{pmatrix}))$, let's consider the following two identities:

$$1)\ \lim_{\varepsilon\to 0}\det(\tilde{C}^{\star 2}(\begin{pmatrix} x^{(0)} \\ y \\ y+\varepsilon\vec{1}_{\dim(y)} \\ t \end{pmatrix}) + \mathrm{Diag}(\begin{pmatrix} \alpha^{(0)} \\ \varepsilon^{-2}\vec{1}_{\dim(y)}+\varepsilon^{2}\alpha^{(1)} \\ \varepsilon^{-2}\vec{1}_{\dim(y)} \\ \beta \end{pmatrix})) =$$

$$= \det(D\tilde{C}^{\star 2}(\begin{pmatrix} x^{(0)} \\ y \\ t \end{pmatrix})D + \mathrm{Diag}(\begin{pmatrix} \alpha^{(0)} \\ \gamma \\ \beta \end{pmatrix}))$$

$$\text{where } D = \mathrm{Diag}(\begin{pmatrix} I_{m_0} \\ \mathrm{Diag}(\{\frac{\partial}{\partial y_k}\}_{\dim(y)}) \\ I_n \end{pmatrix})$$

and

$$2)\ \det(\breve{D}\tilde{C}^{\star 2}(\begin{pmatrix} x^{(0)} \\ x^{(1)} \\ x^{(2)} \\ t \end{pmatrix})\breve{D} + \mathrm{Diag}(\begin{pmatrix} \alpha^{(0)} \\ \alpha^{(1)} \\ \alpha^{(2)} \\ \beta \end{pmatrix})) =$$

$$= \lim_{\varepsilon\to 0}(\varepsilon^{2m_2}\det(D\tilde{C}^{\star 2}(\begin{pmatrix} x^{(0)} \\ x^{(1)} \\ x^{(2)} \\ x^{(2)}+\varepsilon\vec{1}_{m_2} \\ t \end{pmatrix})D + \mathrm{Diag}(\begin{pmatrix} \alpha^{(0)} \\ \alpha^{(1)} \\ \varepsilon^{-4}\vec{1}_{m_2}+\varepsilon^{2}\alpha^{(2)} \\ \varepsilon^{-4}\vec{1}_{m_2} \\ \beta \end{pmatrix})))$$

$$\text{where } \breve{D} = \mathrm{Diag}(\begin{pmatrix} I_{m_0} \\ \mathrm{Diag}(\{\frac{\partial}{\partial x_k^{(1)}}\}_{m_1}) \\ \mathrm{Diag}(\{\frac{\partial^2}{\partial (x_k^{(2)})^2}\}_{m_2}) \\ I_n \end{pmatrix}),\ D = \mathrm{Diag}(\begin{pmatrix} I_{m_0} \\ \mathrm{Diag}(\{\frac{\partial}{\partial x_k^{(1)}}\}_{m_1}) \\ \mathrm{Diag}(\{\frac{\partial}{\partial x_k^{(2)}}\}_{m_2}) \\ \mathrm{Diag}(\{\frac{\partial}{\partial (x_k^{(2)}+\varepsilon)}\}_{m_2}) \\ I_n \end{pmatrix}).$$

$$\dim(x^{(q)}) = \dim(\alpha^{(q)}) = m_q \text{ for } q = 0,1,2$$

Hence, for proving the theorem's second part, it's sufficient to replace, as by a partial case of $x, \alpha$ (yielding a generalization, though), its first part's system of equations by the following system:

for $i = 1,2,3,4,\ j = 1, \dots, n$

$$\begin{cases} \frac{1}{\Delta}\det(\breve{D}\tilde{C}^{\star 2}(\begin{pmatrix} x^{(0)} \\ x^{(1)} \\ x^{(2)} \end{pmatrix}, v_i, t_j)\breve{D} + \mathrm{Diag}(\begin{pmatrix} \alpha^{(0)} \\ \alpha^{(1)} \\ \alpha^{(2)} \\ 0 \end{pmatrix})) = r_i^T M p_j/(v_i - t_j)^5 \\ \frac{1}{\Delta}\frac{\partial}{\partial t_j}\det(\breve{D}\tilde{C}^{\star 2}(\begin{pmatrix} x^{(0)} \\ x^{(1)} \\ x^{(2)} \end{pmatrix}, v_i, t_j)\breve{D} + \mathrm{Diag}(\begin{pmatrix} \alpha^{(0)} \\ \alpha^{(1)} \\ \alpha^{(2)} \\ 0 \end{pmatrix})) = r_i^T M \hat{p}_j/(v_i - t_j)^5 \\ \frac{1}{\Delta}\frac{\partial}{\partial t_j}\det(\breve{D}\tilde{C}^{\star 2}(\begin{pmatrix} x^{(0)} \\ x^{(1)} \\ x^{(2)} \\ t_j \end{pmatrix})\breve{D} + \mathrm{Diag}(\begin{pmatrix} \alpha^{(0)} \\ \alpha^{(1)} \\ \alpha^{(2)} \\ t_j\beta_j \end{pmatrix})) = g_j \end{cases}$$

for $i_1, i_2 = 1,2,3,4$

$$\frac{1}{\Delta}\det(\breve{D}\tilde{C}^{\star 2}(\begin{pmatrix} x^{(0)} \\ x^{(1)} \\ x^{(2)} \end{pmatrix}, v_{i_1}, v_{i_2})\breve{D} + \mathrm{Diag}(\begin{pmatrix} \alpha^{(0)} \\ \alpha^{(1)} \\ \alpha^{(2)} \\ 0 \end{pmatrix})) = r_{i_1}^T M r_{i_2}/(v_{i_1} - v_{i_2})^5 \quad \text{if} \quad i_1 \neq i_2$$

where $\breve{D} = \mathrm{Diag}(\begin{pmatrix} I_{m_0} \\ \mathrm{Diag}(\{\frac{\partial}{\partial x_k^{(1)}}\}_{m_1}) \\ \mathrm{Diag}(\{\frac{\partial}{\partial x_k^{(2)}}\}_{m_2}) \\ I_n \end{pmatrix})$, $p_j, \hat{p}_j$ are the j-th rows of $P, \hat{P}$ correspondingly, $v_1, v_2, v_3, v_4$ are generic scalars, $r_1, r_2, r_3, r_4$ are generic 4-vectors, $\Delta = \det(\breve{D}\tilde{C}^{\star 2}(\begin{pmatrix} x^{(0)} \\ x^{(1)} \\ x^{(2)} \end{pmatrix})\breve{D} + \mathrm{Diag}(\begin{pmatrix} \alpha^{(0)} \\ \alpha^{(1)} \\ \alpha^{(2)} \end{pmatrix}))$

Now let's choose a partial case of the vectors (sets) $\alpha^{(0)}, \alpha^{(1)}, \alpha^{(2)}$ where all of them can be partitioned into subvectors (subsets) of sizes divisible by 5 each of whom consists of entries (elements) $\lambda_k/(\varepsilon d_{k,r})$ (with $\lambda_k$ distinct for different subsets), where $\varepsilon$ is a formal infinitesimal, k is the index of the subset and $r$ is the index of the element in the subset. We'll call each $\lambda_k$ a ***uniting value***, while by $U(\lambda_k) = \{x_{k,1}, \dots, x_{k,|U(\lambda_k)|}\}$ we'll denote the subvector (subset) of $\begin{pmatrix} x^{(0)} \\ x^{(1)} \\ x^{(2)} \end{pmatrix}$ corresponding to $\lambda_k$ and we'll call it the family of denominator-values united by $\lambda_k$ (thus each $x_{k,r}$ is an entry of either $x^{(0)}$ or $x^{(1)}$ or $x^{(2)}$).

Let's show that this partial case turns, for a generic fixed $\begin{pmatrix} x^{(0)} \\ x^{(1)} \\ x^{(2)} \end{pmatrix}$, the left parts of our system of equations into a system of functions in the uniting values $\lambda_k$ and the entries of $\beta$ whose algebraic rank is 11n + 6. Because the absence-weight equation for the variable $\beta_j$

$$\frac{1}{\Delta}\frac{\partial}{\partial t_j}\det(\breve{D}\tilde{C}^{\star 2}(\begin{pmatrix} x^{(0)} \\ x^{(1)} \\ x^{(2)} \\ t_j \end{pmatrix})\breve{D} + \mathrm{Diag}(\begin{pmatrix} \alpha^{(0)} \\ \alpha^{(1)} \\ \alpha^{(2)} \\ t_j\beta_j \end{pmatrix})) = g_j$$

is solvable for any fixed $\begin{pmatrix} x^{(0)} \\ x^{(1)} \\ x^{(2)} \end{pmatrix}$, $\begin{pmatrix} \alpha^{(0)} \\ \alpha^{(1)} \\ \alpha^{(2)} \end{pmatrix}$ such that $\Delta \neq 0$, it's sufficient to prove that the algebraic rank of the system of functions received from the above-mentioned system via excluding all the absence-weights $\frac{1}{\Delta}\frac{\partial}{\partial t_j}\det(\breve{D}\tilde{C}^{\star 2}(\begin{pmatrix} x^{(0)} \\ x^{(1)} \\ x^{(2)} \\ t_j \end{pmatrix})\breve{D} + \mathrm{Diag}(\begin{pmatrix} \alpha^{(0)} \\ \alpha^{(1)} \\ \alpha^{(2)} \\ t_j\beta_j \end{pmatrix}))$ is 7n + 6.

Due to the divisibility of each united family's size by 5, the derivatives of the functions

$$f_{i,j}(\lambda) = \frac{1}{\Delta}\det(\breve{D}\tilde{C}^{\star 2}(\begin{pmatrix} x^{(0)} \\ x^{(1)} \\ x^{(2)} \end{pmatrix}, v_i, t_j)\breve{D} + \mathrm{Diag}(\begin{pmatrix} \alpha^{(0)} \\ \alpha^{(1)} \\ \alpha^{(2)} \\ 0 \end{pmatrix})) \quad ,$$

$$\hat{f}_{i,j}(\lambda) = \frac{1}{\Delta}\frac{\partial}{\partial t_j}\det(\breve{D}\tilde{C}^{\star 2}(\begin{pmatrix} x^{(0)} \\ x^{(1)} \\ x^{(2)} \end{pmatrix}, v_i, t_j)\breve{D} + \mathrm{Diag}(\begin{pmatrix} \alpha^{(0)} \\ \alpha^{(1)} \\ \alpha^{(2)} \\ 0 \end{pmatrix}))$$

and $h_{i_1,i_2}(\lambda) = \frac{1}{\Delta}\det(\breve{D}\tilde{C}^{\star 2}(\begin{pmatrix} x^{(0)} \\ x^{(1)} \\ x^{(2)} \end{pmatrix}, v_{i_1}, v_{i_2})\breve{D} + \mathrm{Diag}(\begin{pmatrix} \alpha^{(0)} \\ \alpha^{(1)} \\ \alpha^{(2)} \\ 0 \end{pmatrix}))$

on any uniting value $\lambda_k$ are of ε-order 1 or bigger. Hence it's sufficient to prove the equality $\mathrm{rank}(\lim_{\varepsilon\to 0}(\varepsilon^{-1}\,\mathfrak{J}(\begin{pmatrix} f \\ \hat{f} \\ h \end{pmatrix},\lambda))) = 7n+6$ where $\mathfrak{J}(\begin{pmatrix} f \\ \hat{f} \\ h \end{pmatrix},\lambda)$ is the above function system's Jacobian matrix on the uniting values.

We'll say that a uniting value $\lambda_k$ is of **auxiliary differentiation order** q if $U(\lambda_k)$ is a subset of $x^{(q)}$, for q = 0,1,2. Then we receive for $\lambda_k$ of auxiliary differentiation order q, for i = 1,2,3,4, j = 1,…,n:

$$\lim_{\varepsilon\to 0}\frac{\partial}{\partial\lambda_k}(\varepsilon^{-1}f_{i,j}(\lambda)) = \sum_{r=1}^{|U(\lambda_k)|} d_{k,r}(\frac{\partial^q}{\partial x_{k,r}^q}\frac{1}{(v_i-x_{k,r})^2})(\frac{\partial^q}{\partial x_{k,r}^q}\frac{1}{(x_{k,r}-t_j)^2})$$

$$\lim_{\varepsilon\to 0}\frac{\partial}{\partial\lambda_k}(\varepsilon^{-1}\hat{f}_{i,j}(\lambda)) = \frac{\partial}{\partial t_j}\sum_{r=1}^{|U(\lambda_k)|} d_{k,r}(\frac{\partial^q}{\partial x_{k,r}^q}\frac{1}{(v_i-x_{k,r})^2})(\frac{\partial^q}{\partial x_{k,r}^q}\frac{1}{(x_{k,r}-t_j)^2})$$

$$\lim_{\varepsilon\to 0}\frac{\partial}{\partial\lambda_k}(\varepsilon^{-1}h_{i_1,i_2}(\lambda)) = \sum_{r=1}^{|U(\lambda_k)|} d_{k,r}(\frac{\partial^q}{\partial x_{k,r}^q}\frac{1}{(v_{i_1}-x_{k,r})^2})(\frac{\partial^q}{\partial x_{k,r}^q}\frac{1}{(x_{k,r}-v_{i_2})^2})$$

For i = 1,2,3,4, j = 1,…,n, the first two above sums are linear combinations of the sums $\sum_{r=1}^{|U(\lambda_k)|}\frac{d_{k,r}}{(v_i-x_{k,r})^w}$ and $\sum_{r=1}^{|U(\lambda_k)|}\frac{d_{k,r}}{(t_j-x_{k,r})^w}$ for w = 2,3,4,5, while the third one is a linear combination of $\sum_{r=1}^{|U(\lambda_k)|}\frac{d_{k,r}}{(v_{i_1}-x_{k,r})^w}$ and $\sum_{r=1}^{|U(\lambda_k)|}\frac{d_{k,r}}{(v_{i_2}-x_{k,r})^w}$ for w = 2,3,4,5. Let's consider the case when for each uniting value $\lambda_k$ $\sum_{r=1}^{|U(\lambda_k)|}\frac{d_{k,r}}{(v_i-x_{k,r})^w} = \sum_{r=1}^{|U(\lambda_k)|}\frac{d_{k,r}}{(t_j-x_{k,r})^w} = 0$ for all i, j except either exactly one index i($\lambda_k$) or exactly one index j($\lambda_k$) and for all w except exactly one degree w($\lambda_k$). Let's call such a uniting value $\lambda_k$ a $v_i$,w-**supporting** and $t_j$,w-**supporting** uniting value correspondingly. We additionally put $\sum_{r=1}^{|U(\lambda_k)|}\frac{d_{k,r}}{(c-x_{k,r})^w} = 1$ and $\sum_{r=1}^{|U(\lambda_k)|}\frac{d_{k,r}}{(v_i-x_{k,r})^w} = 1$ if $\lambda_k$ supports $t_j$,w and $v_i$,w correspondingly. Then, for j = 1,…n, we take

one $t_j$,w-supporting uniting value of auxiliary differentiation order q for each of the following pairs (w,q): (5,2), (4,1), (3,0), (1,1), (2,1), (1,0), (2,0);

one $v_1$,w-supporting uniting value of auxiliary differentiation order 1 for each of w = 1,2,3;

one $v_2$,1-supporting uniting value of auxiliary differentiation order 0 for each of w = 1,2;

one $v_3$,1-supporting uniting value of auxiliary differentiation order 0.

Then, upon multiplying its columns by non-zero constants from the set {1,2,3,4}, $\lim_{\varepsilon\to 0}(\varepsilon^{-1}\,\mathfrak{I}(\begin{pmatrix} f \\ \hat{f} \\ h \end{pmatrix},\lambda))$ will be the block-triangular matrix

$$\begin{pmatrix} \mathrm{Diag}(\{T(t_j)\}_n) & P \\ 0_{6\times 7n} & \begin{pmatrix} \{\frac{1}{(v_i-v_1)^{5-s}}\}_{\substack{i=2,3,4 \\ s=1,2,3}} & B_{12} & B_{13} \\ 0_{2\times 3} & \{\frac{1}{(v_i-v_2)^{3-s}}\}_{\substack{i=3,4 \\ s=1,2}} & B_{23} \\ 0_{1\times 3} & 0_{1\times 2} & \frac{1}{(v_3-v_4)^3} \end{pmatrix} \end{pmatrix}$$

where for j = 1,…,n $T(t_j) = \begin{pmatrix} \{\frac{1}{(v_i-t_j)^{5-s}}\}_{\substack{i=1,2,3,4 \\ s=1,2,3}} & A_j \\ 0_{4\times 3} & \{\frac{1}{(v_i-t_j)^{5-s}}\}_{\substack{i=1,2,3,4 \\ s=1,2,3,4}} \end{pmatrix}$, $A_j$ is a 4×4-matrix, P is an 7n×6-matrix, $B_{12}$ is a 3×2-matrix, $B_{13}$ is a 3×1-matrix, $B_{23}$ is a 2×1-matrix.

Taking into account the fact that each of the above matrix's first n diagonal 8×7-blocks is of rank 7 and the last 3 ones are of ranks 3,2,1 correspondingly, we complete the theorem's proof.

The above theorem implies that in characteristic 5 we can polynomial-time reduce computing the alternate determinant of a symmetric singularized matrix of Cauchy-rank 4 such that all its denominator-values are alternate-wise doubled and $p_j^T M \hat{p}_j = 0$ for j = 1,…,n to computing a Pfaffian via the use of the neighboring computation principle (because the alternate determinant of such a matrix is a polynomial in the entries of its numerator-vectors, alternate numerator-vectors and its alternate-wise block-diagonal entries) and, hence, the alternate determinant of such a matrix is computable in polynomial time.

**Definition:**

We'll call **proper** a directed cycle that isn't a loop.

**Definition:** let dim(y) = dim(g) = n, $A$ be an n×n-matrix, $B^{(1)},\dots,B^{(n)}$ be m×m-matrices. Then we define the **trace-determinant** of A on $B^{(1)},\dots,B^{(n)}$ as

$$\det_{tr}\left(A,\{B^{(i)}\}_n\right) := \sum_{\pi\in S_n}\prod_{\mathcal{C}\in PC(\pi)}((-1)^{|\mathcal{C}|+1}\mathrm{tr}(\prod\nolimits_{q=1}^{|\mathcal{C}|} B^{\left(i_q(\mathcal{C})\right)}))\prod_{i=1}^{n} a_{i,\pi_i}$$

where $PC(\pi)$ is the set of $\pi$'s proper cycles, while for each proper cycle $\mathcal{C} = (i_1(\mathcal{C}),\dots,i_{|\mathcal{C}|}(\mathcal{C})) \in PC(\pi)$ (represented with the lexicographically minimal starting vertex $i_1(\mathcal{C})$) the multiplication order in the matrix product $\prod_{q=1}^{|\mathcal{C}|} B^{\left(i_q(C)\right)} = B^{(i_1(\mathcal{C}))} \dots B^{\left(i_{|C|}(\mathcal{C})\right)}$ is $\mathcal{C}$'s order.

In case if $A = \tilde{C}(x) + \mathrm{Diag}(g)$, we'll call $B^{(i)}$ the ***matrix-weight*** and $g_i$ the absence-weight of the denominator-value $x_i$. Further we'll often consider, when dealing with the trace-determinant, any denominator-value together with these two parameters as the triple $(x_i, B^{(i)}, g_i)$, while assuming the notions of infinitesimal-closeness and limit on an infinitesimal only for the values of $x_i$ themselves when speaking about infinitesimal-close denominator-values or/and their limit.

**Theorem III.10:** Let $P, \widehat{P}$ be 4×n-matrices, t, g be n-vectors, $M = \begin{pmatrix} 0_{2\times2} & \begin{matrix} 0 & 1 \\ -1 & 0 \end{matrix} \\ \begin{matrix} 0 & 1 \\ -1 & 0 \end{matrix} & 0_{2\times2} \end{pmatrix}$.

Then

$$\det_{tr}\left(\tilde{C}(t) + \mathrm{Diag}(g), \{(p_i\hat{p}_i^T + \hat{p}_i p_i^T)M\}_n\right) =$$

$$= \mathrm{altdet}(\begin{pmatrix} P^T MP \star \tilde{C}(t^{\star5}) & P^T M\widehat{P} \star \tilde{C}(t^{\star5}) \\ \widehat{P}^T MP \star \tilde{C}(t^{\star5}) + \mathrm{Diag}(g) & \widehat{P}^T M\widehat{P} \star \tilde{C}(t^{\star5}) \end{pmatrix})$$

As a corollary, due the fact that for any 4×2-matrix such that $F^T MF = 0_{2\times2}$ the matrix $FF^T$ can be represented as $p\hat{p}^T + \hat{p}p^T$ for some 4-vectors p, $\hat{p}$ such that $pM\hat{p}^T = 0$, we get

**Theorem III.10.1:**

$\det_{tr}\left(\tilde{C}(t) + \mathrm{Diag}(g), \{F^{(i)}(F^{(i)})^T M\}_n\right)$ is polynomial-time computable for any 4×2-matrices $F^{(1)},\dots,F^{(n)}$ such that $(F^{(i)})^T MF^{(i)} = 0_{2\times2}$ for i = 1,…,n.

**Definition:** let dim(y) = dim(g) = n, A be an n×n-matrix, $B^{(1)},\dots,B^{(n)}$ be m×m-matrices. Then we define the **open trace-determinant** of A on $B^{(1)},\dots,B^{(n)}$ as

$$\widetilde{\det}_{\mathrm{tr}}\left(\mathrm{A},\left\{\mathrm{B}^{(\mathrm{i})}\right\}_{\mathrm{n}}\right):=$$

$$\sum_{\pi\in\check{\mathrm{S}}_{\mathrm{n}}}\Big(\frac{(-1)^{|\mathcal{P}_\pi|+1}}{\mathrm{a}_{\mathrm{i}_{|\mathcal{P}_\pi|}(\mathcal{P}_\pi),\mathrm{i}_1(\mathcal{P}_\pi)}}\prod_{\mathrm{q}=1}^{|\mathcal{P}_\pi|}\mathrm{B}^{\left(\mathrm{i}_{\mathrm{q}}(\mathcal{P}_\pi)\right)}\Big)\prod_{\mathcal{C}\in\widetilde{\mathrm{PC}}(\pi)}\Big((-1)^{|\mathcal{C}|+1}\mathrm{tr}\Big(\prod_{\mathrm{q}=1}^{|\mathcal{C}|}\mathrm{B}^{\left(\mathrm{i}_{\mathrm{q}}(\mathcal{C})\right)}\Big)\Big)\prod_{\mathrm{i}=1}^{\mathrm{n}}\mathrm{a}_{\mathrm{i},\pi_{\mathrm{i}}}$$

where $\check{\mathrm{S}}_{\mathrm{n}}$ is the set on n-permutations where exactly one cycle is considered as broken and turned into a path $\mathcal{P}_\pi=(i_1(\mathcal{P}_\pi),\dots,i_{|\mathcal{P}_\pi|}(\mathcal{P}_\pi))$ (including, as an option, the case of a loop $\mathcal{P}_\pi=(i_1(\mathcal{P}_\pi))$ ), $\widetilde{\mathrm{PC}}(\pi)$ is the set of π's unbroken proper cycles, while the multiplication order in the matrix product $\prod_{\mathrm{q}=1}^{|\mathcal{C}|}\mathrm{B}^{\left(\mathrm{i}_{\mathrm{q}}(\mathrm{C})\right)}=\mathrm{B}^{\left(\mathrm{i}_1(\mathcal{C})\right)}\dots\mathrm{B}^{\left(\mathrm{i}_{|\mathrm{C}|}(\mathcal{C})\right)}$ is $\mathcal{C}$'s order for each unbroken proper cycle $\mathcal{C}=(\mathrm{i}_1(\mathcal{C}),\dots,\mathrm{i}_{|\mathcal{C}|}(\mathcal{C}))\in\widetilde{\mathrm{PC}}(\pi)$ (represented with the lexicographically minimal starting vertex $\mathrm{i}_1(\mathcal{C})$) and the multiplication order in the matrix product $\prod_{\mathrm{q}=1}^{|\mathcal{P}_\pi|}\mathrm{B}^{\left(\mathrm{i}_{\mathrm{q}}(\mathcal{P}_\pi)\right)}$ is $\mathcal{P}_\pi$'s order.

***Comment***: the above definition remains actual also in the case $\mathrm{a}_{\mathrm{i}_1(\mathcal{P}_\pi),\mathrm{i}_{|\mathcal{P}_\pi|}(\mathcal{P}_\pi)}=0$ due to the presence of the term $\mathrm{a}_{\mathrm{i}_1(\mathcal{P}_\pi),\mathrm{i}_{|\mathcal{P}_\pi|}(\mathcal{P}_\pi)}$ in $\prod_{\mathrm{i}=1}^{\mathrm{n}}\mathrm{a}_{\mathrm{i},\pi_{\mathrm{i}}}$.

**Theorem III.10.2 (in an arbitrary characteristic)**

For i = 1,…,n, let $\mathrm{x}_{\mathrm{i}}$ be a scalar, $\chi_{\mathrm{i}}$, $\mathrm{h}_{\mathrm{i}}$ be $\mathrm{k}_{\mathrm{i}}$-vectors and $\mathrm{L}^{(\mathrm{i},1)}$,…, $\mathrm{L}^{(\mathrm{i},\mathrm{k}_{\mathrm{i}})}$ be m×m-matrices of ε-order 0 or bigger such that $\lim_{\varepsilon\to0}\frac{\det_{\mathrm{tr}}\left(\tilde{\mathrm{C}}(\chi_{\mathrm{i}})+\mathrm{Diag}(\varepsilon^{-1}\mathrm{h}_{\mathrm{i}}),\begin{pmatrix}\mathrm{L}^{(\mathrm{i},1)}\\ \dots\\ \mathrm{L}^{(\mathrm{i},\mathrm{n}_{\mathrm{i}})}\end{pmatrix}\right)}{\varepsilon^{-\mathrm{k}_{\mathrm{i}}}}=0$ for i =1,…n. Then

$$\det_{\mathrm{tr}}(\tilde{\mathrm{C}}(\mathrm{x})+\mathrm{Diag}(\{\mathrm{g}_{\mathrm{i}}\}_n),\{\mathrm{B}^{(\mathrm{i})}\}_n)=$$

$$=\lim_{\varepsilon\to0}\frac{\det_{\mathrm{tr}}\left(\tilde{\mathrm{C}}\left(\left\{\mathrm{x}_{\mathrm{i}}\vec{1}_{\mathrm{n}_{\mathrm{i}}}+\varepsilon\chi_{\mathrm{i}}\right\}_{\mathrm{n}}\right)+\mathrm{Diag}(\{\varepsilon^{-1}\mathrm{h}_{\mathrm{i}}\}_{\mathrm{n}}),\left\{\begin{pmatrix}\mathrm{L}^{(\mathrm{i},1)}\\ \dots\\ \mathrm{L}^{(\mathrm{i},\mathrm{k}_{\mathrm{i}})}\end{pmatrix}\right\}_{\mathrm{n}}\right)}{\varepsilon^{n-(\mathrm{k}_1+\dots+\mathrm{k}_{\mathrm{n}})}}$$

where for i = 1,…,n $\quad \mathrm{g}_i=\lim_{\varepsilon\to0}\frac{\det_{\mathrm{tr}}\left(\tilde{\mathrm{C}}(\chi_{\mathrm{i}})+\mathrm{Diag}(\varepsilon^{-1}\mathrm{h}_{\mathrm{i}}),\begin{pmatrix}\mathrm{L}^{(\mathrm{i},1)}\\ \dots\\ \mathrm{L}^{(\mathrm{i},\mathrm{k}_{\mathrm{i}})}\end{pmatrix}\right)}{\varepsilon^{1-\mathrm{k}_{\mathrm{i}}}}$ ,

$$\mathrm{B}^{(\mathrm{i})}=\widetilde{\det}_{\mathrm{tr}}\left(\tilde{\mathrm{C}}(\chi_{\mathrm{i}})+\mathrm{Diag}\left(\lim_{\varepsilon\to0}\mathrm{h}_{\mathrm{i}}\right),\lim_{\varepsilon\to0}\begin{pmatrix}\mathrm{L}^{(\mathrm{i},1)}\\ \dots\\ \mathrm{L}^{(\mathrm{i},\mathrm{n}_{\mathrm{i}})}\end{pmatrix}\right)$$

Proof:

This statement follows from the definitions of the trace-determinant and the open trace-determinant because the "common" limit $\lim\limits_{\varepsilon\to 0}$ provides, for i = 1,…,n, the "opening" of the weighted sub-digraph corresponding to the ε-close denominator-value family $x_i\vec{1}_{n_i} + \varepsilon\chi_i$ (whose arcs are of ε-order -1, while all the matrix-weights are of ε-order 0) , while the i-th absence weight $g_i$ is obtained under the "common" limit as the limit of this sub-digraph's trace-determinant divided by $\varepsilon^{1-k_i}$ (i.e. via the case of this sub-digraph remaining "closed" in the trace-determinant's transversals).

By the above theorem, we hence introduced one more type of compression (actual for the trace-determinant) when a family of pair-wise infinitesimal-close denominator-values contracts, via the limit on the infinitesimal, into a new denominator-value that is their common limit on the infinitesimal and, accordingly, this family generates its limit's matrix-weight and absence-weight. Therefore it's also a compression of a family of m×m-matrices into another m×m-matrix. Hence, given a family of m×m-matrices, it contracts (for all the possible "accompanying" families of denominator-values and absence-weights providing the corresponding matrix's trace-determinant's equality to zero, i.e. the "opening" of the trace-determinant) into a set of new m×m-matrices (depending on the chosen families of denominator-values and absence-weights). Let's call it the open trace-determinant compression of a family of m×m-matrices. Hence, given a class of m×m-matrices, it generates, via the open trace-determinant compression of its subsets (families), a wider class and accordingly we can also speak, once again, about the compression-closure of the class for this operator.

**Comment:** in the above theorem, the ε-orders can also be considered fractional or/and yield a non-existing (infinitely big) expression.

**Theorem III.11 (in characteristic 5):**

Let $B^{(1)}, \dots, B^{(n)}$ be symmetric 4×4-matrices such that for i = 1,…,n $B^{(i)}M$ has not more than two eigenvalues, $M = \begin{pmatrix} 0_{2\times 2} & \begin{matrix} 0 & 1 \\ -1 & 0 \end{matrix} \\ \begin{matrix} 0 & 1 \\ -1 & 0 \end{matrix} & 0_{2\times 2} \end{pmatrix}$, $F^{(i)}, F^{(n+i)}$ be 4×2-matrices for i = 1,…n, x, g be n-vectors. Then

1) $\det_{tr}\left(\tilde{C}(x) + \mathrm{Diag}(g), \{B^{(i)}M\}_n\right) = \lim\limits_{\varepsilon\to 0} \dfrac{\det_{tr}\left(\tilde{C}\left(\begin{pmatrix} x \\ x \ddot{+} \varepsilon \end{pmatrix}\right) + \mathrm{Diag}\left(\begin{pmatrix} \varepsilon^{-1/2} g \\ \varepsilon^{-1/2}\vec{1}_n \end{pmatrix}\right), \begin{pmatrix} \{F^{(i)}(F^{(i)})^T M\}_n \\ \{F^{(n+i)}(F^{(n+i)})^T M\}_n \end{pmatrix}\right)}{\varepsilon^{-n}}$

$$\text{if for i = 1,…,n}\quad \begin{cases} Y^{(i)}\begin{pmatrix} 0_{2\times2} & Z^{(i)} \\ (Z^{(i)})^T & 0_{2\times2} \end{pmatrix}(Y^{(i)})^T = B^{(i)} \\ (Y^{(i)})^T M Y^{(i)} = \begin{pmatrix} 0_{2\times2} & -Z^{(i)} \\ (Z^{(i)})^T & 0_{2\times2} \end{pmatrix} \\ Y^{(i)} = \begin{pmatrix} F^{(i)} & F^{(n+i)} \end{pmatrix} \end{cases}$$

for a 2×2-matrix $Z^{(i)}$ such that $\operatorname{tr}(Z^{(i)}(Z^{(i)})^T) = 0$.

The above system of equations for the variables $Y^{(i)}, Z^{(i)}, F^{(i)}$ is solvable for an arbitrary symmetric 4×4-matrix $B^{(i)}$ such that $B^{(i)}M$ has not more than two eigenvalues.

2) this theorem's Part (1)'s conditions imply $(F^{(i)})^T M F^{(i)} = 0_{2\times2}$ for i = 1,…,2n and, accordingly, $\det_{tr}\left(\tilde{C}(x) + \operatorname{Diag}(g), \{B^{(i)}M\}_n\right)$ is polynomial-time computable for arbitrary symmetric 4×4-matrices $B^{(1)}, \dots, B^{(n)}$ such that for i = 1,…,n $B^{(i)}M$ has not more than two eigenvalues (including the partial case of $B^{(i)}M = F^{(i)}(F^{(i)})^T M$ having exactly one eigenvalue equal to zero, when $(F^{(i)})^T M F^{(i)} = 0_{2\times2}$).

Proof:
The main formula of this theorem (in Part (1)) is a direct implication of Theorem III.10.2 as its conditions imply the "opening" of the corresponding weighted sub-digraph for each ε-close pair of denominator-values $x_i, x_i + \varepsilon$. And now let's prove the theorem's statements on solvability and computability.
Let's use the fact that the system of equations $\begin{cases} YAY^T = B \\ Y^T\breve{A}Y = \breve{B} \end{cases}$ for the m×m-matrix variable $Y$, where $A, B$ are symmetric m×m-matrices and $\breve{A}, \breve{B}$ are skew-symmetric m×m-matrices, is solvable if and only if $A\breve{B}$ and $B\breve{A}$ have equal eigenvalue spectrums, while the product of a symmetric matrix and a skew-symmetric one (of the same size) has the eigenvalue spectrum of a skew-symmetric matrix, i.e. partitionable into pairs of opposite eigenvalues.

We hence conclude that the eigenvalue spectrums' equality for the matrices

$$\begin{pmatrix} 0_{2\times2} & Z^{(i)} \\ (Z^{(i)})^T & 0_{2\times2} \end{pmatrix}\begin{pmatrix} 0_{2\times2} & -Z^{(i)} \\ (Z^{(i)})^T & 0_{2\times2} \end{pmatrix} = \begin{pmatrix} Z^{(i)}(Z^{(i)})^T & 0_{2\times2} \\ 0_{2\times2} & -(Z^{(i)})^T Z^{(i)} \end{pmatrix} \text{ and } B^{(i)}M$$

is equivalent to the solvability of the system of equations
$$\begin{cases} Y^{(i)}\begin{pmatrix} 0_{2\times2} & Z^{(i)} \\ (Z^{(i)})^T & 0_{2\times2} \end{pmatrix}(Y^{(i)})^T = B^{(i)} \\ (Y^{(i)})^T M Y^{(i)} = \begin{pmatrix} 0_{2\times2} & -Z^{(i)} \\ (Z^{(i)})^T & 0_{2\times2} \end{pmatrix} \\ Y^{(i)} = \begin{pmatrix} F^{(i)} & F^{(n+i)} \end{pmatrix} \end{cases}$$
for the variables $Y^{(i)}, Z^{(i)}, F^{(i)}$, while the condition $tr(Z^{(i)}(Z^{(i)})^T) = 0$ implies that $\begin{pmatrix} Z^{(i)}(Z^{(i)})^T & 0_{2\times2} \\ 0_{2\times2} & -(Z^{(i)})^T Z^{(i)} \end{pmatrix}$ has not more than two eigenvalues.

**Theorem III.11.1 (in characteristic 5):**

Let $B^{(1)}, \dots, B^{(n)}$ be symmetric 4×4-matrices such that for i = 1,…,n $B^{(i)}M$ has not more than two eigenvalues, $M = \begin{pmatrix} 0_{2\times2} & \begin{matrix} 0 & 1 \\ -1 & 0 \end{matrix} \\ \begin{matrix} 0 & 1 \\ -1 & 0 \end{matrix} & 0_{2\times2} \end{pmatrix}$, $F^{(1)}, \dots, F^{(n)}$ be 4×2-matrices, t, g be n-vectors. Then

$$\det_{tr}\left(\tilde{C}(x) + Diag(g), \left\{(B^{(i)} + F^{(i)}(K^{(i)} - tr\left(K^{(i)}\right)I_2)(F^{(i)})^T)M\right\}_n\right) =$$
$$= \lim_{\varepsilon\to0}\lim_{\varepsilon_1\to0}\frac{\det_{tr}(\tilde{C}(\begin{pmatrix} x\overset{\dots}{+}\varepsilon \\ x \\ x\overset{\dots}{+}\varepsilon_1 \end{pmatrix}) + Diag(\begin{pmatrix} -g \\ \varepsilon^{-1}\vec{1}_n \\ -\varepsilon^{-1}\vec{1}_n \end{pmatrix}), \begin{pmatrix} \left\{B^{(i)}M\right\}_n \\ \left\{F^{(i)}(F^{(i)})^T M\right\}_n \\ \left\{F^{(i)}(F^{(i)})^T M\right\}_n \end{pmatrix})}{\varepsilon^{-2n}}$$

where for i = 1,…,n $K^{(i)} = (F^{(i)})^T M B^{(i)} M F^{(i)}$

Proof:
This statement follows from Theorem III.10.2 for the same reasons as Theorem III.11, with the only difference that first we take $\lim_{\varepsilon_1\to0}$ for to receive, for i =1,…,n, the pair of equal denominator-values $x_i, x_i$ with equal matrix-weights and opposite absence-weights that “make” them to be present and absent only “together”. And, when “appearing” together, they “join” the denominator-value $x_i + \varepsilon$ for to form an ε-close denominator-value family (with two identical “twin”-members) whose trace-determinant “opens”.

**Theorem III.11.2 (in characteristic 5):**

1) $\det_{tr}\left(\tilde{C}(x) + \mathrm{Diag}(g), \left\{(-B^{(i)} + F^{(i)}(K^{(i)} - \mathrm{tr}\left(K^{(i)}\right)I_2)(F^{(i)})^T)M\right\}_n\right)$, where for i = 1,…,n $K^{(i)} = (F^{(i)})^T M B^{(i)} M F^{(i)}$, is polynomial-time computable if for i = 1,…,n $B^{(i)}$ is an arbitrary symmetric 4×4-matrix such that $B^{(i)}M$ has not more than two eigenvalues and $F^{(i)}$ is an 4×2-matrix such that $(F^{(i)})^T M F^{(i)} = 0_{2\times 2}$, $M = \begin{pmatrix} 0_{2\times 2} & \begin{matrix} 0 & 1 \\ -1 & 0 \end{matrix} \\ \begin{matrix} 0 & 1 \\ -1 & 0 \end{matrix} & 0_{2\times 2} \end{pmatrix}$.
2) $\det_{tr}\left(\tilde{C}(x) + \mathrm{Diag}(g), \left\{G^{(i)}M\right\}_n\right)$ is polynomial-time computable for arbitrary symmetric 4×4-matrices $G^{(1)}, \dots, G^{(n)}$.

Proof:

Part (2). This statement follows from Theorem III.11.1 and the fact that any symmetric 4×4-matrix $G$ can be represented as $-B + F(K - \mathrm{tr}(K)I_2)F^T$ where $B$ is a symmetric 4×4-matrix such that $B^{(i)}M$ has not more than two eigenvalues, $F$ is an 4×2-matrix such that $F^T M F = 0_{2\times 2}$, and $K = F^T M B M F$.

Let's also notice that the class of matrices of the form $BM$, where $B$ is a symmetric 4×4-matrix and $M = \begin{pmatrix} 0_{2\times 2} & \begin{matrix} 0 & 1 \\ -1 & 0 \end{matrix} \\ \begin{matrix} 0 & 1 \\ -1 & 0 \end{matrix} & 0_{2\times 2} \end{pmatrix}$, is closed under the open trace-determinant compression operator. Hence above we've proven that its subclass of matrices of the form $FF^T M$, where $F$ is an 4×2-matrix such that $F^T M F = 0_{2\times 2}$ , generates the whole class via this operator's closure.

**Theorem III.12 (in characteristic 5):** let dim(x) = n, dim(y) = dim(λ) = m. Then

$$\mathrm{per}\left(C^{\star 2}(x, y)\mathrm{Diag}(\lambda)\right) =$$

$$= \left(\prod_{j=1}^{m} \lambda_j\right)\det_{tr}\left(\tilde{C}\left(\begin{pmatrix} x \\ y \end{pmatrix}\right) + \mathrm{Diag}\left(\begin{pmatrix} \vec{0}_n \\ \lambda^{\star(-1)} \end{pmatrix}\right), \begin{pmatrix} \{GM\}_n \\ \{\widehat{G}M\}_m \end{pmatrix}\right)$$

where

$$M=\begin{pmatrix} 0_{2\times 2} & \begin{matrix} 0 & 1 \\ -1 & 0 \end{matrix} \\ \begin{matrix} 0 & 1 \\ -1 & 0 \end{matrix} & 0_{2\times 2} \end{pmatrix},\ G=\begin{pmatrix} 0_{2\times 2} & \begin{matrix} 0 & 1 \\ \sqrt{-1} & 0 \end{matrix} \\ \begin{matrix} 0 & \sqrt{-1} \\ 1 & 0 \end{matrix} & 0_{2\times 2} \end{pmatrix},\ \widehat{G}=\begin{pmatrix} 0_{2\times 2} & \begin{matrix} 0 & -1 \\ \sqrt{-1} & 0 \end{matrix} \\ \begin{matrix} 0 & \sqrt{-1} \\ -1 & 0 \end{matrix} & 0_{2\times 2} \end{pmatrix}$$

Proof:

This identity follows from the fact that the matrices $GM$ and $\widehat{G}M$ are diagonal and hence commute under the matrix multiplication what makes the sum of the trace-weights of all the cycles covering a vertex set $K$ equal to the trace of the product of its vertices' matrix-weights multiplied by $\mathrm{ham}(\tilde{C}(\binom{x}{y}_K))$, while the latter expression isn't zero if and only if $|K| = 2$ and, in the meantime, $\mathrm{tr}((GM)^2) = \mathrm{tr}((\widehat{G}M)^2) = 0,\ \mathrm{tr}(GM\widehat{G}M) = 1$.

This theorem hence provides, due to the previous theorem regarding the polynomial-time computability of $\det_{tr}\left(\tilde{C}(x) + \mathrm{Diag}(g), \{G^{(i)}M\}_n\right)$ for any symmetric matrices $G^{(i)}$ in characteristic 5, the polynomial-time computability of $\mathrm{per}\left(C^{*2}(x,y)\mathrm{Diag}(\lambda)\right)$ in characteristic 5. Further, in Theorem III.32, we'll prove its $\#_5P$-completeness.

**Lemma III.12.1 (in a prime characteristic p):**

Let $\omega$ be a p-vector whose entries are all the elements of GF(p). Then

1) Let $f(u_1, \ldots, u_{p-k})$ be a symmetric homogeneous polynomial in p-k variables of degree q. Then $\sum_{I\subseteq\{1,\ldots,p\},|I|=k} \det\left(\tilde{C}(\omega_I)\right) f\left(\omega_{\backslash I}\right) = 0$ if $q \leq k < p-1$;
2) $\sum_{i=1}^{p} \det\left(\tilde{C}\left(\omega_{\backslash i}\right)\right) \omega_i^q = \begin{bmatrix} -1, q = p-1 \\ 0, q < p-1 \end{bmatrix}$ , for q = 0,…,p-1

Proof:

Part (1). This lemma can be proven via the use of Lemma III.4 as in characteristic p

$$\det\left(\tilde{C}(\omega_I)\right) = \det(\tilde{C}(\omega_I) + \mathrm{Diag}(\{\textstyle\sum_{j\in I, j\neq i}\frac{1}{\omega_i-\omega_j} - (p-1)\sum_{j\notin I}\frac{1}{\omega_i-\omega_j}\}_{i\in I})) =$$

$$= \mathrm{per}(C(\omega_I, \omega_{\backslash I} \otimes \vec{1}_{p-1})$$

where $\{\sum_{j\in I, j\neq i}\frac{1}{\omega_i-\omega_j} - (p-1)\sum_{j\notin I}\frac{1}{\omega_i-\omega_j}\}_{i\in I}$ is the |I|-vector indexed by the elements of I and having, for $i \in I$, its i-th entry equal to $\sum_{j\in I, j\neq i}\frac{1}{\omega_i-\omega_j} - (p-1)\sum_{j\notin I}\frac{1}{\omega_i-\omega_j}$. This vector is zero if $\omega$ is a p-vector whose entries are all the elements of GF(p).

Let’s also take into account the fact that for independent indeterminates $u_1, \dots, u_m$ there hold the identities

$$(*)\ \sum_{r=1}^{m} \frac{u_r^d}{\prod_{w, w \neq r}(u_r - u_w)} = 0 \ \text{ if } d < m-1$$

and

$$(**)\ \sum_{r=1}^{m} \frac{u_r^{m-1}}{\prod_{w, w \neq r}(u_r - u_w)} = 1.$$

For proving the lemma’s first part it’s sufficient to consider just a symmetric polynomial of the form $\frac{\sum_{\pi \in S_{p-k}} u_{\pi_1}^{q_1} \dots u_{\pi_{p-k}}^{q_{p-k}}}{(-1)^q q_1! \dots q_{p-k}!}$, where $q_1, \dots, q_{p-k}$ are non-zero integers such that $q_1 + \dots + q_{p-k} = q$, because any symmetric polynomial is a linear combination of such polynomials. Hence for proving Part (1) it’s sufficient to show that the identity

$$\sum_{I \subseteq \{1,\dots,p\}, |I| = k} \operatorname{per}(C(\omega_I, \omega_{\setminus I} \otimes \vec{1}_{p-1}) \sum_{\pi \in S_{GF(p) \setminus I}} \omega_{\pi_1}^{q_1} \dots \omega_{\pi_{p-k}}^{q_{p-k}} = 0$$

(where for a set J $S_J$ denotes the set of permutations on it)

holds for $q \leq k < p-1$.

Due to the above-mentioned facts, for q < k it equals zero because of (*) and for q = k, because of (**), it’s the number of GF(p)’s partitions into subsets of cardinalities $q_1 + 1, \dots, q_{p-k} + 1$ that is zero when k < p-1.

(Part 2) . It follows from (**).

**Theorem III.14 (in characteristic 5):**
let $P, \widehat{P}$ be 4×n-matrices, 5dim(u) + dim(v) + dim(w) = n. Then

$$\varphi_{5,2}(u, w, v, \gamma) = \frac{(-1)^{\dim(w)}}{2^{\dim(u)}} \left(\prod_{j=1}^{\dim(v)} \gamma_j\right) \lim_{\varepsilon \to 0} \operatorname{coef}_{\lambda^{\dim(v) - \dim(w)}}$$

$$\det_{tr}\left(\tilde{C}\left(\begin{pmatrix} u \overset{\cdots}{+} \varepsilon\omega \\ v \\ w \end{pmatrix}\right) + \operatorname{Diag}\left(\begin{pmatrix} -C\left(u \otimes \vec{1}_5, w\right)\vec{1}_{\dim(w)} \\ \lambda\gamma^{\star(-1)} \\ 0_{\dim(v)} \end{pmatrix}\right), \left\{\left(p_i \hat{p}_i^T + \hat{p}_i p_i^T\right) M\right\}_n\right)$$

where: $\omega = \begin{pmatrix} 0 \\ 1 \\ 2 \\ 3 \\ 4 \end{pmatrix}$; $M = \begin{pmatrix} 0_{2\times2} & \begin{matrix} 0 & 1 \\ -1 & 0 \end{matrix} \\ \begin{matrix} 0 & 1 \\ -1 & 0 \end{matrix} & 0_{2\times2} \end{pmatrix}$;

all the numerator-vectors corresponding to $u\dddot{\mp}\varepsilon\omega$ are $f$ and all the alternate numerator-vectors corresponding to $u\dddot{\mp}\varepsilon\omega$ are $\hat{f}$; all the numerator-vectors corresponding to w and v are equal to their alternate numerator-vectors and for w they are f, for v they are $\hat{f}$, where f, $\hat{f}$ are arbitrary 4-vectors satisfying the relation $f^T M\hat{f} = 1$

Proof:

In the present proof, with the considered trace-determinant we'll associate a weighted digraph whose vertices will be associated with the denominator-values $x = \begin{pmatrix} u\dddot{\mp}\varepsilon\omega \\ v \\ w \end{pmatrix}$, while by the weight of a proper cycle $\mathcal{C} = (x_{i_1}, \dots, x_{i_{|\mathcal{C}|}})$ we'll understand the cycle's ***trace-weight*** $(-1)^{|\mathcal{C}|+1}\mathrm{tr}\left(\prod_{q=1}^{|\mathcal{C}|} B^{\left(i_q(\mathcal{C})\right)}\right)\prod_{q=1}^{|\mathcal{C}|}\frac{1}{x_{i_q}-x_{\pi_{i_q}}}$, by a loop $(x_{i_1})$'s weight -- the corresponding absence-weight $g_i$ (i.e. the corresponding diagonal entry of the matrix $\tilde{C}\left(\begin{pmatrix} u\dddot{\mp}\varepsilon\omega \\ v \\ w \end{pmatrix}\right) + \mathrm{Diag}\left(\begin{pmatrix} -C\left(u\otimes\vec{1}_5, w\right)\vec{1}_{\dim(w)} \\ \lambda\gamma^{\star(-1)} \\ 0_{\dim(v)} \end{pmatrix}\right)$), and by a cycle system's weight -- the product of its cycles' weights. We'll also call a vertex absent (in a spanning cycle system) if it's covered by its loop, and ***present*** otherwise.

As we have $B^{(i)} = p_i\hat{p}_i^T + \hat{p}_i p_i^T$, the considered trace-determinant is the corresponding alternate determinant

$$\mathrm{altdet}\left(\begin{pmatrix} P^T MP \star \tilde{C}(t^{\star 5}) & P^T M\hat{P} \star \tilde{C}(t^{\star 5}) \\ \hat{P}^T MP \star \tilde{C}(t^{\star 5}) + \mathrm{Diag}(g) & \hat{P}^T M\hat{P} \star \tilde{C}(t^{\star 5}) \end{pmatrix}\right) =$$

with the numerator-vectors and alternate numerator-vectors $p_i$ and $\hat{p}_i$ correspondingly and the absence-weights $g_i$.

Let's call the denominator-values (as well as the corresponding vertices of the weighted digraphs we're going to build in this proof) of $\hat{u} = u\dddot{\mp}\varepsilon\omega$ **regular**, of w **active**, of v **passive**.

In each transversal summand of the considered matrix, let's consider its proper cycle system spanning its set of present vertices. We'll call a regular vertex **busy** if it's located in a cycle having not only regular vertices (and we'll call such a cycle **non-regular**), and **free** otherwise (and we'll accordingly call a cycle regular if it consists of regular vertices only). Besides, a pair of vertices whose denominator-values' difference's ε-order is bigger than zero will be called **ε-close** (of a specified ε-order, if necessary to detail).

First of all, let's notice that we can consider only regular cycles of length 2 because the sum of the weights of all the regular cycles covering a set of regular vertices of a cardinality bigger than 2 equals zero due to the fact that, for each regular vertex, its numerator-vector is f and its alternate numerator-vector is $\hat{f}$ and hence, due to the theorem's condition $f^T M\hat{f} = 1$, the weight of a cycle $\mathcal{C}$ covering a regular vertex set $\hat{u}_I$ equals $((\hat{f}^T Mf)^l + (\hat{f}Mf^T)^l)\prod_{(i_1,i_2)\in\mathcal{C}}\frac{1}{\hat{u}_{i_1}-\hat{u}_{i_2}}$ (where $l$ is the cycle's length) and hence it is zero when $l > 2$ because of the earlier proven fact that $\mathrm{ham}\left(\tilde{C}(x)\right) = 0$ if dim(x) > 2. Besides, due to the identity $\mathrm{ham}\left(\tilde{C}(x,y,z)\right) = \frac{1}{y-z}\prod_{i=1}^{\dim(x)}(\frac{1}{y-x_i} - \frac{1}{z-x_i})$ for dim(y) = dim(z) = 1 (that is a partial case of Lemma III.4) and the values of the numerator-vectors and alternate numerator-vectors given in the theorem (providing that active and passive vertices should alternate in any cycle of a non-zero weight if we don't take into account the regular vertices between them), any non-regular cycle of a non-zero weight should contain equal quantities of active and passive vertices and its weight equals the weight of the cycle received from it by removing all its regular vertices multiplied by $\prod_{i\in I}(\sum_{k\in K}\frac{2}{w_k-\hat{u}_i} - \sum_{j\in J}\frac{2}{v_j-\hat{u}_i})$ where I, J, K are the sets of its regular, passive and active vertices correspondingly. The latter relation will remain true if we replace the word "cycle" by "cycle system". Altogether, due to the given absence-weights of our denominator-values and upon taking the given coefficient at $\lambda^{\dim(v)-\dim(w)}$ (providing exactly dim(w) present passive vertices, while all the active ones are present due to having zero absence-weights), we receive the expression

$$\lim_{\varepsilon\to 0}\sum_K 2^{\frac{\dim(\hat{u}_{\setminus K})}{2}}\det\left(\tilde{C}(\hat{u}_{\setminus K})\right)\sum_{J,|J|=\dim(w)}(\prod_{i\in I}(\sum_{k=1}^{\dim(w)}\frac{2}{w_k-\hat{u}_i} - \sum_{j\in J}\frac{2}{v_j-\hat{u}_i}))\cdot$$

$$\cdot\det^2(C(w,v_J))\prod_{l\in\{1,\dots,\dim(v)\}\setminus J}\frac{1}{\gamma_l}$$

Let's now show that, in the above expression, for each i = 1,…,dim(u) the family of denominator-values $u_i\vec{1}_5 + \varepsilon\omega$ yields, under this limit, the Cauchy-base multiplier $\sum_{j\in J}\frac{1}{(u_i-v_j)^5}$ multiplied by $-1$. It follows from Theorem III.12.1 because the minimal ε-order we receive for this family is zero and we get it either when four denominator-values of the family are free and one is busy or when all its denominator-values are busy -- and those two cases together give us the multiplier $\sum_{k=1}^{\dim(w)}\frac{1}{(u_i-w_k)^5}-\sum_{j\in J}\frac{1}{(u_i-v_j)^5}$, -- or when all of them are absent, i.e. covered by their loops each of whom has the weight $-\sum_{k=1}^{\dim(w)}\frac{1}{(u_i-w_k)^5}$.

We hence obtain, altogether, the expression

$$\sum_{J,|J|=\dim(w)}\left(\prod_{i\in I}\left(-\sum_{j\in J}\frac{1}{(u_i-v_j)^5}\right)\right)\det^2\left(C(w,v_J)\right)\prod_{r\in\{1,\dots,\dim(v)\}\setminus J}\frac{1}{\gamma_r}$$

what completes the proof.

**Theorem III.15 (in a prime characteristic p):**

Let A be an n×n-matrix, h be an even number. Then $\mathrm{per}(A)=$
$\varphi_{p,h}\left(z\otimes\vec{1}_{p^{q-1}},w,v,\gamma\right)=\varphi_{1,h}\left(z\otimes\vec{1}_{p^{q}},w,v,\gamma\right)$

where: dim(w) = dim(z) = n, dim(v) = dim($\gamma$) > $h(n^2+n)$

$C^{\star p^q}(z,v)\mathrm{Diag}(\gamma)C^{\star s}(v,w)=0_{n\times n}$ for s = 1,...,h-1

$C^{\star p^q}(z,v)\mathrm{Diag}(\gamma)C^{\star h}(v,w)=A$

$C^{\star s}(w,v)\mathrm{Diag}(\gamma)=\vec{0}_n$ for s = 1,...,h

while the above system of linear equations for $\gamma$ is nonsingular in the generic case if $p^q>h(n^2+n)$ .

**Comment:** $\varphi_{p,h}\left(z\otimes\vec{1}_{p^{q-1}},w,v,\gamma\right)$ can be polynomial-time computed as $\lim_{\varepsilon\to 0}\varphi_{p,h}\left(z\otimes\vec{1}_{p^{q-1}}+\varepsilon\zeta,w,v,\gamma\right)$ where $\zeta$ is an arbitrary $p^{q-1}\dim(z)$-vector with pair-wise distinct entries.

Proof:

This theorem is based on the following generalization of the Cauchy-Binet identity (about the determinant of the product of two matrices), valid in an arbitrary characteristic:

Let $A^{(1)},\dots,A^{(h)}$ be n×m-matrices, h be non-zero even, B be a k×m-matrix. Then $\left(\prod_{v=1}^{h}\det\left((A^{(v)})^{(\{1,\dots,n\},J)}\right)\right)\prod_{r=1}^{k}\sum_{j\in J}b_{r,j}=$

$$=\sum_{\substack{\pi^{(2)},\dots,\pi^{(h)}\in S_n,\\ (R_1,\dots,R_n)\in\mathcal{P}_n(\{1,\dots,k\})}}\sigma(\pi^{(2)}\dots\pi^{(h)})\prod_{i=1}^{n}\sum_{j=1}^{m}a_{i,j}^{(1)}a_{\pi_i^{(2)},j}^{(2)}\dots a_{\pi_i^{(h)},j}^{(h)}\prod_{r\in R_i}b_{r,j}$$

where $\mathcal{P}_n(\{1,\dots,k\})$ is the set of partitions of the set $\{1,\dots,k\}$ into n subsets (some of them possibly empty) and $\sigma(\pi^{(2)}\dots\pi^{(h)})$ is the sign of the permutation $\pi^{(2)}\dots\pi^{(h)}$.

In our case we have, by the definition, $\varphi_{p,h}\left(z\otimes\vec{1}_{5^{q-1}}, w, v, \gamma\right) =$
$$= \sum_{\substack{J\subseteq\{1,\dots,m\}\\ |J|=n}} \det\left((C(w,v))^{(\{1,\dots,n\},J)}\right)\dots\det\left((C(w,v))^{(\{1,\dots,n\},J)}\right)\prod_{r=1}^{k}\sum_{j\in J}\frac{\gamma_j}{(z_r-v_j)^{p^q}}.$$
Hence, while considering the entries of the vectors w, z as “constants” and the entries of the vectors v, γ as “variables”, we can say that in our case each expression $\sum_{j=1}^{m} a_{1,j}^{(1)} a_{\pi_i^{(2)},j}^{(2)}\dots a_{\pi_i^{(n)},j}^{(h)}\prod_{r\in R_i} b_{r,j}$ is a linear combination of the sums $\sum_{j=1}^{m}\frac{\gamma_j}{(w_i-v_j)^s}$ with i = 1,…,n, s = 1,…,h and $\sum_{j=1}^{m}\frac{\gamma_j}{(z_r-v_j)^s}$ with r = 1,…,n, s = 1,…,$p^q$ . Due to the fact that, according to the theorem’s conditions, all the former ones are equal to zero, in our case each $\sum_{j=1}^{m} a_{1,j}^{(1)} a_{\pi_i^{(2)},j}^{(2)}\dots a_{\pi_i^{(n)},j}^{(h)}\prod_{r\in R_i} b_{r,j}$ is a linear combination of the sums $\sum_{j=1}^{m}\frac{\gamma_j}{(z_r-v_j)^s}$ only and isn’t zero only if $R_i$ isn’t empty; hence we can consider only partitions $(R_1,\dots,R_n)$ where all the subsets $R_i$ are of cardinality 1. Therefore we can consider only the expressions $\sum_{j=1}^{m}\frac{\gamma_j}{(w_{i_1}-v_j)\dots(w_{i_h}-v_j)(z_r-v_j)^{p^q}}$ that are linear combinations of the sums $\sum_{j=1}^{m}\frac{\gamma_j}{(w_i-v_j)^s(z_r-v_j)^{p^q}} = \sum_{t=0}^{p^q-1}\frac{(-s)\dots(-s-t+1)}{t!(w_i-z_r)^{s+t}}\sum_{j=1}^{m}\frac{\gamma_j}{(z_r-v_j)^{p^q-t}}$ with s =1,…,h, i = 1,…,n, r = 1,…,n and form a non-singular system of $hn^2$ linear functions in the sums $\sum_{j=1}^{m}\frac{\gamma_j}{(z_r-v_j)^{p^q-t}}$ with r = 1,…,n, t = 1,…, $p^q-1$. According to the theorem’s conditions, the sum $\sum_{j=1}^{m}\frac{\gamma_j}{(w_i-v_j)^s(z_r-v_j)^{p^q}}$ is zero when $s < h$ and equals $a_{i,r}$ when s = h. The latter case implies $w_{i_1}=\dots=w_{i_h}$ and hence we can consider only the case $\pi^{(2)}=\dots=\pi^{(h)}=\begin{pmatrix}1,\dots,n\\1,\dots,n\end{pmatrix}$ and, because $\sigma\left(\pi^{(2)}\dots\pi^{(h)}\right)=1$ in such a case, we eventually get per(A), while the sums $\sum_{j=1}^{m}\frac{\gamma_j}{(z_r-v_j)^{p^q-t}}$ and $\sum_{j=1}^{m}\frac{\gamma_j}{(w_i-v_j)^s}$ with r, i = 1,…,n and t = 0,…,$p^q-1$ generically form a nonsingular system of linear functions in $\gamma_1,\dots,\gamma_m$ provided $p^q > h(n^2+n)$.

We’ve hence proven the $\#_p$P-completeness of the Cauchy determinant base-sum for any odd prime p and the Cauchy base-degree 1. In fact, a similar proof can be arranged for any natural Cauchy base-degree.

Let’s also formulate the Cauchy-Binet identity’s generalization we used in this proof, even in a wider form:

**Theorem III.15.1 (in any characteristic):**
Let $A^{(1)}, \dots, A^{(h)}, A^{(h+1)}, \dots, A^{(h+d)}$ be n×m-matrices, h be non-zero even, B be a k×m-matrix. Then

$$\sum_{\substack{J\subseteq\{1,\dots,m\}\\ |J|=n}} \left(\prod_{v=1}^{h} \det\left(\left(A^{(v)}\right)^{(\{1,\dots,n\},J)}\right)\right)\left(\prod_{v=h+1}^{h+d} \operatorname{per}\left(\left(A^{(v)}\right)^{(\{1,\dots,n\},J)}\right)\right)\prod_{r=1}^{k}\sum_{j\in J} b_{r,j} =$$

$$= \sum_{\substack{\pi^{(2)},\dots,\pi^{(h+d)}\in S_n,\\ (R_1,\dots,R_n)\in\mathcal{P}_n(\{1,\dots,k\})}} \sigma(\pi^{(2)}\dots\pi^{(h)}) \prod_{i=1}^{n}\sum_{j=1}^{m} a^{(1)}_{i,j}\, a^{(2)}_{\pi^{(2)}_i,j} \dots a^{(h+d)}_{\pi^{(h+d)}_i,j} \prod_{r\in R_i} b_{r,j}$$

where $\mathcal{P}_n(\{1,\dots,k\})$ is the set of partitions of the set $\{1,\dots,k\}$ into n subsets (some of them possibly empty) and $\sigma(\pi^{(2)}\dots\pi^{(h)})$ is the sign of the permutation $\pi^{(2)}\dots\pi^{(h)}$.

Additionally, we can also formulate

**Theorem III.16 (in characteristic 5):** let dim(z) = n, dim(y) = m. Then

$$\left(\prod_{i=1}^{n}\left(1+d_i\frac{\partial}{\partial z_i}\right)\right)\det\left(\left\{\frac{\alpha_i-\alpha_j+(\beta_i+\beta_j)(z_i-z_i)}{(z_i-z_j)^5}\right\}_{n\times n}+\operatorname{Diag}(h)\right)$$

$$= \operatorname{coef}_{\lambda^n}\lim_{\varepsilon_1\to 0}\lim_{\varepsilon\to 0}\frac{\operatorname{Pf}\left(K\left(\begin{pmatrix} y\dddot{+}\begin{pmatrix}0\\ \varepsilon\end{pmatrix}\dddot{+}\begin{pmatrix}0\\ \varepsilon_1\end{pmatrix}\\ z\end{pmatrix},\begin{pmatrix}\vec{0}_{4m}\\ d\end{pmatrix},\begin{pmatrix}(\varepsilon\varepsilon_1)^2(\lambda g)^{\star(-1)}\dddot{+}\left(\varepsilon^{-2}+\frac{\varepsilon^2}{2}\varepsilon_1^{-4}\right)\vec{1}_4\\ h\end{pmatrix}\right)\right)}{(-2)^{n+4m}\varepsilon_1^{-2m}}$$

where $\alpha_i = \alpha(z_i) = \sum_{k=1}^{m}\frac{g_k}{(y_k-z_i)^3}$ , $\beta_i = \beta(z_i) = \sum_{k=1}^{m}\frac{g_k}{(y_k-z_i)^4}$ for i = 1,…,n

****************************************************

**Theorem III.17 (in characteristic 5):** Let dim(x) = n, A be a nonsingular skew-symmetric 4n×4n-matrix, $\dim(\zeta) = m$ .Then

$$\operatorname{per}\left(C\left(x\otimes\vec{1}_4,\zeta\otimes\vec{1}_2\right)\operatorname{Diag}\left(d\otimes\begin{pmatrix}1\\ -1\end{pmatrix}\right)\right) =$$

$$= \frac{\operatorname{coef}_{\gamma^{4n}}\det^4(\operatorname{Van}(x))\operatorname{Pf}\left(\begin{pmatrix}\gamma D\tilde{C}(\zeta)D & I_n\\ -I_n & \left(\operatorname{Van}^{[4n]}(\zeta)\right)^T A\operatorname{Van}^{[4n]}(\zeta)\end{pmatrix}\right)}{\operatorname{Pf}(A)}$$

where $D = \mathrm{Diag}(\left\{\sqrt{\frac{-d_j}{pol(\zeta_j,x)}}\right\}_m)$

Proof:

This statement is due to Lemma III.4. The numerator of the theorem's equality's right side is

$$\det^4(Van(x)) \sum_{J,|J|=2n} Pf(\tilde{C}(\zeta_J))Pf(Van^{[4n]}(\zeta_J))^T A Van^{[4n]}(\zeta_J) \prod_{j\in J} \frac{-d_j}{pol(\zeta_j,x)} =$$

$$= \det^4(Van(x)) \sum_{J,|J|=2n} Pf(\tilde{C}(\zeta_J))Pf(Van(\zeta_J))^T A Van(\zeta_J) \prod_{j\in J} \frac{-d_j}{pol(\zeta_j,x)}$$

$$= \det^4(Van(x)) \sum_{J,|J|=2n} Pf(\tilde{C}(\zeta_J))\det(Van(\zeta_J))\, Pf(A) \prod_{j\in J} \frac{-d_j}{pol(\zeta_j,x)} =$$

$$= \det^4(Van(x)) \sum_{J,|J|=2n} \frac{\det(W^{[4]}(\binom{\zeta_J}{\zeta_J}))}{\det(Van(\zeta_J))} \det(Van(\zeta_J))\, Pf(A) \prod_{j\in J} \frac{-d_j}{pol(\zeta_j,x)} =$$

$$= Pf(A) \sum_{J,|J|=2n} \det^4(Van(x))\det(W^{[4]}(\binom{\zeta_J}{\zeta_J})) \prod_{j\in J} \frac{-d_j}{pol(\zeta_j,x)} =$$

$$= Pf(A) \sum_{J,|J|=2n} per(C(x \otimes \vec{1}_4, \zeta_j \otimes \vec{1}_2) \prod_{j\in J}(-d_j)\ ,$$

while the left side is $\sum_{J,|J|=2n} per(C(x \otimes \vec{1}_4, \zeta_j \otimes \vec{1}_2) \prod_{j\in J}(-d_j)$

*******************************************************

**Theorem III.18 (the Binet-Minc identity, for any characteristic)**

Let A be an n×m-matrix, then

$$per(A) = (-1)^n \sum_{P\in Part(\{1,\dots,n\})} \prod_{I\in P} (-(|I|-1)! \sum_{j=1}^{m} \prod_{i\in I} a_{ij})$$

where Part({1,,…,n}) is the set of partitions of the set {1,…,n} into non-empty subsets.

# Sparse compressions in characteristic 3

**Theorem III.19** (in any characteristic). Let dim(z) = 2dim(x). Then

$$\det(C^{\star(1,2)}(x,z)) \;=(-1)^{\dim(x)}\frac{\det^4\big(\mathrm{Van}(x)\big)\det\big(\mathrm{Van}(z)\big)}{\mathrm{pol}^2(x,z)}$$

**Theorem III.20** (in any characteristic). Let dim(z) = 2dim(x). Then

$$\det(C^{\star(2,3)}(x,z)) \;=(1/2)^{\dim(x)}\mathrm{per}(C(\binom{x}{x},z)\det(C^{\star(1,2)}(x,z))$$

Proof:

This statement follows directly from the Borchardt identity as

$$\det(C^{\star(2,3)}(x,z))=\lim_{\varepsilon\to 0}\frac{\det(C^{\star 2}(\begin{pmatrix} x \\ x+\varepsilon\end{pmatrix},z))}{(-2\varepsilon)^{\dim(x)}}=\lim_{\varepsilon\to 0}\frac{\det(C(\begin{pmatrix} x \\ x+\varepsilon\end{pmatrix},z))\mathrm{per}(C(\begin{pmatrix} x \\ x+\varepsilon\end{pmatrix},z))}{(-2\varepsilon)^{\dim(x)}}=$$

$$=\frac{\mathrm{per}(C(\binom{x}{x},z)(-1)^{\dim(x)}\det(C^{\star(1,2)}(x,z))}{(-2)^{\dim(x)}}$$

## *A conjectured polynomial-time algorithm for computing the permanent in characteristic 3*

**Theorem III.21 (in characteristic 3):**

Let A be a nonsingular skew-symmetric 2n×2n-matix, dim(x) = n, dim(y) = m. Then

$$\mathrm{per}^3(C(\binom{x}{x},y)\mathrm{Diag}(d))=2^n\frac{\mathrm{coef}_{\gamma^n}\mathrm{Pf}(\begin{pmatrix}\gamma D\tilde{C}(y)D & I_m \\ -I_m & (C^{\star(2,3)}(x,y))^T A C^{\star(2,3)}(x,y)\end{pmatrix})}{\mathrm{Pf}(A)}$$

where $D=\mathrm{Diag}^3(d)$ .

Proof:

The proof of this theorem is based on Lemma III.4 as $\mathrm{Pf}(\tilde{C}(y_J))=\frac{\mathrm{per}^2(W^{[2]}(y_J))}{\det(\mathrm{Van}(y_J))}$ and $\mathrm{Pf}((C^{\star(2,3)}(x,y_J))^T A C^{\star(2,3)}(x,y_J))=\det(C^{\star(2,3)}(x,y_J))\mathrm{Pf}(A)=$

$$=2^n\mathrm{per}(C(\binom{x}{x},y_J)\det(C^{\star(1,2)}(x,y_J))\mathrm{Pf}(A)=$$

$$= \mathrm{per}(C(\binom{x}{x}, y_J) \frac{\det^4(\mathrm{Van}(x)) \det(\mathrm{Van}(y_J))}{\mathrm{pol}^2(x, y_J)} \mathrm{Pf}(A) \;,$$

while the numerator of the theorem's equality's right side is
$\sum_{J,|J|=2n} \mathrm{Pf}(\tilde{C}(y_J))\mathrm{Pf}((C^{\star(2,3)}(x, y_J))^T A C^{\star(2,3)}(x, y_J)) \prod_{j\in J} d_j^3 =$

$$= \sum_{J,|J|=2n} \frac{\mathrm{per}^2(W^{[2]}(y_J))}{\det(\mathrm{Van}(y_J))} \mathrm{per}(C(\binom{x}{x}, y_J) \frac{\det^4(\mathrm{Van}(x))\det(\mathrm{Van}(y_J))}{\mathrm{pol}^2(x, y_J)} \prod_{j\in J} d_j^3$$

Taking into account the fact that, according to Lemma III.4, $\mathrm{per}^2(C(\binom{x}{x}, y_J) = \frac{\det^4(\mathrm{Van}(x))}{\mathrm{pol}^2(x,y_J)} \mathrm{per}^2(W^{[2]}(y_J))$, we complete the proof.

Accordingly, via the reduction

$$\lim_{\varepsilon\to 0}(\varepsilon^{\dim(x)} \mathrm{per}(C(\binom{x}{x}, \binom{x + \varepsilon\vec{1}_{\dim(x)}}{z})\mathrm{Diag}(\binom{\vec{1}_{\dim(x)}}{\lambda}))) =$$

$= 2^{\dim(x)}\mathrm{per}(C(x, z)\mathrm{Diag}(\lambda))$, we receive also

**Theorem III.22 (in characteristic 3):**

$\mathrm{per}(C(x, z)\mathrm{Diag}(\lambda))$ is computable in polynomial time for arbitrary .

**Definition:** for dim(x) = dim(d) = dim($a$),

$$\rho(x, d, a) \coloneqq (\prod_{i=1}^{n} (1 + d_i \frac{\partial}{\partial r_i}))\det(\tilde{C}(x) + \mathrm{Diag}(a))$$

**Theorem III.23 (in characteristic 3):** $\rho(x, d, a) = \mathrm{per}(C(x^{\star 3}, z)\mathrm{Diag}(\lambda))$

where $C(x^{\star 3}, z)\lambda^{\star q} = \delta(q-1)a - \delta(q-2)(x + d)$ for q = 1,2,3 and this system of equations for z, λ is generically algebraically nonsingular.

**Theorem III.24 (in characteristic 3):**

$$(\prod_{i=1}^{\dim(t)} \frac{\partial}{\partial t_i})\det(\tilde{C}(\binom{x}{t}) + \mathrm{Diag}\binom{\alpha}{t \star \beta}) = \det(\tilde{C}(x) + \mathrm{Diag}(\alpha)) \cdot$$

$$\cdot \operatorname{altdet}(\mathrm{Schur}_{T,T}\begin{pmatrix} \tilde{C}(x) + \mathrm{Diag}(\alpha) & C(x, t) & C^{\star 2}(x, t) \\ C(t, x) & \tilde{C}(t) + D_{11} & \tilde{C}^{\star 2}(t) + D_{12} \\ -C^{\star 2}(t, x) & -\tilde{C}^{\star 2}(t) + D_{21} & \tilde{C}^{\star 3}(t) + D_{22} \end{pmatrix}) =$$

$$= \mathrm{altdet}\begin{pmatrix} P_1^T P_2 \star \tilde{C}(t^{\star 9}) + \breve{D}_{11} & P_1^T \hat{P}_2 \star \tilde{C}(t^{\star 9}) + \breve{D}_{12} \\ \hat{P}_1^T P_2 \star \tilde{C}(t^{\star 9}) + \breve{D}_{21} & \hat{P}_1^T \hat{P}_2 \star \tilde{C}(t^{\star 9}) + \breve{D}_{22} \end{pmatrix} =$$

$$= \det_{tr}(\tilde{C}(t) + \mathrm{Diag}(g), \{p_{1,i}\hat{p}_{2,i}^T + \hat{p}_{1,i}p_{2,i}^T\}_n)$$

where T={1,…,dim(x)} , $P_1, \hat{P}_1, P_2, \hat{P}_2$ are some n×9-matrices such that for i = 1,…n $p_{1,i}^T p_{2,i} = \hat{p}_{1,i}^T p_{2,i} = p_{1,i}^T \hat{p}_{2,i} = \hat{p}_{1,i}^T \hat{p}_{2,i} = 0$; for $k, l$ = 1,2 $D_{k,l}, \breve{D}_{k,l}$ are diagonal,

$D_{12} + D_{21} = \mathrm{Diag}(\beta)$, $\breve{D}_{12} + \breve{D}_{21} = \mathrm{Diag}(g)$.

**Conjecture III.25**: let T={1,…,dim(v)}, $D_{k,l}$ be diagonal matrices for $k, l$ = 1,2. Then the class of matrices of the form

$$\mathrm{Schur}_{\{1,\dots,\dim(x)\},\{1,\dots,\dim(x)\}}\begin{pmatrix} \tilde{C}(x) + \mathrm{Diag}(\alpha) & C(x,t) & C^{\star 2}(x,t) \\ C(t,x) & \tilde{C}(t) + D_{11} & C^{\star 2}(t) + D_{12} \\ -C^{\star 2}(t,x) & -C^{\star 2}(t) + D_{21} & C^{\star 3}(t) + D_{22} \end{pmatrix})$$

(where $D_{k,l}$ are diagonal for $k, l = 1,2$)

is generically the class

$$\begin{pmatrix} P_1^T P_2 \star \tilde{C}(t^{\star 9}) + \breve{D}_{11} & P_1^T \hat{P}_2 \star \tilde{C}(t^{\star 9}) + \breve{D}_{12} \\ \hat{P}_1^T P_2 \star \tilde{C}(t^{\star 9}) + \breve{D}_{21} & \hat{P}_1^T \hat{P}_2 \star \tilde{C}(t^{\star 9}) + \breve{D}_{22} \end{pmatrix}$$

(where $P_1, \hat{P}_1, P_2, \hat{P}_2$ are n×9-matrices, $\breve{D}_{k,l}$ are diagonal for $k, l$=1,2)

of singularized matrices of Cauchy-rank 9 all whose denominator-values are alternate-wise doubled and $p_{1,i}^T p_{2,i} = \hat{p}_{1,i}^T p_{2,i} = p_{1,i}^T \hat{p}_{2,i} = \hat{p}_{1,i}^T \hat{p}_{2,i} = 0$ for i = 1,…,n.

The above conjecture is an analogue of Theorem III.9 in characteristic 5 and it's based on Theorem III.23. If it's true we can, analogically, generate certain families of infinitesimal-close denominator-values with their non-symmetric (due to the considered matrix's non-symmetry in this case) matrix-weights and absence-weights, while using Theorem III.10.2 for compressing those families into denominator-values whose matrix-weights and absence-weights can be *conjectured* arbitrary. If the later conjecturing doesn't fail too then we can use the fact that in any characteristic we have an exact analog of Theorem III.14 for left and right numerator-vectors and alternate numerator-vectors (row- and column-numerators and alternate ones) of dimension 9, with any nonsingular multiplication matrix M. There holds also an analog of Theorem III.12 providing polynomial-time generating, by the trace-determinant of $\tilde{C}(t) + \mathrm{Diag}(g)$ on arbitrary 9×9-matrix-weights, $\operatorname{per}\begin{pmatrix} C^{\star 2}(y^{(1)}, y)\operatorname{Diag}(\lambda^{(1)}) \\ \dots \\ C^{\star 2}(y^{(4)}, y)\operatorname{Diag}(\lambda^{(4)}) \end{pmatrix}$ that is reducible, as it will further be shown in Theorem III.31, to $\operatorname{per}\begin{pmatrix} C^{\star 3}(y^{(1)}, y)\operatorname{Diag}(\lambda^{(1)}) \\ \dots \\ C^{\star 3}(y^{(4)}, y)\operatorname{Diag}(\lambda^{(4)}) \end{pmatrix}$ and $\#_3 P$-complete.

Thus we come to the following conclusion:

**Theorem III.26:**

Over fields of characteristic 3, computing the trace-determinant of $(\tilde{C}(t) + \operatorname{Diag}(g))$ on arbitrary 9×9-matrix-weights $\#_3$-P-complete.

We can comment, however, that in fact Theorem III.14 can be re-formulated even for 2×2-matrix-weights (i.e. numerator-vectors and alternate numerator-vectors of dimension 2, with any non-singular skew-symmetric multiplication 2×2-matrix).

**Theorem III.28 (in characteristic p):**

Let h be an even natural number bigger than 2 that isn't a square modulo p. Then

$$\varphi_{p,h}(u, w, v, \gamma) = \lim_{\varepsilon_1 \to 0} \lim_{\varepsilon \to 0} \frac{\varphi_{0,h}\left(\emptyset, \binom{w}{\hat{u}}, \overset{v}{\left(\hat{u} \dddot{\mp} \varepsilon \binom{1}{-1}\right)}, \overset{\gamma}{\left(\vec{1}_{\dim(\hat{u})} \otimes \binom{1}{-1} + \varepsilon h C(\hat{u}, w) \vec{1}_{\dim(w)} \otimes \binom{1}{0}\right)}\right)}{(2h)^{\dim(u)} \varepsilon^{\dim(\hat{u})}}$$

where $\hat{u} = u \dddot{\mp} \varepsilon_1 \begin{pmatrix} 0 \\ 1 \\ \dots \\ p-1 \end{pmatrix}$

Proof:

The proof of this theorem is based on Theorem III.12.1 and Lemma III.4.

The first limit $\lim_{\varepsilon \to 0}$ turns the fraction in the theorem's equality's right side into

$$\frac{1}{(2h)^{\dim(u)}} \sum_{\substack{J \subseteq \{1, \dots, \dim(v)\} \\ |J| = \dim(w)}} \sum_{I, I^{(1)}, \dots, I^{(h)}} \left(\prod_{i \in I} \left(h C(\hat{u}_i, w) \vec{1}_{\dim(w)}\right)\right) \left(\prod_{q=1}^{h} \det{}^h\left(\tilde{C}\left(\hat{u}_{I^{(q)}}, w, v_J\right)\right)\right) \prod_{j \in J} \gamma_j$$

where the summation is over all the (h+1)-tuples $I, I^{(1)}, \dots, I^{(h)}$ that are partitions of the set {1,…,dim($\hat{u}$)} into h+1 subsets, some of them possibly empty.

According to the statement (3) of Lemma III.4, this expression is equal to $\frac{1}{(2h)^{\dim(u)}} \sum_{\substack{J \subseteq \{1,\dots,\dim(v)\} \\ |J| = \dim(w)}} \sum_{K \subseteq \{1,\dots,\dim(\hat{u})\}} h^{\frac{\dim(\hat{u}) - |I|}{2}} \det(\tilde{C}(\hat{u}_{\setminus I}))(\prod_{i \in I} \sum_{j \in J} \frac{h}{\hat{u}_i - v_j}) \det^h(\tilde{C}(w, v_J)) \prod_{j \in J} \gamma_j$ .

Therefore, due to the structure of the vector $\hat{u}$, the second limit $\lim_{\varepsilon_1 \to 0}$ provides the correctness of the theorem's identity because of the same argument (referring to Theorem III.12.1) that was applied in the proof of Theorem III.14.

**Corollary III.30:**
For an arbitrary prime characteristic p, $\varphi_{0,h}(\emptyset, w, v, \beta)$ is $\#_p P$-complete for any even h > 2 that isn't a square modulo p.
Proof: this corollary from Theorem III.28 is based on Theorem III.15 proving the $\#_p P$-completeness of $\varphi_{p,h}(u, w, v, \gamma)$.

**Theorem III.31:** let p be a prime number bigger than 5. Then computing $\mathrm{per}(C(x,z)\mathrm{Diag}(\lambda))$ over fields of characteristic p is $\#_p P$-complete.

Proof:
For the case when $-1$ isn't a square modulo p, it follows immediately from Corollary III.30 and the fact that

$$\varphi_{0,p-1}(\emptyset, w, v, \lambda^{\star(p-1)}) = \mathrm{per}(C(w \otimes \vec{1}_{p-1}, v \otimes \vec{1}_{p-1})\mathrm{Diag}(\lambda \otimes \begin{pmatrix} 1 \\ \dots \\ p-1 \end{pmatrix}))$$

However, this theorem can be proven in a different way (common for all the prime characteristics bigger than 5) based on the Binet-Minc identity.

We'll say that the left denominator-value $x_i$ is of multiplicity $\mathrm{mult}(x_i)$ if it's repeated $\mathrm{mult}(x_i)$ times in the vector x. Then, according to the Binet-Minc identity for characteristic p, $\mathrm{per}(C(x,z)\mathrm{Diag}(\lambda))$ is a polynomial in the values $\sum_{j=1}^{\dim(z)} \frac{\lambda_j^r}{(x_i - z_j)^s}, s = 1, \dots, \mathrm{mult}(x_i), r = 1, \dots, p$, that are a system of algebraically independent functions in z,λ upon excluding those of them where r and s are both divisible by p. Let's call the sum $\sum_{j=1}^{\dim(z)} \frac{\lambda_j^r}{(x_i - z_j)^s}$ the ***r,s-row-weight*** of the left denominator-value $x_i$ and the maximum set $\{x_1, \dots, x_m\}$ of pair-wise distinct left (row) denominator-values the ***left denominator-value spectrum*** of $C(x,z)$. When the vectors $z, \lambda$ are presented as $\begin{pmatrix} z^{(1)} \\ z^{(2)} \end{pmatrix}$ and $\begin{pmatrix} \lambda^{(1)} \\ \lambda^{(2)} \end{pmatrix}, \dim(z^{(1)}) = \dim(\lambda^{(1)})$, we'll call $z^{(1)}, \lambda^{(1)}$ the main parts of $z, \lambda$ correspondingly and $z^{(1)},\lambda^{(1)}$ the ***prolonged*** parts, while a row-weight will accordingly be the sum of its main part $\sum_{j=1}^{\dim(z)}\frac{(\lambda_j^{(1)})^r}{(x_i-z_j^{(1)})^s}$ and its prolonged part $\sum_{j=1}^{\dim(z)}\frac{(\lambda_j^{(2)})^r}{(x_i-z_j^{(2)})^s}$, and the latter we'll also call the ***prolonged row-weight***.

Hence, upon putting $\sum_{j=1}^{\dim(z)}\frac{(\lambda_j^{(2)})^{mult(x_i)}}{(x_i-z_j^{(2)})^{mult(x_i)}}=\frac{1}{d_i}$ and all the other prolonged row-weights equal to zero, we receive the relation

$$per\left(C\left(\{x_i\vec{1}_{mult(x_i)}\}_{i\in I},z\right)Diag(\lambda)\right)=\frac{\sum_{I\subseteq\{1,\dots,m\}}(\prod_{i\in I}d_i)per(C\left(\{x_i\vec{1}_{mult(x_i)}\}_{i\in I},z^{(1)}\right)Diag(\lambda^{(1)}))}{\prod_{i=1}^{\dim(x)}d_i}$$

where $d_i$ we'll call the ***summation weight*** of $x_i$. If p > 5 this expression polynomial-time yields, upon taking a number of infinitely close (on some infinitesimal) pairs of left-spectral denominator-values of multiplicity p-2 and their summation weights of infinitesimal-order -1, the expression

$$\left(\prod_{t\in T}\frac{\partial}{\partial x_t}\right)per\left(C\left(x\otimes\vec{1}_{p-2},z\right)Diag(\lambda)\right)$$

where T is a subset of {1,…,m}. Let's now consider, as a generalization of the above expression (up to multiplying its rows by constants), the expression:

$$per\left(\begin{pmatrix}C^{\star\gamma_1}(x_1,z)\\ \dots\\ C^{\star\gamma_m}(x_m,z)\end{pmatrix}Diag(\lambda)\right)$$

where $\gamma_1$, …, $\gamma_m$ are natural sequences (Hadamard vector-degrees) which we'll call the ***valences*** of the left-spectral denominator-values $x_1,\dots,x_m$ correspondingly.

Particularly, the already above-considered expression $\left(\prod_{t\in T}\frac{\partial}{\partial x_t}\right)per\left(C\left(\{x\otimes\vec{1}_{p-2},z\right)Diag(\lambda)\right)$ can be written as the expression $per\left(\begin{pmatrix}C^{\star\gamma_1}(x_1,z)\\ \dots\\ C^{\star\gamma_m}(x_m,z)\end{pmatrix}Diag(\lambda)\right)$ (multiplied by a constant) where some valences are $(\vec{1}_{p-2}^T)$ and others (i.e. those from the set T) are $(\vec{1}_{p-3}^T,2)$.

Given a ∈ $F(\varepsilon)$, where F is the ground field and ε is an infinitesimal, let's further say that we apply the limit technique to a when we compute $\lim_{\varepsilon\to 0}\frac{a}{e^{order_\varepsilon a}}$.

If p > 5 then, due to the algebraic independence of all the r,s-row-weights where r and s are not both divisible by p, we can transform the valence $(\vec{1}_{p-3}^T,2)$ into (2) via taking the prolonged 1,1-row-weights of those "differentiated" left denominator-values equal to a formal variable (while taking all the other prolonged row-weights equal to zero) and calculating the coefficient at its power of the maximal degree. Then, via the limit technique, we can polynomial-time receive (for an arbitrary subset of the left denominator-value spectrum) the valence $(2\vec{1}_{p-1}^{T}, \dots, q\vec{1}_{p-1}^{T})$, q = 2,…,p, (via taking infinitely close denominator-value families of sizes (q-1)(p-1)) which, in turn, can correspondingly polynomial-time generate the valences (2), (3), …, (5) (via taking, for the valence (q), the prolonged 2,(q+1)-weight equal to a formal variable and all the other prolonged row-weights equal to zero). Each of the above-mentioned steps is provided by one polynomial-time reduction via calculating either the limit on a new infinitesimal of the expression multiplied by an appropriate power of the infinitesimal (i.e. via using the limit technique) or the maximal degree power's coefficient of a new formal variable. Eventually, taking into account the non-differentiated denominator-values whose valences are $(\vec{1}_{4}^{T})$ which we can turn into $(1)$ via taking the corresponding prolonged (p-2),(p-2)-row-weights equal to a formal variable, we can polynomial-time compute, for any left denominator-value spectrum x = $\{x_1, \dots, x_m\}$, the expression $per(\begin{pmatrix} C^{\star\gamma_1}(x_1, z) \\ \dots \\ C^{\star\gamma_m}(x_m, z) \end{pmatrix} Diag(\lambda))$ for any valences whose elements are taken from the set {1,…,p}.

Let's consider its partial case when the denominator-value spectrum-vector x can be partitioned into subvectors $y^{(1)}, \dots, y^{(p)}$ of valences (1),…,(p) correspondingly and y of valence (p) such that for j = 1,…,dim(y) the 1,p-row-weight of $y_j$ is 1 and its 2,s-row-weight is $\lambda_j^{(s)}\omega$, s = 1,…,p (where ω is another formal variable), while all the other row-weights of all the left denominator-values are zeros (hence $y^{(1)}, \dots, y^{(p)}$ have all their row-weights equal to zero). Then, upon calculating the coefficient at ω's power of the minimal degree, we'll polynomial-time get the permanent $per\begin{pmatrix} C(y^{(1)}, y) Diag(\lambda^{(1)}) \\ \dots \\ C(y^{(p)}, y) Diag(\lambda^{(p)}) \end{pmatrix}$ which, in turn, generates (upon taking infinitely close families of left denominator-values of size p) $per\begin{pmatrix} C(y^{(1)}, y) Diag(\lambda^{(1)}) \\ \dots \\ C(y^{(p)}, y) Diag(\lambda^{(p)}) \end{pmatrix}^{\star(\vec{1}_{p-1}^{T}, 2)}$. If we take $y^{(1)} = \dots = y^{(p)} = v$ then we'll obtain $per\begin{pmatrix} C(v, y) Diag(\lambda^{(1)}) \\ \dots \\ C(v, y) Diag(\lambda^{(p)}) \end{pmatrix}^{\star(\vec{1}_{p-1}^{T}, 2)}$ that is a polynomial in the sums $\sum_{j=1}^{\dim(y)}\frac{(\lambda_j^{(1)})^{r_1}\ldots(\lambda_j^{(p)})^{r_p}}{(v_i-y_j)^s}$ (let's call such a sum, by analogy, the $\boldsymbol{r_1,\ldots,r_p}$, ***s-row-weight*** of $v_i$, while correspondingly defining the prolonged $r_1,\ldots,r_p$, s-row-weight). We can also analogically notice that the set of row-weights such that not all of the numbers $r_1,\ldots,r_p$, s are multiples of p is an algebraically independent system of functions. And, at last, upon taking, for a formal variable ω, the $v_i$-th prolonged 1,…,1,2p-row-weight equal to $\alpha_i\omega^p$ and all the prolonged $r_1,\ldots,r_p,(p+1)$ −row-weights where one of the numbers $r_1,\ldots,r_p$ is p and all the others are zeros equal to ω (while all the other prolonged row-weights are to be taken equal to zero), we'll obtain, after calculating the coefficient at ω's power of the maximal degree, the sum

$$\sum_I\left(\prod_{i\in I}\alpha_i\right)\operatorname{per}\begin{pmatrix}C(v_I,y)\mathrm{Diag}(\lambda^{(1)})\\ \ldots\\ C(v_I,y)\mathrm{Diag}(\lambda^{(p)})\end{pmatrix}^{\star(\vec{1}^T_{p-1})}.$$

For any prime characteristic, by the limit technique this sum yields, via taking pairs of infinitely close denominator-values and summation-weights of infinitesimal-order -2 (opposite for each pair), the partial derivative

$$\left(\prod_{t=1}^{\dim(v)}\frac{\partial}{\partial v_t}\right)\operatorname{per}\begin{pmatrix}C(v,y)\mathrm{Diag}(\lambda^{(1)})\\ \ldots\\ C(v,y)\mathrm{Diag}(\lambda^{(p)})\end{pmatrix}^{\star(\vec{1}^T_{p-1})}$$

which is equal (due to Lemma III.4, and the next passage is due to it too) to the expression

$$\sum_{J,|J|=p(p-1)\dim(v)}\operatorname{per}\begin{pmatrix}C(v,y_J)\mathrm{Diag}(\lambda^{(1)})\\ \ldots\\ C(v,y_J)\mathrm{Diag}(\lambda^{(p)})\end{pmatrix}^{\star(\vec{1}^T_{p-1})}\prod_{i=1}^{\dim(v)}\sum_{j\in J}\frac{1}{v_i-y_j}=$$

$$=\frac{\sum_{J,|J|=p(p-1)\dim(v)}\operatorname{per}\begin{pmatrix}C(v^{(1)},y_J)\mathrm{Diag}\left(\lambda^{(1)}\star\left\{\frac{\mathrm{pol}(v^{(1)},y_j)}{\mathrm{pol}(v,y_j)}\right\}_{\dim(y)}\right)\\ \ldots\\ C(v^{(p)},y_J)\mathrm{Diag}\left(\lambda^{(p)}\star\left\{\frac{\mathrm{pol}(v^{(p)},y_j)}{\mathrm{pol}(v,y_j)}\right\}_{\dim(y)}\right)\end{pmatrix}^{\star(\vec{1}^T_{p-1})}\prod_{i=1}^{\dim(v)}\sum_{j\in J}\frac{1}{v_i-y_j}}{\left(\prod_{q=1}^{p}\det^{p-1}\left(\mathrm{Van}(v^{(q)})\right)\right)/\det^{p(p-1)}\left(\mathrm{Van}(v)\right)}$$

$$=\sum_{J,|J|=p(p-1)\dim(v)}\operatorname{per}\begin{pmatrix}C(v^{(1)},y_J)\mathrm{Diag}(\hat{\lambda}^{(1)})\\ \ldots\\ C(v^{(p)},y_J)\mathrm{Diag}(\hat{\lambda}^{(p)})\end{pmatrix}^{\star(\vec{1}^T_{p-1})}\prod_{i=1}^{\dim(v)}\sum_{j\in J}\frac{1}{v_i-y_j}$$

(for arbitrary dim(v)-vectors $v^{(1)},\ldots,v^{(p)}$ and, because of the arbitrariness of $\lambda^{(1)},\ldots,\lambda^{(p)}$, arbitrary $\hat{\lambda}^{(1)},\ldots,\hat{\lambda}^{(p)}$). The latter expression is easy to turn, via taking, for an infinitesimal $\varepsilon$, y of the generic form $\begin{pmatrix} \mathrm{w} \\ \mathrm{v}^{(1)}_{\backslash\{1,\dots,\dim(\mathrm{u})\}} \otimes \vec{1}_{\mathrm{p}-1} \\ \mathrm{v}^{(2)} \otimes \vec{1}_{\mathrm{p}-1} \\ \dots \\ \mathrm{v}^{(\mathrm{p})} \otimes \vec{1}_{\mathrm{p}-1} \end{pmatrix} + \mathrm{O}(\varepsilon)$ (with no other indeterminates involving $\varepsilon$ ) and applying the limit technique, into

$$\sum_{J,|J|=(p-1)\dim(u)} \mathrm{per}\big(\mathrm{C}(\mathrm{u},\mathrm{w}_J)\mathrm{Diag}(\alpha)\big)^{\star(\vec{1}^{T}_{p-1})} \prod_{i=1}^{\dim(v)} \sum_{j\in J} \frac{1}{\mathrm{v}_i - \mathrm{w}_j}$$

where $\mathrm{u} = \mathrm{v}^{(1)}_{\{1,\dots,\dim(u)\}}$ , $\alpha$ is the first dim(w) entries of $\hat{\lambda}^{(1)}$,

$\sum_{q=1}^{p}\sum_{j=1}^{\dim(v)}\frac{1}{\mathrm{v}_i - \mathrm{v}^{(q)}_j} - \sum_{k=1}^{\dim(u)}\frac{1}{\mathrm{v}_i - \mathrm{u}_k} = 0$ for i = 1,…,dim(v).

Upon putting $\mathrm{w} = \hat{\mathrm{w}} \otimes \vec{1}_{\mathrm{p}-1}$ and $\alpha = \hat{\alpha} \otimes \begin{pmatrix} 1 \\ \dots \\ p-1 \end{pmatrix}$, the latter expression eventually turns into $\varphi_{1,p-1}\big(\mathrm{v},\mathrm{u},\hat{\mathrm{w}},\hat{\alpha}^{\star(p-1)}\big)$ (for arbitrary $\mathrm{v},\mathrm{u},\hat{\mathrm{w}},\hat{\alpha}$) what completes the proof due to Theorem III.15 regarding the Cauchy determinant base-sum's $\#_{\mathrm{p}}\mathrm{P}$-completeness for the Cauchy base-degree 1.

We can also add that all the above proof's polynomial-time reductions from $\mathrm{per}\begin{pmatrix} \mathrm{C}(\mathrm{y}^{(1)},\mathrm{y})\mathrm{Diag}(\lambda^{(1)}) \\ \dots \\ \mathrm{C}(\mathrm{y}^{(\mathrm{p})},\mathrm{y})\mathrm{Diag}(\lambda^{(\mathrm{p})}) \end{pmatrix}$ to the very end remain valid in characteristics 3 and 5 as well what makes this permanent $\#_{\mathrm{p}}\mathrm{P}$-complete for any odd prime. This fact is equivalent to the $\#_{\mathrm{p}}\mathrm{P}$-completeness of the permanent of a rectangular matrix of Cauchy-rank p in any odd prime characteristic p.

Hence we've shown that the permanent of a "column-weighted" rectangular Cauchy matrix is polynomial-time computable in characteristic 3 and $\#_{\mathrm{p}}\mathrm{P}$-complete for any prime p > 5. In the case of all the column-weights equal to unity (i.e. of a "non-column-weighted" rectangular Cauchy matrix) this permanent is polynomial-time computable in any characteristic according to Lemma III.1. The question also arises whether in characteristic 3 there is a likewise polynomial-time manipulation with denominator-values' grouping and row-weights that generates the valence (2).

Besides, as it was said earlier, Theorem III.12 provides the polynomial-time computability of $\mathrm{per}\left(C^{\star 2}(x,y)\mathrm{Diag}(\lambda)\right)$ in characteristic 5. In this regard, let's prove the following theorem:

**Theorem III.32.**

$\mathrm{per}\left(C^{\star 2}(x,y)\mathrm{Diag}(\lambda)\right)$ is $\#_5$P-complete.

Proof:

The expression $\mathrm{per}\left(C^{\star 2}(x,y)\mathrm{Diag}(\lambda)\right)$ polynomial-time generates $\mathrm{per}\begin{pmatrix} C(y^{(1)},y)\mathrm{Diag}(\lambda^{(1)}) \\ \dots \\ C(y^{(4)},y)\mathrm{Diag}(\lambda^{(4)}) \end{pmatrix}$ via a process analogical to the one described in the above proof of Theorem III.31: first we receive, likewise,

$$\mathrm{per}\left(\begin{pmatrix} C^{\star 2}\left(y^{(1)},z\right) \\ C^{\star 3}\left(y^{(2)},z\right) \\ C^{\star 4}\left(y^{(3)},z\right) \\ C^{\star 5}\left(y^{(4)},z\right) \\ C^{\star 5}(y,z) \end{pmatrix}\mathrm{Diag}(\lambda)\right) = \mathrm{coef}_{\omega^d}\mathrm{per}\left(C\begin{pmatrix} C^{\star 2}\left(y^{(1)},\check{z}\right) \\ C^{\star\left(2\vec{1}_4^T,3\right)}\left(y^{(2)},\check{z}\right) \\ C^{\star\left(2\vec{1}_4^T,3\vec{1}_4^T,4\right)}\left(y^{(3)},\check{z}\right) \\ C^{\star\left(2\vec{1}_4^T,3\vec{1}_4^T,4\vec{1}_4^T,5\right)}\left(y^{(4)},\check{z}\right) \\ C^{\star\left(2\vec{1}_4^T,3\vec{1}_4^T,4\vec{1}_4^T,5\right)}(y,\check{z}) \end{pmatrix}\mathrm{Diag}\left(\check{\lambda}\right)\right) =$$

$$= \mathrm{coef}_{\omega^d}\lim_{\varepsilon\to 0}\frac{\mathrm{per}\left(C^{\star 2}\left(\begin{pmatrix} y^{(1)} \\ y^{(2)}\ddot{+}\varepsilon\alpha \\ y^{(3)}\dddot{+}\varepsilon\beta \\ y^{(4)}\dddot{+}\varepsilon\gamma \\ y\dddot{+}\varepsilon\gamma \end{pmatrix},\check{z}\right)\mathrm{Diag}\left(\check{\lambda}\right)\right)}{\mathrm{per}^{n_2}\left((\varepsilon\alpha^T)^{\star\left(0\vec{1}_4^T,1\right)}\right)\mathrm{per}^{n_3}\left((\varepsilon\beta^T)^{\star\left(0\vec{1}_4^T,\vec{1}_4^T,2\right)}\right)\mathrm{per}^{n_3+m}\left((\varepsilon\gamma^T)^{\star\left(0\vec{1}_4^T,\vec{1}_4^T,2\vec{1}_4^T,3\right)}\right)}$$

where $\alpha,\beta,\gamma$ are arbitrary 5-, 9-, 14-vectors correspondingly, $\check{z} = \binom{z}{\hat{z}}$, $\check{\lambda} = \binom{\lambda}{\hat{\lambda}}$, $d = 4n_2+8n_3+16n_4+16m$, for q =1,2,3,4 $\sum_{j=1}^{\dim(\hat{z})}\frac{\hat{\lambda}_j^r}{(y_i^{(q)}-\hat{z}_j)^s} = \begin{cases} \omega, r=2, s=3+q \\ 0, \text{else} \end{cases}$ for $i=1,\dots,n_q$, $\dim(y^{(q)}) = n_q$,

$\sum_{j=1}^{\dim(\hat{z})}\frac{\hat{\lambda}_j^r}{(y_k-\hat{z}_j)^s} = \begin{cases} \omega \text{ if } r=2, s=3+q \\ 0, \text{else} \end{cases}$ for k = 1 ,…,m, dim(y) = m;

and then we get

$$\mathrm{per}\begin{pmatrix} C(y^{(1)},y)\mathrm{Diag}(\lambda^{(1)}) \\ \dots \\ C(y^{(4)},y)\mathrm{Diag}(\lambda^{(4)}) \end{pmatrix} = \mathrm{coef}_{\omega^{m-\sum_{q=1}^{4} n_q}} \mathrm{per}(\begin{pmatrix} C^{\star 2}(y^{(1)},z) \\ C^{\star 3}(y^{(2)},z) \\ C^{\star 4}(y^{(3)},z) \\ C^{\star 5}(y^{(4)},z) \\ C^{\star 5}(y,z) \end{pmatrix} \mathrm{Diag}(\lambda))$$

where

for q = 1,2,3,4 $\sum_{j=1}^{\dim(z)} \frac{\lambda^r}{(y_i^{(q)}-z_j)^s} = 0$ for $i = 1,\dots,n_q$,

$$\sum_{j=1}^{\dim(z)} \frac{\lambda^r}{(y_k-z_j)^s} = \left[\begin{matrix} \lambda_k^{(q)} \text{ if } r=2,\ s=q \\ \omega \text{ if } r=1,\ s=5 \\ 0\text{, else} \end{matrix}\right. \quad \text{for } k = 1,\dots,m$$

The expression $\mathrm{per}\begin{pmatrix} C(y^{(1)},y)\mathrm{Diag}(\lambda^{(1)}) \\ \dots \\ C(y^{(4)},y)\mathrm{Diag}(\lambda^{(4)}) \end{pmatrix}$ generates, also by the technique shown in Theorem III.31's proof while replacing p by p-1 (i.e. through considering its partial case $\mathrm{per}\begin{pmatrix} C(v,y)\mathrm{Diag}(\lambda^{(1)}) \\ \dots \\ C(v,y)\mathrm{Diag}(\lambda^{(4)}) \end{pmatrix}^{\star(1,1,1,1,2)} = \lim_{\varepsilon\to 0} \frac{\mathrm{per}\begin{pmatrix} C(v\ddot{\mp}\beta,y)\mathrm{Diag}(\lambda^{(1)}) \\ \dots \\ C(v\ddot{\mp}\beta,y)\mathrm{Diag}(\lambda^{(4)}) \end{pmatrix}}{\mathrm{per}^{4\dim(v)}((\varepsilon\beta^T)^{\star(0,0,0,0,1)})}$, where $\beta$ is an arbitrary 5-vector, and putting all its 5,0,0,0,6-, 0,5,0,0,6-, 0,0,5,0,6- and 0,0,0,5,6-row-weights equal to $\omega$ and, for i = 1,…,dim(v), its i-th 1,1,1,1,8-row-weight equal to $\frac{\alpha_i}{-3!}\omega^4$ for computing the coefficient at $\omega^{4\dim(v)}$, where $\omega$ is a formal variable), the sum $\sum_I (\prod_{i\in I} \alpha_i)\, \mathrm{per}\begin{pmatrix} C(v_I,y)\mathrm{Diag}(\lambda^{(1)}) \\ \dots \\ C(v_I,y)\mathrm{Diag}(\lambda^{(4)}) \end{pmatrix}^{\star(\vec{1}_4^T)}$. This sum generates, via its denominator-values' grouping into infinitesimal-close pairs and applying the limit technique, the sum $\sum_I (\prod_{t\in I} \frac{\alpha_i\,\partial}{\partial v_t})\, \mathrm{per}\begin{pmatrix} C(v_I,y)\mathrm{Diag}(\lambda^{(1)}) \\ \dots \\ C(v_I,y)\mathrm{Diag}(\lambda^{(4)}) \end{pmatrix}^{\star(\vec{1}_4^T)}$ where we can put all the prolonged 1,1,1,1,5-row-weights equal to a formal variable for computing its maximal degree power and receive $\sum_I (\prod_{t\in I} \alpha_i)\, \mathrm{per}\begin{pmatrix} C(v_I,y)\mathrm{Diag}(\lambda^{(1)}) \\ \dots \\ C(v_I,y)\mathrm{Diag}(\lambda^{(4)}) \end{pmatrix}^{\star(\vec{1}_3^T)}$. This sum, in turn, generates (also via its denominator-values' infinitesimal-close pairing and applying the limit technique, like it was done for the valence $\vec{1}_4^T$) the partial derivative $(\prod_{t=1}^{\dim(v)} \frac{\partial}{\partial v_t})\mathrm{per}\begin{pmatrix} C(v,y)\mathrm{Diag}(\lambda^{(1)}) \\ \dots \\ C(v,y)\mathrm{Diag}(\lambda^{(4)}) \end{pmatrix}^{\star(\vec{1}_3^T)}$. The latter expression is the sum of Cauchy-like permanents where for each of the denominator-values its ***multiple valence*** (for $\lambda^{(1)},\lambda^{(2)},\lambda^{(3)},\lambda^{(4)}$) is one of the following four:

((1,1,2),(1,1,1),(1,1,1),(1,1,1)),

((1,1,1),(1,1,2),(1,1,1),(1,1,1)),

((1,1,1),(1,1,1),(1,1,2),(1,1,1)),

((1,1,1),(1,1,1),(1,1,1),(1,1,2)),

what allows to receive the multiple valence $((1,1,2),\emptyset,\emptyset,\emptyset)$ (via taking, as a formal variable, the prolonged 0,1,0,0,1-, 0,0,1,0,1- and 0,0,0,1,1-row-weights and computing the coefficient at its maximal degree power) that, in turn, generates either $((1),\emptyset,\emptyset,\emptyset)$ or $((2),\emptyset,\emptyset,\emptyset)$ via taking, as a formal variable, either the prolonged 2,0,0,0,3-row-weight or the prolonged 1,0,0,0,1-row-weight correspondingly (while choosing one of these two options for each denominator-value) and computing the coefficient at its maximal degree power. Thus we obtain the expression $\mathrm{per}\begin{pmatrix} C^{\star\gamma_1}(v_1,y)\mathrm{Diag}(\lambda^{(1)}) \\ \dots \\ C^{\star\gamma_m}(v_m,y)\mathrm{Diag}(\lambda^{(4)}) \end{pmatrix}$ for the left denominator-value spectrum $\{v_1,\dots,v_m\}$ whose valences are either (1) or (2) and this expression polynomial-time generates, according to the scheme of Theorem III.31's proof we already referred earlier in the present proof, $\mathrm{per}\begin{pmatrix} C(y^{(1)},y)\mathrm{Diag}(\lambda^{(1)}) \\ \dots \\ C(y^{(p)},y)\mathrm{Diag}(\lambda^{(5)}) \end{pmatrix}$ that is, as it also was shown in the referred proof, $\#_5$P-complete.

We hence can conclude that the above theorem implies once more, independently of Theorem III.14, the $\#_5$P-completeness of the trace-determinant $\det_{tr}\left(\tilde{C}(x) + \mathrm{Diag}(g), \{G^{(i)}M\}_n\right)$ (where $M = \begin{pmatrix} 0_{2\times 2} & \begin{matrix} 0 & 1 \\ -1 & 0 \end{matrix} \\ \begin{matrix} 0 & 1 \\ -1 & 0 \end{matrix} & 0_{2\times 2} \end{pmatrix}$) for arbitrary symmetric $G^{(1)}, \dots, G^{(n)}$ and absence-weights and, eventually, the permanent's polynomial-time computability in characteristic 5.

**Definition:**

Let $x_1, \dots, x_n$ be independent variables, $A = A(x) = \{a_{i,j}(x_i, x_j)\}_{n\times n}$ be an n×n-matrix such that for k = 1,…,n its k-th row and column are functions in the variable $x_k$ and for k = 1,…,n $\alpha_k = \{\alpha_{k,u}\}_{\dim(\alpha_k)}$, $\beta_k = \{\beta_{k,v}\}_{\dim(\beta_k)}$ be non-decreasing sequences (optionally empty) of natural numbers of lengths $\dim(\alpha_k)$ and $\dim(\beta_k)$ correspondingly.

Then we define the ***α, β-valence power*** of $A$ as the $(\sum_{i=1}^{n} \dim(\alpha_i)) \times (\sum_{j=1}^{n} \dim(\beta_j))$-matrix $A^{\langle\alpha,\beta\rangle} :=$

$$\{\{\frac{1}{(\alpha_{i,u}-1)!(\beta_{j,v}-1)!} \frac{\partial^{\alpha_{i,u}-1}}{\partial x_i^{\alpha_{i,u}-1}} \frac{\partial^{\beta_{j,v}-1}}{\partial x_j^{\beta_{j,v}-1}} a_{i,j}(x_i, x_j)\}_{\substack{u=1,\dots,\dim(\alpha_i) \\ v=1,\dots,\dim(\beta_j)}}\}_{\substack{i=1,\dots,n \\ j=1,\dots,n}}$$

and we'll call $\alpha_k$ and $\beta_k$ the row and column (or left and right) valences of $x_k$ (or just the k-th row and column valences) correspondingly, while the pair $val(x_k) := (\alpha_k, \beta_k)$ will be called the valence of $x_k$, the vectors $\alpha = \{\alpha_k\}_n$ and $\beta = \{\beta_k\}_n$ – the left and right valence-vectors correspondingly, and the vector $\{(\alpha_k, \beta_k)\}_n$ – the valence-vector.

Let's notice that the valences introduced in the proofs of Theorem III.31 and III.32 are in fact such two-sided valences with empty right sides, i.e. actually **left-sided** valences, and, accordingly, $\begin{pmatrix} C^{\star\gamma_1}(x_1, z) \\ \dots \\ C^{\star\gamma_m}(x_m, z) \end{pmatrix} = C^{\langle\gamma,\vec{\emptyset}_m\rangle}(x, z)$. Moreover, this definition also implies that all the Cauchy-like matrices we considered in the present article's chapter III are either the $\alpha, \beta$-valence powers of a Cauchy-wave matrix for some left and right valence-vectors $\alpha, \beta$ or can be expressed through them via the operations of vector-composition and the left- and right-multiplication by diagonal matrices.

**Definition:**

Let for k = 1,…,n $\alpha_k^{(1)}, \dots, \alpha_k^{(t_k)}, \beta_k^{(1)}, \dots, \beta_k^{(t_k)}$ be non-decreasing sequences of natural numbers, $d_k^{(1)}, \dots, d_k^{(t_k)}$ be elements of the ground field and $A = A(x) = \{a_{i,j}(x_i, x_j)\}_{n\times n}$ be an n×n-matrix.

Then we define for variables $x_1, \dots, x_n$:

the formal sum $\mathfrak{v}_k := \mathrm{valsum}(x_k) := \sum_{r_k=1}^{t_k} d_k^{(r_k)}(\alpha_k^{(r_k)}, \beta_k^{(r_k)})$ as the **valence-sum** of $x_k$ (or the k-th valence-sum) where $d_k^{(r_k)}$ will be called the **summation-weight** of the valence $(\alpha_k^{(r_k)}, \beta_k^{(r_k)})$

and

the expression $\mathrm{per}_{\mathfrak{v}}(A) := \mathrm{per}_{\mathfrak{v}_1,\dots,\mathfrak{v}_n}(A) :=$

$$:= \sum_{r_1=1}^{t_1} \dots \sum_{r_n=1}^{t_n} d_1^{(r_1)} \dots d_n^{(r_n)} \delta\left(\sum_{k=1}^{n} \left(\dim(\alpha_k^{(r_k)}) - \dim(\beta_k^{(r_k)})\right)\right) \mathrm{per}\left(A^{\langle\{\alpha_k^{(r_k)}\}_n, \{\beta_k^{(r_k)}\}_n\rangle}\right)$$

where for a real number m $\delta(m) = \begin{cases} 0, m \neq 0 \\ 1, m = 0 \end{cases}$ (and hence the summation is over all the possible n-tuples $r_1, \dots, r_n$ making the matrix $A^{\langle\{\alpha_k^{(r_k)}\}_n, \{\beta_k^{(r_k)}\}_n\rangle}$ square) as the **$\mathfrak{v}$-sum permanent** of A (or the **valence-sum permanent** of A on the vector $\mathfrak{v}$ that we'll call the valence-sum vector of this permanent).

According to the two latter definitions, we can state that $\mathrm{per}_{\mathfrak{v}}(\tilde{C}(\binom{y}{z}))$ is a polynomial in the row-weights $\sum_{j=1}^{\dim(z)} \frac{\lambda_j^r}{(y_i - z_j)^s}$ if for j = 1,…,dim(z) $\mathrm{valsum}(z_j) = \lambda_j(\emptyset, (1)) + (\emptyset, \emptyset)$. This fact provides the opportunity of taking those row-weights (due to their algebraic independence in characteristic p for all the pairs r,s that are not both divisible by p) equal to polynomials in a formal variable and receiving, upon computing the formal variable's maximal degree power's coefficient of $\mathrm{per}_{\mathfrak{v}}(\tilde{C}(\binom{y}{z}))$ (as this permanent would become, in such a case, a polynomial in this formal variable as well), $\mathrm{per}_{\hat{\mathfrak{v}}}(\tilde{C}(y))$ where the new valence-sum vector $\hat{\mathfrak{v}}$ is obtained from the subvector of $\mathfrak{v}$ corresponding to y via transforming, for i = 1,…,dim(y), the old valence-sum of $y_i$ in the accordance with the polynomials in the formal variable the corresponding (i.e. having the denominator-value $y_i$) row-weights are equal to. We can call such a transformation ***a prolongation derivative*** and its partial cases were actually applied in the proofs of Theorems III.31 and III.32 for left-sided (having only empty right parts in all their valences) valence-sums. Taking into account the fact that $\mathrm{per}_{\mathfrak{v}}(A) = \mathrm{per}_{\mathfrak{v}^T}(A^T)$ where $\mathfrak{v}^T$ denotes the valence-sum vector received from $\mathfrak{v}$ via permuting the left and right valences in each involved valence and for a skew-symmetric A $\mathrm{per}_{\mathfrak{v}}(A) = \mathrm{per}_{\mathfrak{v}^*}(A)$ where $\mathfrak{v}^*$ is received from $\mathfrak{v}^T$ via multiplying each involved valence's summation-weight by -1 if its left valence's length is odd, we get an option to apply a prolongation derivative to any valence-sum on "both its sides".

Besides, grouping the denominator-values (i.e. the variables) into infinitesimal-close families generates, accordingly, a contraction of a family of valence-sums into a new valence-sum analogical to the contraction occurring in families of matrix-weights for the trace-determinant of a Cauchy-wave matrix via the open trace-determinant compression operator that was discussed earlier. Let's call such a contraction an infinitesimal-close contraction and we can also notice that quite a number of partial cases of this contraction were applied in the proofs of Theorems III.31 and III.32.

If we speak about characteristic 3, we can notice that the above-proven polynomial-time computability of $\mathrm{per}(\mathrm{C}(\mathrm{x},\mathrm{z})\mathrm{Diag}(\lambda))$ allows, in this characteristic, to polynomial-time compute the valence-sum permanent of a Cauchy-wave matrix where the valence-sum vector can contain any singleton (i.e. having just one valence in its sum) left-sided valence-sum $((1,1,2,2,\ldots,\mathrm{k},\mathrm{k}),\emptyset)$ for any natural k that is able to generate, via its various prolongation-derivatives, quiet a complex variety of valence-sums including a number of two-sided and non-singleton ones. However, as it was shown in Theorem III.31's proof, computing the valence-sum permanent of a Cauchy-wave matrix on a valence-sum vector whose entries are taken from a set including the singleton valence-sums $((1),\emptyset)$ and $((2),\emptyset)$ is $\#_3$-P-complete and this fact arises the question of determining the closure of the set of singleton left-sided valence-sums $\{((1,1,2,2,\ldots,\mathrm{k},\mathrm{k}),\emptyset),\mathrm{k}\in\mathrm{N}\}$ by all the existing prolongation-derivative and infinitesimal-close contraction operators.